\documentclass[12pt]{article}
\pdfoutput=1
\usepackage{jheppub}
\usepackage[skip=2pt,font=footnotesize]{caption}
\usepackage[utf8]{inputenc}
\usepackage{verbatim}
\usepackage{amsmath}
\usepackage{amssymb}
\usepackage{amsthm}
\usepackage{slashed}
\usepackage{amsfonts}
\usepackage{dsdshorthand}
\usepackage{makecell}
\usepackage{tabularx}
\usepackage{comment}
\usepackage{subfigure}
\usepackage{float}
\usepackage{color}

\begin{document}

\hbox{{CALT-TH-2020-005}}

\title{The Lorentzian inversion formula and the spectrum of the 3d O(2) CFT}
\author[a,b]{Junyu Liu,}
\author[a]{David Meltzer,}
\author[c]{David Poland,}
\author[a]{David Simmons-Duffin}

\affiliation[a]{Walter Burke Institute for Theoretical Physics, California Institute of Technology,\\ Pasadena, California 91125, USA}
\affiliation[b]{Institute for Quantum Information and Matter, California Institute of Technology,\\ Pasadena, California 91125, USA}
\affiliation[c]{Department of Physics, Yale University,\\ New Haven, CT 06520, USA}

\abstract{We study the spectrum and OPE coefficients of the three-dimensional critical O(2) model, using four-point functions of the leading scalars with charges 0, 1, and 2 ($s$, $\phi$, and $t$). We obtain numerical predictions for low-twist OPE data in several charge sectors using the extremal functional method. We compare the results to analytical estimates using the Lorentzian inversion formula and a small amount of numerical input. We find agreement between the analytic and numerical predictions. We also give evidence that certain scalar operators lie on double-twist Regge trajectories and obtain estimates for the leading Regge intercepts of the O(2) model.
}
\maketitle
\section{Large-scale bootstrap: analytics and numerics}

The conformal bootstrap is a powerful tool for exploring the space of conformal field theories. It can reveal universal properties of all allowed unitary CFTs, and it can predict some conformal data with high precision; see~\cite{Poland:2016chs,Poland:2018epd} for recent reviews. However, we still possess limited knowledge of the precise spectrum of essentially all nontrivial CFTs, particularly in $d>2$ dimensions.

Recently, improved efficiency of the semidefinite program solver \texttt{SDPB} \cite{Simmons-Duffin:2015qma,Landry:2019qug} has enabled new numerical bootstrap investigations. Together with a novel algorithm for scanning through OPE space \cite{Chester:2019ifh}, we have new tools for computing high-precision OPE data involving large systems of crossing equations. In \cite{Chester:2019ifh}, a subset of the present authors applied these tools to obtain new results for critical exponents of the 3d $\text{O}(2)$ model.

At the same time, novel analytic bootstrap tools are emerging. One of the most powerful is the Lorentzian inversion formula \cite{Caron-Huot:2017vep}, which unifies and extends the lightcone bootstrap methods of \cite{Komargodski:2012ek,Fitzpatrick:2012yx,Alday:2015eya,Alday:2015ewa,Alday:2016njk,Simmons-Duffin:2016wlq}.\footnote{Other recent developments include conformal dispersion relations \cite{Carmi:2019cub} and analytic functionals \cite{Paulos:2019gtx,Mazac:2019shk}.} The Lorentzian inversion formula yields (among other results) interrelationships between the low-twist spectrum of a CFT, which leads to predictions for low-twist Regge trajectories. By supplementing these analytical predictions with a small amount of numerical input, one can often obtain excellent agreement with numerical data~\cite{Caron-Huot:2017vep,Cornagliotto:2017snu,Albayrak:2019gnz}.

The recent developments on both the numerical and analytic sides provide further opportunities for obtaining precise spectra of specific CFTs. In addition, one can perform tests of analytic methods using numerical data. In this work, we use the 3d O(2) CFT as a playground for the following investigations:
\begin{itemize}
\item We use the numerical data provided in \cite{Chester:2019ifh} and the extremal functional method~\cite{Poland:2010wg,ElShowk:2012hu,Simmons-Duffin:2016wlq} to obtain detailed numerical approximations for dimensions and OPE coefficients of low twist operators. Using allowed points of the $\Lambda=35$ computation given in \cite{Chester:2019ifh}, we compute upper and lower bounds on the magnitude of OPE coefficients involving the external operators. The resulting extremal functionals give a high-precision picture of the spectrum of the 3d O(2) CFT.

\item We describe how to apply the Lorentzian inversion formula to estimate low-twist OPE data. Performing the estimate correctly requires a nontrivial synthesis of methods in the literature, including the inversion of 3d conformal blocks, the use of ``twist Hamiltonians", and the incorporation of sums over double-twist operators.
 Although many of the required ideas have appeared in previous literature, especially in the context of the lightcone bootstrap \cite{Simmons-Duffin:2016wlq}, their application using the Lorentzian inversion formula is new.
 
\item We justify these approaches by comparing numerical and analytic results for low-twist data in the charge $0^{\pm},1,2,3$ sectors of the 3d O(2) CFT, finding good agreement.

For charge 4 operators of even spin, we find decent agreement for one of the low lying trajectories between analytics and numerics, despite the fact that we have relatively limited numerical data. We make additional predictions for the charge 4 odd sector, which is currently inaccessible in the numerics. Using this conformal data, we also provide an initial analysis of the Regge intercepts for the leading trajectories.

\item Some byproducts of our study include a more detailed analysis of the O(2) representation theory and crossing equations, some computations of the Mean Field Theory (MFT) OPE coefficients in different O(2) charge sectors, a study of different expansions for 3d conformal blocks, new results for sums of $\text{SL}_2$ blocks, and the introduction of a new concept, the \emph{sharing effect}, which must be overcome for precise spectrum extraction in the numerical bootstrap.
\end{itemize}

\begin{figure}
  \centering	
  \includegraphics[width=0.8\textwidth]{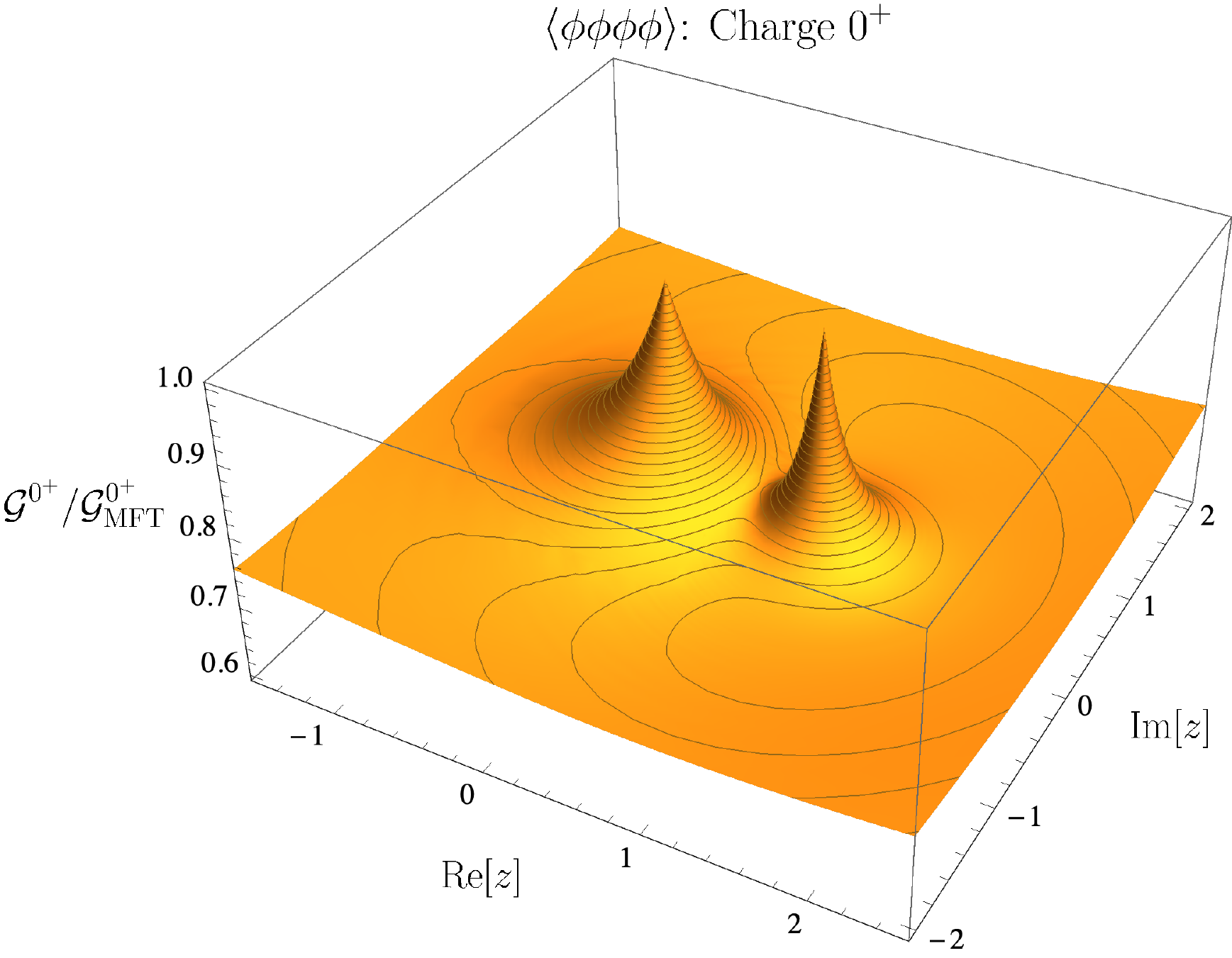}
  \caption{\label{fig:Euc4pt0p} We plot the Euclidean four-point function $\<\f\f\f\f\>$ projected onto $0^+$ exchange in the $s$-channel and normalized by the corresponding MFT four-point function. We include both the isolated and double-twist operators and expand to 5$^{\text{th}}$ order in dimensional reduction. The regions around $z=0$ and $z=1$ are computed using the $s$ and $t$-channel, respectively.
  }
\end{figure}

Using the data for the low-twist operators, we can also verify that crossing symmetry of $\<\f\f\f\f\>$ holds to a high precision when we only include operators of low twist in the block expansion. Using this data, we can also obtain an estimate for this four-point function in Euclidean configurations, that is $\bar{z}=z^{*}$, as shown in figure \ref{fig:Euc4pt0p}. Specifically, we project onto singlet, or charge $0^+$, exchange in the $s$-channel and normalize by the corresponding MFT correlator. We see that for generic Euclidean configurations, away from the $s$ or $t$-channel OPE limits, the full O(2) correlator has significant non-gaussianities.

This paper is organized as follows. 
In section \ref{numerics}, we present a numerical analysis for the 3d O(2) CFT using the extremal functional method. 
In section \ref{formalism}, we discuss our framework for spectrum computation using the Lorentzian inversion formula.
In section \ref{InvForO2}, we specialize the analytic techniques to the 3d O(2) CFT.
In section \ref{comparing}, we discuss the comparison between the low-lying numerical spectra and analytic predictions, and also make further predictions using the analytic bootstrap for quantities where the numerics are insufficient.
In section \ref{future}, we discuss future directions.

\section{Numerical computations in the O(2) model}\label{numerics}
\subsection{Bootstrap setup}
To begin, let us summarize our setup for bootstrapping the O(2) model, which is the same as the setup described in \cite{Chester:2019ifh}. We consider quantum field theories with conformal invariance and O(2) global symmetry. The group O(2) has two-dimensional irreducible representations labeled by an integer $q\in \mathbb{Z}_{\geq 1}$, together with one-dimensional irreducible representations $0^+$ and $0^-$. The integer label is the charge of the highest weight state with respect to the U(1) subgroup. In the charge-0 case, the superscripts $\pm$ denote parity under the $\mathbb{Z}_2$ reflection subgroup of O(2).  A more detailed introduction can be found in \cite{Chester:2019ifh}. We call the lowest Lorentz scalars with O(2) representations $0^+$, 1, 2, 3, 4 as $s,\phi,t,\chi,\tau$.

In the work \cite{Chester:2019ifh} (see also \cite{Go:2019lke}), we computed all possible crossing equations involving four-point functions of $s$, $\phi$ and $t$. (A review of this derivation, our conventions, and the crossing equations needed for analytic computations, can be found in appendix \ref{cross}.)

An allowed island for the scaling dimensions and OPE coefficients involving $\f$, $s$, and $t$ was obtained using \texttt{SDPB}~\cite{Simmons-Duffin:2015qma,Landry:2019qug} at derivative order $\Lambda = 27,35,43$. In this work, we will use the 20 allowed (primal) points given in table \ref{tab:the20points}, with the parameters and gap assumptions given in tables \ref{tab:extremalparameters} and \ref{tab:spectrumassumptions}, to compute both upper and lower bounds on the OPE coefficient $f_{\f\f s}$. The resulting extremal functionals are then used to extract the extremal spectra using method outlined in \cite{Poland:2010wg,ElShowk:2012hu,Simmons-Duffin:2016wlq} and the code \texttt{spectrum.py} \cite{Komargodski:2016auf,spectrum}.

In formulating the optimization problem, we impose gaps in the internal operator sectors, given in table \ref{tab:extremalparameters}. As we noted in \cite{Chester:2019ifh}, sometimes the solver \texttt{SDPB} expends unnecessary effort to find functionals that are positive for operators close to the unitarity bounds, causing a steady decay of \texttt{dualError} without vanishing. To avoid this, we set a small gap $\delta \Delta=10^{-4}$ in those sectors above the unitarity bound. In the work \cite{Chester:2019ifh}, we used the same points and setup to estimate the scalar dimensions in various charge sectors.
\begin{table}
\centering
\begin{tabular}{@{}c|c|c|c|c|c@{}}
\hline
$\De_\f$ & $\De_s$ & $\De_t$ & $\frac{f_{sss}}{f_{\f\f s}}$ & $\frac{f_{tts}}{f_{\f\f s}}$ & $\frac{f_{\f\f t}}{f_{\f\f s}}$ \\
\hline
0.519130434&1.51173444&1.23648971&1.20977354&1.82254374&1.76606470\\
0.519135171&1.51172427&1.23649356&1.20947477&1.82245370&1.76605159\\
0.519076518&1.51110487&1.23620503&1.20766586&1.82191247&1.76584197\\
0.519115548&1.51167580&1.23642873&1.21014420&1.82257643&1.76603227\\
0.519113909&1.51170936&1.23646025&1.21013097&1.82272756&1.76607582\\
0.519096732&1.51147972&1.23636344&1.20944426&1.82251617&1.76600087\\
0.519128801&1.51168098&1.23648846&1.20929738&1.82252856&1.76605495\\
0.519119255&1.51170685&1.23646324&1.21007964&1.82275976&1.76606055\\
0.519109342&1.51150256&1.23640031&1.20891847&1.82236481&1.76600112\\
0.519087647&1.51141667&1.23630721&1.20963450&1.82247476&1.76594440\\
0.519105802&1.51141826&1.23635621&1.20856734&1.82219520&1.76595563\\
0.519125142&1.51173460&1.23646472&1.21012577&1.82250871&1.76605236\\
0.519107610&1.51164424&1.23640715&1.21022297&1.82258938&1.76603036\\
0.519115226&1.51174173&1.23647414&1.21033291&1.82281805&1.76609054\\
0.519084390&1.51137895&1.23628833&1.20979136&1.82229748&1.76593252\\
0.519096529&1.51153244&1.23635748&1.20995866&1.82250999&1.76599060\\
0.519122718&1.51168123&1.23647847&1.20940108&1.82261368&1.76607344\\
0.519138689&1.51177044&1.23653770&1.20947377&1.82262309&1.76609008\\
0.519057668&1.51097950&1.23611240&1.20794762&1.82181966&1.76576836\\
0.519074424&1.51116298&1.23616082&1.20864157&1.82181577&1.76579563\\
\hline
\end{tabular}
\caption{\label{tab:the20points}Allowed points in the $\Lambda=35$ island used for the extremal functional method.}
\end{table}

\begin{table}
\begin{center}
\begin{tabular}{@{}c|c@{}}
	\hline
$\Lambda$ &  27 \\
{\small\texttt{keptPoleOrder}}&  12 \\
{\small\texttt{order}}&  60 \\
{\small\texttt{spins}} & $S_{27}$ \\
{\small\texttt{precision}} &  900 \\
{\small\texttt{dualityGapThreshold}} & $10^{-80}$ \\
{\small\texttt{primalErrorThreshold}}& $10^{-200}$ \\
{\small\texttt{dualErrorThreshold}} & $10^{-100}$\\ 
{\small\texttt{initialMatrixScalePrimal}} & $10^{20}$\\
{\small\texttt{initialMatrixScaleDual}} & $10^{20}$\\
{\small\texttt{feasibleCenteringParameter}} & 0.1\\
{\small\texttt{infeasibleCenteringParameter}} & 0.3\\
{\small\texttt{stepLengthReduction}} & 0.7\\
{\small\texttt{maxComplementarity}} & $10^{200}$\\
{\small\text{Threshold for \texttt{spectrum.py}}} & $10^{-30}$\\
 \hline
\end{tabular}
\caption{Parameters used for the computations of the extremal functional method, as defined in \cite{Simmons-Duffin:2015qma}. We define $S_{27}= \{0,\dots,31\}\cup \{49,50\}$.}
\label{tab:extremalparameters}
\end{center}
\end{table}

\begin{table}
\begin{center}
\begin{tabular}{@{}c|c|c@{}}
\hline
charge & spin & dimensions \\
\hline
0 & 0 & $\De_s$ or $\De\geq 3$ \\
1 & 0 & $\De_\f$ or $\De\geq 3$ \\
2 & 0 & $\De_t$ or $\De \geq 3$ \\
3 & 0 & $\De \geq 1$ \\
4 & 0 & $\De \geq 3$ ( $\De \geq \frac{1}{2}$ when predicting charge 4 scalars) \\
0 & 1 & $\De=2$ or $\De \geq 2+\de_\tau$ \\
0 & 2 & $\De=3$ or $\De\geq 3 + \de_\tau$ \\
$\cR$ & $\ell$ & $\De \geq \ell + 1 + \de_\tau$\\
\hline
\end{tabular}
\end{center}
\caption{\label{tab:spectrumassumptions}Assumptions about the spectrum of the O(2) model when doing the numerical bootstrap. $\cR,\ell$ represent any other choices of representation $\cR$ and spin $\ell$. In this work we set $\de_\tau=10^{-4}$.}
\end{table}

\subsection{Numerical spectrum}
Based on the above construction and 20 primal points, we run \texttt{SDPB} to construct 40 different spectra and their corresponding OPE coefficients. We follow the approach of \cite{Simmons-Duffin:2016wlq} and plot the spectra by identifying all operators in a given small interval and using the density of operators in that interval as an indicator of stability. We find that the spectra is nicely organized along double-twist trajectories, which are families of operators $[O_1 O_2]_{n}(\ell)$ whose twist, $\tau=\Delta-\ell$, asymptotically approaches $\tau_1+\tau_2+2n$ as $\ell\rightarrow\infty$~\cite{Fitzpatrick:2012yx,Komargodski:2012ek}. When double-twist operators can have different O(2) representations, corresponding to different irreducible factors in the tensor product of the O(2) representations of $O_1$ and $O_2$, we distinguish them with a superscript:
\be
\label{eq:superscriptnotation}
[O_1 O_2]^q_n(\ell): \qquad \parbox{4.5in}{operators on the trajectory with $\lim_{\ell\to\oo}\tau(\ell)= \tau_1+\tau_2+2n$, \\ with O(2) representation $q$.}
\ee

For the leading double-twist operators ($n=0$), the results are stable, and we notice remarkably clear numerical curves for all charge sectors. Thus, these trajectories provide a good playground for testing analytic methods against numerics. In section \ref{comparing} we will make a detailed comparison between the analytic and numerical predictions for these double-twist trajectories.

Before we start comparing numerical data with analytics, let us make the following observations about the spectrum:
\begin{itemize}
\item We can clearly identify $n=0$ double-twist families in each sector. However, for higher-twist families, $n\geq1$, we do not have enough accuracy in identifying clean curves. This is different from the 3d Ising model case, where one can identify at least one higher double-twist family relatively precisely. This could be improved at higher derivative order, but it might also be due to the more complicated nature of the O(2) model. For example, we may also need to include the leading charge 3 scalar $\chi$ as an external operator in the crossing equation in order to probe higher-twist towers built using $\chi$.

The data in the charge 4 even-spin sector is relatively limited. We believe this is partially due to the fact that our crossing relations do not contain $\chi$ as an external operator, so it is harder to probe charge 4 double-twist operators built using $\chi$. In addition, there are large mixing effects in this sector. Nevertheless, our numerics show reasonable agreement with the low-lying ${\left[ {\chi \phi } \right]_{n = 0}}$ trajectory. We will revisit this problem in section \ref{sec:ch4predictions}.

\item We have also noticed that numerical predictions for the leading scalar operators in each charge sector are quite stable. We can then make predictions for the dimensions and OPEs of these operators and compare them with existing Monte Carlo data. However, we have also noticed that the predictions for the low-lying operator dimensions might be affected by the gap we have imposed in the corresponding charge sector. If the CFT operator has a dimension very close to the gap we impose, there might be errors in the numerical predictions for the corresponding CFT data due to the presence of unphysical contributions at the gap dimension. We call this the {\it sharing effect}, and we will describe it in more detail below.
\end{itemize}

\subsection{Predictions for scalar CFT data and the sharing effect}
The leading charge 4 scalar operator, $\tau$, has a conformal dimension around $3.1$, with recent Monte Carlo computations giving the value $3.114(2)$ \cite{Shao:2019dbi}. In the charge 4 sector we have also imposed that all operators have dimension $\Delta\geq3$. So we see $\Delta_{\tau}$ is relatively close to the imposed gap. In the extremal functional method, we often find that the functional develops zeros around the gap we impose. These zeros are usually not physical, and we will refer them as fake zeros. Fake zeros will also come with nonzero OPE coefficients. Thus, we expect that numerically, there might be some unphysical ``sharing" of the contribution between the fake zeros and the true zeros near the gap. This effect can then make the predictions for the scaling dimension and OPE coefficients inaccurate.

\begin{figure}[t]
  \centering
  \includegraphics[width=0.8\textwidth]{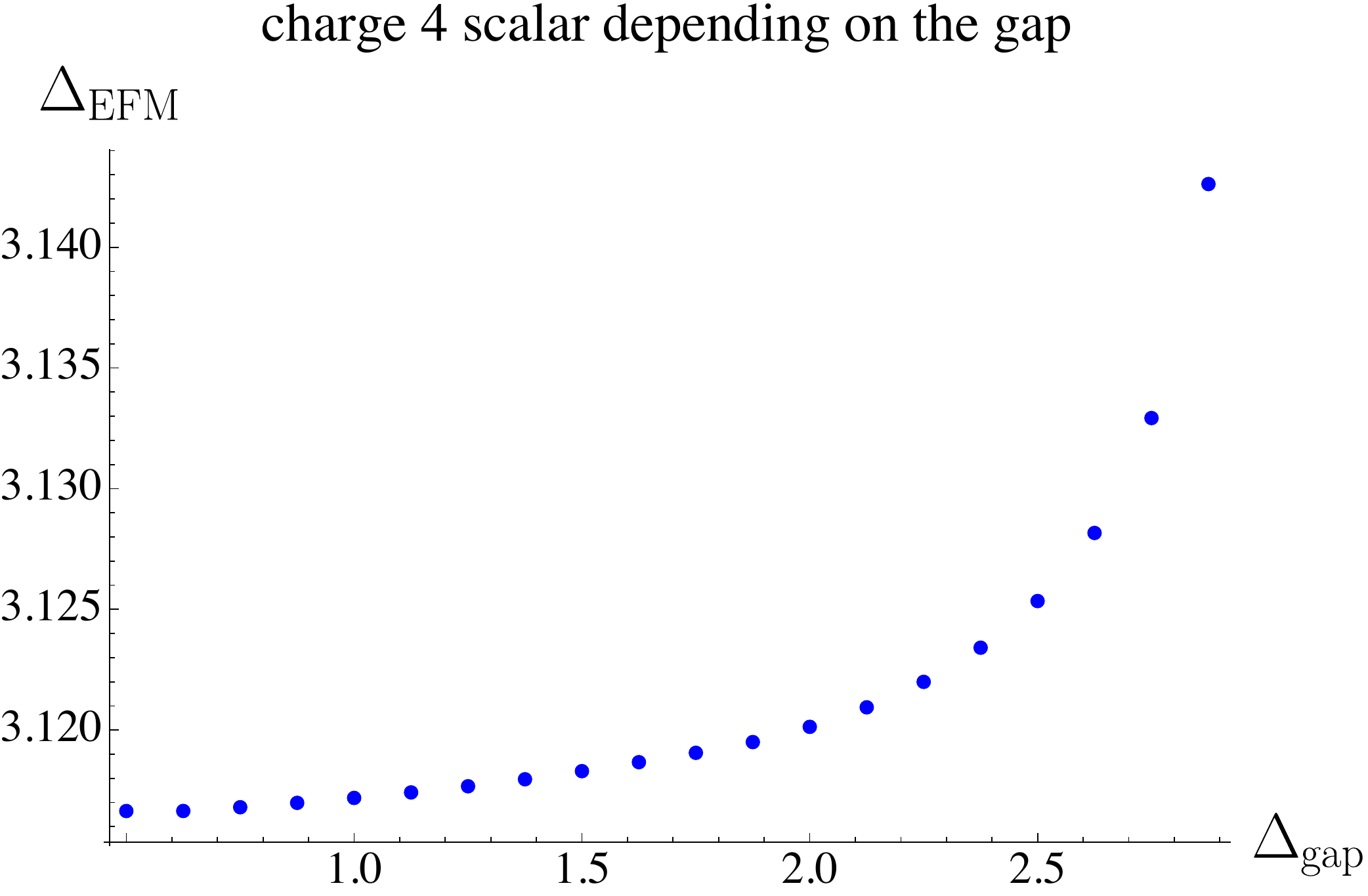}
  \caption{\label{gap}Dimension of the leading charge 4 scalar operator (above the gap) as one changes the imposed gap in the charge 4 sector. Here we have chosen a single primal point and computed the extremal spectra by optimizing the upper bound on the OPE coefficient $f_{\f\f s}$. This plot gives a clear illustration of the sharing effect, where when we impose gaps closer to the target operators, we get less accurate results from the extremal functional method in some cases.}
\end{figure}

To avoid this issue, one can attempt to compute the spectrum by lowering the gap. For instance, here we can lower the gap in the charge 4 scalar sector from 3 down to the unitarity bound of 1/2. As a demonstration, we compute the numerical spectrum by computing the extremal functional with upper bounds of norms of external OPEs, for a single point (the first point in the table \ref{tab:the20points}), given by
\begin{align}
&\left( {{\Delta _\phi }, {\Delta _s}, {\Delta _t}, \frac{{{f _{sss}}}}{{{f _{\phi \phi s}}}}, \frac{{{f _{tts}}}}{{{f _{\phi \phi s}}}}, \frac{{{f _{\phi \phi t}}}}{{{f _{\phi \phi s}}}}} \right)\\
&= ({\text{0.519130434, 1.51173444, 1.23648971, 1.20977354, 1.82254374, 1.76606470}}).\nonumber
\end{align}
We then extract the dimension of leading charge 4 scalar operator above the gap using the extremal functional as we increase the gap from 1/2 to 3. The result is shown in figure \ref{gap}. We can see that when the gap is set to be around 1/2, the scaling dimension decreases to be around 3.116, which is closer to the current Monte Carlo estimate $3.114(2)$.
To conclude this section, we will summarize in table~\ref{tab:resultsefm} the best predictions we have for the subleading (non-external) scalar operators based on the extremal functional method. Some of the results were already reported in the paper \cite{Chester:2019ifh}. For $\tau$, we used the gap of $1/2$ to avoid the sharing effect. 
 \begin{table*}[h!]
\centering
\begin{tabular}{@{}|cc|cc@{}}
	\hline
$\Delta$ &  value \\
	\hline
 $\Delta_{s'}$ &  $3.794(8^*)$ \\
  	 \hline
 $\Delta_{t'}$ &   $3.650(2^*)$ \\
  	 \hline
 $\Delta_{\chi}$ &$2.1086(3^*)$\\
  	 \hline
 $\Delta_{\tau}$ &$3.11535(73^*) $  \\
 \hline
\end{tabular}
	\caption{Conformal bootstrap predictions for subleading scalar operators using the extremal functional method. The superscript $^*$ means that the error bar is estimated based on the extremal functional method.
	\label{tab:resultsefm}}
\end{table*}

\section{Analytic predictions using the inversion formula}\label{formalism}
In this section, we describe the analytic tools used in this paper. This includes an introduction of our usage of the Lorentzian inversion formula \cite{Caron-Huot:2017vep}, brief descriptions of expansions of conformal blocks in 3d, a review of the twist Hamiltonian, and a discussion on inverting infinite sums of double-twist operators \cite{Simmons-Duffin:2016wlq}. This section is mostly a review of existing results, reformulated for the inversion formula.

\subsection{The inversion formula}
The inversion formula developed in \cite{Caron-Huot:2017vep} (see also \cite{Simmons-Duffin:2017nub}) yields precise relationships between low-twist operators in a CFT that are well-suited for extracting OPE data of double-twist operators. In particular, it generalizes and extends the lightcone bootstrap methods of \cite{Komargodski:2012ek,Fitzpatrick:2012yx,Alday:2015eya,Alday:2015ewa,Alday:2016njk,Simmons-Duffin:2016wlq}. Comparisons between the inversion formula and numerics for the 3d Ising and O(2) models were made in \cite{Albayrak:2019gnz}. 

In this paper, we extend the calculations of \cite{Albayrak:2019gnz} in the O(2) model by including more external operators, more numerical data, and by inverting infinite sums of double-twist operators. In the rest of this section, we review the analytic tools that we will use.

We start by describing the Lorentzian inversion formula for a four-point function of distinct scalar operators $\left\langle {{\phi _1}{\phi _2}{\phi _3}{\phi _4}}\right\rangle $ in a generic CFT. Including global symmetries does not modify the analysis significantly and will simply amount to weighting the contribution of operators by some group theory factors. The $s$-channel OPE data for this correlator can be encoded in the OPE function
\begin{align}
c(h, \bar{h})=c^{t}(h, \bar{h})+(-1)^{\bar{h}-h} c^{u}(h, \bar{h})~,
\end{align}
which can be expressed as an integral of the four-point function,
\begin{align}
&{c^t}(h,\bar h) = \frac{{{\kappa _{2\bar h}}}}{4}\int_0^1 d zd\bar z\mu (z,\bar z)g_{d - 1 -h,\bar{h}}^{r,s}(z,\bar z){{\mathop{\text{dDisc}}\nolimits} _t}[{\cal G}(z,\bar z)]~,\nonumber\\
&{\kappa _{2\bar h}} \equiv \frac{{\Gamma (\bar h + r)\Gamma (\bar h - r)\Gamma (\bar h + s)\Gamma (\bar h - s)}}{{2{\pi ^2}\Gamma (2\bar h - 1)\Gamma (2\bar h)}}
~,\nonumber\\
&\mu (z,\bar z) = {\left| {\frac{{z - \bar z}}{{z\bar z}}} \right|^{d - 2}}\frac{{{{((1 - z)(1 - \bar z))}^{s - r}}}}{{{{(z\bar z)}^2}}}~.
\end{align}
Here, $d$ is the spacetime dimension and we have introduced the labels
\begin{align}
h=\frac{\Delta-\ell}{2}=\frac{\tau}{2}~,& \qquad \bar{h}=\frac{\Delta+\ell}{2}=\frac\tau 2+\ell~,
\nonumber\\
r=h_{12}~,& \qquad s=h_{34}~,
\end{align}
where $h_{ij}=h_i-h_j$. These are the natural variables in the large-spin expansion, which is an expansion at large $\bar{h}$ for fixed $h$. The function $\mathcal{G}(z,\bar{z})$ is the four-point function after factoring out some kinematic factors,
\begin{align}
&\<\f_1(x_1)...\f_4(x_4)\>=\frac{1}{x_{12}^{\Delta_1+\Delta_2}x_{34}^{\Delta_3+\Delta_4}}\left(\frac{x_{14}}{x_{24}}\right)^{\Delta_2-\Delta_1}\left(\frac{x_{14}}{x_{13}}\right)^{\Delta_3-\Delta_4}\mathcal{G}(z,\bar{z})~, \label{eq:mGdef}
\nonumber\\
&z\bar{z}=\frac{x_{12}^{2}x_{34}^{2}}{x_{13}^{2}x_{24}^{2}}~, \qquad (1-z)(1-\bar{z})=\frac{x_{14}^{2}x_{23}^{2}}{x_{13}^{2}x_{24}^{2}}~.
\end{align}
We refer to $g_{d - 1 -h,\bar{h}}^{r,s}(z,\bar z)$ in the integrand as the \emph{Weyl-reflected} block. Our conventions for conformal blocks are described in appendix \ref{block}. We use the superscripts $s$, $t$ and $u$ to denote the three OPE channels.\footnote{We apologize for the abuse of notation where $t$ and $s$ also denote external operators in the O(2) model. We hope that there will be minimal confusion based on the context.} The notation $\text{dDisc}_\text{channel}$ denotes the standard double discontinuity in each channel. For instance, in the $s$-channel we have
\begin{align}
{{\mathop{\text{dDisc}}\nolimits} _s}[\mathcal{G}(z,\bar z)] = \cos (\pi (s - r))\mathcal{G}(z,\bar z) - \frac{1}{2}{e^{i\pi (s - r)}}\mathcal{G}\left( {z{e^{2\pi i}},\bar z} \right) - \frac{1}{2}{e^{i\pi (r - s)}}\mathcal{G}\left( {z{e^{ - 2\pi i}},\bar z} \right)~,
\end{align}
in which the analytic continuation is taken around $z=0$. For the $t$ and $u$-channels, the double discontinuity is performed around $z=1$ and $z=\infty$ respectively. Once we compute the function $c(h,\bar{h})$, one can read off the OPE data using
\begin{align}
{\left( { - \frac{1}{2}} \right)^\ell }{f_{12O}}{f_{34O}} =  - {{\mathop{\text{Res}}\nolimits} _{{\Delta} = \Delta_{O} }}c\left( {{\Delta},\ell _{O}} \right)~,
\end{align}
where we have rewritten $c$ as a function of $\Delta$ and $\ell$ to emphasize that we are taking the residue in dimension at fixed spin. 

In practice, it is convenient to define the \emph{generating function} to package the conformal data,
\begin{align}
{\left. {{c^t}(h,\bar h)} \right|_{{\text{poles}}}} = \int_0^1 {\frac{{dz}}{{2z}}} {z^{ - h}}{C^t}(z,\bar h)~,
\end{align}
where $c^u$ is defined in a similar way. Then the powers of $z$ in the generating function $C^{t}(z,\bar{h})$ turn into poles in $h$ for the OPE function.  

Now, we will derive a formula for the generating function. First, it will be convenient to use the identity
\begin{align}
g_{h,\bar{h}}^{ - r, - s}(z,\bar z) = {((1 - z)(1 - \bar z))^{ - r + s}}g_{h,\bar{h}}^{r,s}(z,\bar z)~.
\end{align}
Next, we need the conformal block expansion in the $t$-channel for $\mathcal{G}(z,\bar{z})$:\footnote{Generically, we mainly use $O'$ to denote the internal operators in the crossed channels if we want to emphasize the cross channels, while $O$ denotes the internal operators in the $s$-channel. Similarly, we will sometimes use the notation $h',\bar{h}',\Delta',\ell'$ for the crossed channels.}
\begin{align}
\mathcal{G}(z,\bar z) = \frac{{{{(z\bar z)}^{{h_1} + {h_2}}}}}{{{{[(1 - z)(1 - \bar z)]}^{{h_2} + {h_3}}}}}\sum\limits_{O'} {{{\left( { - \frac{1}{2}} \right)}^{{\ell _{O'}}}}} {f_{32O'}}{f_{14O'}}g_{O'}^{{h_{32}},{h_{14}}}(1 - z,1 - \bar z)~.
\end{align}
Finally, setting $d=3$ we find the generating function is:
\begin{align}
&{C^t}(z,\bar h) = {\kappa _{2\bar h}}\int_z^1 d \bar z\left( {\frac{{\bar z - z}}{{z\bar z}}} \right)\frac{1}{{z{{\bar z}^2}}}g_{2-h,\bar{h}}^{{h_{21}},{h_{43}}}(z,\bar z)\frac{{{{(z\bar z)}^{{h_1} + {h_2}}}}}{{{{[(1 - z)(1 - \bar z)]}^{{h_2} + {h_3}}}}}\nonumber\\
&\sum\limits_{O'} {2\sin \left( {\pi \left( {{h_{O'}} - {h_1} - {h_4}} \right)} \right)\sin \left( {\pi \left( {{h_{O'}} - {h_2} - {h_3}} \right)} \right){{\left( { - \frac{1}{2}} \right)}^{\ell_{O'}}}} {f_{32O'}}{f_{14O'}}g_{O'}^{{h_{32}},{h_{14}}}(1 - z,1 - \bar z)~.
\end{align}
By taking $z\rightarrow 0$ and expanding the 3d block in this limit \cite{Simmons-Duffin:2016wlq,Caron-Huot:2017vep} we find two powers of $z$: $z^{h_1+h_2}$ and $z^{h_3+h_4}$. The inversion of a single block will then produce poles for the double-twist operators $[\phi_1\phi_2]_{n}(\bar{h})$ and $[\phi_3\phi_4]_{n}(\bar{h})$ in the OPE function \cite{Liu:2018jhs}. However, in general, we cannot take the $z\rightarrow 0$ limit under the sum over $O'$ on the right-hand side. Instead, the infinite sum can produce new powers of $z$ corresponding to other multi-twist families. So we then have to distinguish two different cases:
\begin{itemize}
\item One possibility is the low-lying operators appearing in the $t$-channel, for instance the scalars $s$, $\phi$,  and $t$, or the conserved operators $J$ and $T$, give an accurate estimation for the double-discontinuity. In this case, we can directly invert each block and take the sum.
\item One can also include infinite families of operators, for instance, the double-twist trajectory. In this case, the sum and integral may not commute, and we have to perform the infinite summation first. When this happens, a careful regularization needs to be performed.
\end{itemize} 
In the following subsection, we will develop techniques to study how individual operators contribute to the generating function. We will return to the problem of performing the infinite sums in section \ref{sec:DTI}.
\subsection{Expanding conformal blocks in the inversion formula}
Simplifying the generating function typically involves a procedure for expanding the conformal blocks in a simple set of functions, which can be inverted analytically. Below we describe two possible expansions.
\subsubsection{$\text{SL}_2$ expansion for the Weyl-reflected block}
There is no known closed-form formula for conformal blocks in 3d. However, there is a natural expansion for any conformal block in terms of  $\text{SL}_2$ blocks,
\begin{align}
k_{\bar h}^{r,s}(\bar z) \equiv {{\bar z}^{\bar h}}{ \times _2}{F_1}(\bar h - r,\bar h + s,2\bar h,\bar z)~.
\end{align}
Firstly, we wish to apply this expansion to the Weyl-reflected block. We have \cite{Caron-Huot:2017vep}
\begin{align}
\left( {\frac{{\bar z - z}}{{z\bar z}}} \right)\frac{1}{z}g_{2-h,\bar{h}}^{{h_{21}},{h_{43}}}(z,\bar z) = \sum\limits_{n = 0}^\infty  {{z^{ - h + n}}} \sum\limits_{j =  - n}^n {\mathcal{C}_{n,j}^{{h_{21}},{h_{43}}}} (h,\bar h)k_{\bar h + j}^{{h_{21}},{h_{43}}}(\bar z)~.
\end{align}

Plugging this in, we can simplify the generating function as
\begin{align}
&{C^t}(z,\bar h) = {\kappa _{2\bar h}}\sum\limits_{n = 0}^\infty  {{z^n}} \sum\limits_{j =  - n}^n {\mathcal{C}_{n,j}^{{h_{21}},{h_{43}}}(h,\bar h)} \int_z^1 d \bar z\frac{1}{{{{\bar z}^2}}}k_{\bar h + j}^{{h_{21}},{h_{43}}}(\bar z)\frac{{{{(z\bar z)}^{{h_1} + {h_2}}}}}{{{{[(1 - z)(1 - \bar z)]}^{{h_2} + {h_3}}}}}\nonumber\\
&\sum\limits_{O'} {2\sin \left( {\pi \left( {{h_{O'}} - {h_1} - {h_4}} \right)} \right)\sin \left( {\pi \left( {{h_{O'}} - {h_2} - {h_3}} \right)} \right){{\left( { - \frac{1}{2}} \right)}^{\ell_{O'}}}} {f_{32O'}}{f_{14O'}}g_{O'}^{{h_{32}},{h_{14}}}(1 - z,1 - \bar z)~.
\end{align}
In this work, we are only interested in the leading double-twist trajectories, in which case we can focus on the small $z$ dependence of the generating function. Therefore, we only need to keep the $n=0$ piece of the above sum to make predictions for $[\f_1\f_2]_{n=0}(\bar{h})$ and $[\f_3\f_4]_{n=0}(\bar{h})$. In that case we have $\mathcal{C}^{h_{21},h_{43}}_{0,0}=1$ (more examples of these coefficients are given in appendix \ref{expansion}) and then taking the small $z$ limit we find:\footnote{Note that here we have modified the integration range in $\bar{z}$ to go from 0 to 1. The integration from $0$ to $z$ contributes to higher-order corrections $\sim z^{h_i+h_j+1}$, while here we are computing the generating function in a small $z$ expansion. Therefore, at leading order we can drop these terms and extend the integral. However, they will be important when computing higher-twist trajectories. For an alternative way to study the $z$ dependence, see \cite{Caron-Huot:2020ouj}.}
\begin{align}
&{C^t}(z,\bar h) \supset {\kappa _{2\bar h}}\int_0^1 d \bar z\frac{1}{{{{\bar z}^2}}}k_{\bar h}^{{h_{21}},{h_{43}}}(\bar z)\frac{{{{(z\bar z)}^{{h_1} + {h_2}}}}}{{{{[(1 - z)(1 - \bar z)]}^{{h_2} + {h_3}}}}}\nonumber\\
&\sum\limits_{O'} {2\sin \left( {\pi \left( {{h_{O'}} - {h_1} - {h_4}} \right)} \right)\sin \left( {\pi \left( {{h_{O'}} - {h_2} - {h_3}} \right)} \right){{\left( { - \frac{1}{2}} \right)}^{\ell_{O'}}}} {f_{32O'}}{f_{14O'}}g_{O'}^{{h_{32}},{h_{14}}}(1 - z,1 - \bar z)~.
\end{align}

\subsubsection{$\text{SL}_2$ expansion for $\mathcal{G}$}
We can also use the $\text{SL}_2$ expansion to simplify $\mathcal{G}$ itself. We expand the blocks as:
\begin{align}
g_{h,\bar{h}}^{r,s}(z,\bar z) = \sum\limits_{p = 0}^\infty  {\sum\limits_{q =  - p}^p {A_{p,q}^{r,s}} } (h,\bar h){z^{h + p}}k_{\bar h + q}^{r,s}(\bar z)~.
\end{align}
The coefficients $A$ are summarized in appendix \ref{expansion}. This expansion works for small $z$ and $0\leq\bar{z}\leq1$. Alternatively, one can expand the analogous formula at small $\bar{z}$ and arbitrary ${z}$:
\begin{align}
g_{h,\bar h}^{r,s}(z,\bar z) = \sum\limits_{p = 0}^\infty  {\sum\limits_{q =  - p}^p {A_{p,q}^{r,s}} } (h,\bar h){\bar z^{h + p}}k_{\bar h + q}^{r,s}(z)~,
\end{align}
(the conformal block is symmetric in $z$ and $\bar{z}$). We can apply it to the $t$-channel block,
\begin{align}
g_{O'}^{{h_{32}},{h_{14}}}(1 - z,1 - \bar z) = \sum\limits_{p = 0}^\infty  {\sum\limits_{q =  -p}^p {A_{p,q}^{{h_{32}},{h_{14}}}} } (h,\bar h){(1 - \bar z)^{h + p}}k_{{{\bar h}_{O'}} + q}^{{h_{32}},{h_{14}}}(1 - z)~,
\end{align}
in the limit $\bar{z}\to 1^-$, and we can then safely take the small $z$ limit. 

After performing the $\bar{z}$ integral and dropping the OPE coefficients, we find
\begin{align}
&{C^t_{O'}}(z,\bar h) = {\kappa _{2\bar h}} {2\sin \left( {\pi \left( {{h_{O'}} - {h_1} - {h_4}} \right)} \right)\sin \left( {\pi \left( {{h_{O'}} - {h_2} - {h_3}} \right)} \right)}  \nonumber\\
&\sum\limits_{p = 0}^{  \infty } {\sum\limits_{q =  - p}^p {A_{p,q}^{{h_{32}},{h_{14}}}\left( {{h_{O'}},{{\bar h}_{O'}}} \right)} } \frac{{{z^{{h_1} + {h_2}}}k_{{{\bar h}_{O'}} + q}^{{h_{32}},{h_{14}}}(1 - z)}}{{{{(1 - z)}^{{h_2} + {h_3}}}}}R_{\bar h - {h_{21}},\bar h + {h_{43}},2\bar h}^{\bar h + {h_1} + {h_2} - 2,{h_{O'}} + p - {h_2} - {h_3}}~,
\end{align}
where $C^t_{O'}$ is the contribution of $O'$ to the generating function, and $R$ is defined and explained in appendix \ref{integral}.
\subsubsection{Dimensional reduction for $\mathcal{G}$}
Instead of using the decomposition into $\text{SL}_2$ blocks, one can also decompose the three-dimensional blocks into two-dimensional blocks. This treatment is manifestly symmetric for $z$ and $\bar{z}$. 

The decomposition reads
\begin{align}
g_{h,\bar{h}}^{r,s}(z,\bar z) = \sum\limits_{p = 0}^\infty  {\sum\limits_{q =  - p}^p {{\cal A}_{p,q}^{r,s}} } (h,\bar h)k_{h + p}^{r,s}(z)k_{\bar h + q}^{r,s}(\bar z)~,
\end{align}
where the $\mathcal{A}$ coefficients are described in appendix \ref{expansion}. For conformal blocks with integer spin, $\bar{h}-h=\ell$, the coefficients ${\cal A}_{p,q}^{r,s}$ vanish for $q<p-2\ell$.  

We can perform a similar expansion in the $t$-channel, and after plugging it into the inversion formula, we obtain
\begin{align}
&{C^t_{O'}}(z,\bar h) = {\kappa _{2\bar h}} {2\sin \left( {\pi \left( {{h_{O'}} - {h_1} - {h_4}} \right)} \right)\sin \left( {\pi \left( {{h_{O'}} - {h_2} - {h_3}} \right)} \right)}  \nonumber\\
&\sum\limits_{p = 0}^{ \infty } {\sum\limits_{q =  - p}^p {\mathcal{A}_{p,q}^{{h_{32}},{h_{14}}}\left( {{h_{O'}},{{\bar h}_{O'}}} \right)} } \frac{{{z^{{h_1} + {h_2}}}k_{{{\bar h}_{O'}} + q}^{{h_{32}},{h_{14}}}(1 - z)}}{{{{(1 - z)}^{{h_2} + {h_3}}}}}\Omega _{\bar h,{h_{O'}} + p,{h_2} + {h_3}}^{{h_i}}~,\label{eq:inversionOmega}
\end{align}
where $\Omega$ is given in appendix \ref{integral} in terms of ${}_4F_3$ hypergeometrics and we have once again dropped the OPE coefficients. In practice, we evaluate this by truncating the sum over $p$ at some $p_{\text{max}}$.

The two types of expansions are closely related, and one can easily derive one expansion from the other (see appendix \ref{expansion}). Practically, we see that the predictions arising from truncating each expansion give very similar answers. In section \ref{comparing}, we use dimensional reduction to perform our calculations. We give a more detailed comparison between the two approaches in appendix \ref{performance}.

\subsection{Double-twist improvement (DTI)}
\label{sec:DTI}
Until now, we only considered isolated low-lying operators in the $t$ and $u$-channels. One can also consider an infinite tower of operators in the crossed channels and study its effect on OPE data in the $s$-channel. Importantly, taking the $z\rightarrow 0$ limit does not commute with the infinite sum. 

To see how this works in practice, we consider a four-point function of scalars $\<\s\s\s\s\>$ and look at the inversion of a general, double-twist trajectory $[OO]_{0}(\bar{h})$. To keep the notation compact, we denote weights of this double-twist trajectory as $h_\ell$ and $\bar{h}_\ell$, and we will consider the sum over spins $\ell \geq \ell_0$. Their contribution to the generating function is:
\begin{align}
C^t(z,\bar{h})&\supset \sum\limits_{\ell=\ell_0}^{\infty}\left(-\frac{1}{2}\right)^{\ell}f^2_{\s\s[OO]_0}(\bar{h}_\ell)C^{t}_{[OO]_{0}(\bar{h}_{\ell})}(z,\bar{h})
\nonumber \\ &\supset\sum\limits_{\ell=\ell_0}^{\infty} \kappa _{2\bar h} \left(- \frac{1}{2} \right)^{\ell }f^2_{\s\s[OO]_0}(\bar{h}_\ell) 2\sin^2 \left( \pi \left( {h_{\ell} - 2h_\s} \right) \right)  
\nonumber\\
& \quad \sum\limits_{p = 0}^{ \infty } {\sum\limits_{q =  - p}^p {\mathcal{A}_{p,q}^{0,0}\left( {{h_\ell},{{\bar h}_\ell}} \right)} } \frac{{{z^{{2h_\s} }}k_{{{\bar h}_\ell} + q}^{{0},{0}}(1 - z)}}{{(1 - z)}^{2h_\s}}\Omega _{\bar h,h_{\ell} + p,2h_\s}^{h_\s}. \label{eq:sumDoubletwists}
\end{align}
We see there are now two infinite sums to perform, one sum for the double-twist operators we are inverting and another to expand them in 2d blocks. In practice, the sum over 2d blocks converges quickly for $\bar{h}>1$ so we can truncate the sum at some $p_{\max} \sim 10-20$. 

To see why we cannot commute $z\rightarrow 0$ with the infinite sum over $\ell$ we can consider the following sum derived in \cite{Simmons-Duffin:2016wlq}:
\begin{align}
\sum\limits_{\substack{\bar{h}=\bar{h}_0+\ell \\ \ell=0,1,...}}S_{a}(\bar{h})k^{0,0}_{\bar{h}}(1-z)&=\left(\frac{z}{1-z}\right)^a+\sum\limits_{k=0}^{\infty}\partial_{k} \mathcal{B}_{a,-k-1}(\bar{h}_0)\left(\frac{z}{1-z}\right)^k, \label{eq:sum2dBlocks}
\\
S_{a}(h)&\equiv\frac{\Gamma(h)^{2}\Gamma(h-a-1)}{\Gamma(-a)^{2}\Gamma(2h-1)\Gamma(h+a+1)}~,
\\
\mathcal{B}_{a,b}(h_0)&=-\frac{(a+h_0)(b+h_0)\Gamma(2h_0-1)^{2}}{(1+a+b)\Gamma(h_0)^4}S_{a}(h_0)S_b(h_0)~.
\end{align}
If we take the limit $z\rightarrow 0$ under the sum we only see powers of $z$ and $\log(z)$, which correspond to the second term of (\ref{eq:sum2dBlocks}):
\begin{align}
k_{\bar{h}}(1-z)&=-\frac{\Gamma(2h)}{\Gamma(h)^{2}}\sum\limits_{k=0}^{\infty}\partial_{k} T_{-k-1}\left(\frac{z}{1-z}\right)^k, \label{eq:smallzSL2}
\\
T_{a}(h)&\equiv \frac{\Gamma(2h-1)}{\Gamma(h)^{2}}S_{a}(h)~.
\end{align}
The sum over blocks (\ref{eq:sum2dBlocks}) and the series expansion (\ref{eq:smallzSL2}) are consistent because if we expand around $z=0$ first, we find the sum over $\bar{h}$ is divergent. Therefore, we cannot do this series expansion under the sum.

We can now use the sum (\ref{eq:sum2dBlocks}) to evaluate (\ref{eq:sumDoubletwists}). Specifically, we expand around $z=0$, and if we find the sum diverges, we subtract by the left-hand side of (\ref{eq:sum2dBlocks}), for appropriate $S_{a}(\bar{h})$, sufficiently many times until the sum converges. We can then add back in their contribution using the right-hand side of (\ref{eq:sum2dBlocks}). This procedure was spelled out in detail in \cite{Simmons-Duffin:2016wlq}, so here we will discuss new subtleties from using the inversion formula and dimensional reduction.

First, in (\ref{eq:sumDoubletwists}) there are many terms we have to expand at large $\ell$: we have the OPE coefficients and the weights $h_{\ell}$, and $\bar{h}_{\ell}$ which appear as arguments in several places. In general, we need to expand all these terms at large $\ell$, expand the summand around $z=0$, and then perform the subtractions at each $p$ and $q$. In practice, for the problems we will consider here, a single subtraction is needed, and the problem simplifies significantly. 
To see how this works, we need that at asymptotically large spin:
\begin{align}
h_{\ell}&\rightarrow 2h_{O}~,\nonumber\\
\bar{h}_{\ell}&\rightarrow 2h_{O}+\ell~.
\end{align}
If we only need to perform one subtraction, then we can break the sum over $\ell$ in two, we have a sum from $\ell_0$ to $\ell_{*}$ which we can perform directly and a sum from $\ell_*$ to $\infty$ which we can compute analytically to a sufficiently high degree of precision:
\begin{align}
B^t_{[OO],1}&=\sum\limits_{\ell=\ell_0}^{\ell_{*}-1}\left(- \frac{1}{2} \right)^{\ell }f^2_{\s\s[OO]_0}(\bar{h}_\ell)C^{t}_{[OO]_{0}(\bar{h}_{\ell})}(z,\bar{h})~,\nonumber\\
B^t_{[OO],2}&=\sum\limits_{\ell=\ell_{*}}^{\infty}\left(- \frac{1}{2} \right)^{\ell }f^2_{\s\s[OO]_0}(\bar{h}_\ell)C^{t}_{[OO]_{0}(\bar{h}_{\ell})}(z,\bar{h})~.
\end{align}
For the first sum, because we are inverting a finite number of operators, we can take the $z\rightarrow 0$ limit under the sum. To actually perform the sum, we need to know the OPE coefficients $f_{\s\s[OO]}$, which can be calculated either using the inversion formula or through numerics.

For the second sum, we assume $\ell_*$ is large enough such that we can set $h=2h_{O}$ and $\bar{h}=2h_{O}+\ell$. For asymptotically large $\ell$ we also have that the dimensional reduction coefficients, $\mathcal{A}$, become independent of $\ell$,
\begin{align}
\mathcal{A}^{0,0}_{p,q}(2h_O,\bar{h}_\ell)\sim \hat{\mathcal{A}}^{0,0}_{p,q}(2h_O)~,
\end{align}
for all $p$ and $q$. At leading order, we can then ignore corrections to $\mathcal{A}$ itself. Then the second sum becomes
\begin{align}
B^t_{[OO],2}&\approx\sum\limits_{\substack{\bar{h}_\ell=2h_O+\ell \\ \ell=\ell_*, \ell_*+1, ...}}^{\infty} \kappa _{2\bar h} \left(- \frac{1}{2} \right)^{\ell }f^2_{\s\s[OO]_0}(\bar{h}_\ell)2\sin^2 \left(2 \pi \left( {h_O- h_\s} \right) \right)  
\nonumber\\
& \quad \sum\limits_{p = 0}^{ \infty } {\sum\limits_{q =  - p}^p {\hat{\mathcal{A}}_{p,q}^{0,0}\left( {2h_O} \right)} } \frac{{{z^{{2h_\s} }}k_{{{\bar h}_\ell} + q}^{{0},{0}}(1 - z)}}{{(1 - z)}^{2h_\s}}\Omega _{\bar h,2h_O+ p,2h_\s}^{h_\s}~. \label{eq:B2_approx}
\end{align}
We see that if we take $\ell_*$ large enough we get an integrally spaced sum over $\bar{h}$ and we can straightforwardly apply identities like (\ref{eq:sum2dBlocks}). Finally, we can compute the large $\bar{h}$ asymptotics of the OPE coefficients $\left(- \frac{1}{2} \right)^{\ell }f_{\s\s[OO]_0}(\bar{h}_\ell)^2$ by studying the inversion formula for $\<\s\s OO\>$. If we assume the OPE coefficients at large $\ell$ are given by,
\begin{align}
\left(- \frac{1}{2} \right)^{\ell }f^2_{\s\s[OO]_0}(\bar{h}_\ell)= c_{a} S_{a}(\bar{h}_{\ell})~,
\end{align}
we then need to evaluate the sum
\begin{align}
\sum\limits_{\substack{\bar{h}_{\ell}=2h_O+\ell \\ \ell=\ell_*, \ell_*+1,...}}S_{a}(\bar{h}_{\ell})k_{{{\bar h}_\ell} + q}^{{0},{0}}(1 - z)~.
\end{align}
The last complication before we can use (\ref{eq:sum2dBlocks}) is the hypergeometric has twist displaced by $q$ in comparison to the $S_{a}(\bar{h}_{\ell})$, due to using dimensional reduction for the 3d block. To take this into account we use that for asymptotically large $\bar{h}$:
\begin{align}
S_{a}(\bar{h})\sim 2^{2q}S_{a}(\bar{h}+q) \label{eq:shiftS}~.
\end{align}
Using this we can now evaluate the sum in (\ref{eq:B2_approx}),
\begin{align}
B^t_{[OO],2}\approx \sum\limits_{p = 0}^{ \infty } \sum\limits_{q =  - p}^p & 2\kappa _{2\bar h} \sin^2 \left(2 \pi \left( {h_O- h_\s} \right) \right)  
 \hat{\mathcal{A}}_{p,q}^{0,0}\left( 2h_O \right)  \frac{{{z^{2h_\s }}}}{{(1 - z)}^{2h_\s}}\Omega _{\bar h,2h_O+ p,2h_\s}^{h_\s}2^{2q}c_{a}
\nonumber \\ & \left[\left(\frac{z}{1-z}\right)^a+\sum\limits_{k=0}^{\infty}\partial_{k} \mathcal{B}_{a,-k-1}(\bar{h}_0)\left(\frac{z}{1-z}\right)^k\right]~.
\end{align}
We can now expand the sum at small $z$ and read off the twist of the exchanged operators from the powers of $z$. For example, the extra power of $z^a$ signals the presence of a multi-twist operator with $h=2h_\s+a$.

In \cite{Liu:2018jhs,Sleight:2018ryu,Cardona:2018qrt,Albayrak:2019gnz,Li:2019dix} it was revealed that performing the full inversion formula gives rise to new non-perturbative corrections in the large-spin expansion. One question is if these corrections play any important role in the large spin sums considered here. We find that they are only relevant for calculating the OPE coefficients in $B^t_{[OO],1}$, the bounded sum over spin. They do not play any role in $B^t_{[OO],2}$ since we are taking $\ell_*$ sufficiently large such that these non-perturbative effects can be neglected. Equivalently, the new corrections do not lead to any divergences in the sum over spin, so they do not lead to any new subtractions.  

Finally, one special case is when $[OO]_{n}=[\sigma\sigma]_{n}$.  In that case in (\ref{eq:B2_approx}) we expand the $\sin^2$ factor to leading order in the anomalous dimension and use the MFT OPE coefficients for $f_{\s\s[\s\s]_{0}}(\bar{h}_{\ell})$. Besides that, the analysis remains unchanged if we again assume we only need to perform one subtraction. 

\subsection{Exact vs.\ approximate generating function}
\label{subsec:ExactvsApprox}
Next, we discuss how we can use the generating function to calculate anomalous dimensions and OPE coefficients at finite spin. For simplicity we will again restrict to a four-point function of identical scalars $\<\sigma\sigma\sigma\sigma\>$. 
In the small $z$ limit the generating function for this correlator scales as\footnote{For identical scalars we have $C^t(z,\bar{h})=C^u(z,\bar{h})$ so we will focus on the $t$-channel.} 
\begin{align}\label{anomal}
C^t(z,\bar h) \approx {C_{{{[\sigma \sigma ]}_0}}}(\bar h){z^{2{h_\sigma } + \delta {h_{{{[\sigma \sigma ]}_0}}}(\bar h)}}+...~,
\end{align}
where ${{{[\sigma \sigma ]}_0}}$ denotes the leading double-twist family and we have dropped higher twist families.
The anomalous dimension ${\delta {h_{{{[\sigma \sigma ]}_0}}}(\bar h)}$ can be computed by taking the exact generating function and evaluating
\begin{align}
\delta {h_{{{[\sigma \sigma ]}_0}}}(\bar h) =\lim\limits_{z\rightarrow 0} \frac{{\partial C(z,\bar h)}}{{C(z,\bar h)}} - 2{h_\sigma }~, \label{eq:dhFromC}
\end{align}
where we take the $z\rightarrow 0$ limit to extract the minimal-twist operators. Here we also defined the partial derivative symbol
\begin{align}
\partial  \equiv z{\partial _z} = \frac{\partial }{{\partial \log z}}~.
\end{align}

In practice, we cannot calculate the exact generating function in theories like the Ising or O(2) model. Instead, we try to get an estimate for the generating function by inverting operators of bounded twist. For example, if we invert a finite number of operators, including the identity, we find the generating function takes the form:
\begin{align}
C^t(z,\bar h)\approx  z^{2h_{\sigma}}C_{[\sigma\sigma]_{0}}(\bar{h})\left(1+\delta h_{{\sigma\sigma}_{0}}(\bar{h})\log(z)\right)~.
\end{align}
Comparing to (\ref{anomal}) we see that the inversion of the light, isolated operators are capturing the leading approximation to the generating function from expanding in the anomalous dimension. If we expand (\ref{anomal}) to higher orders in $\delta h$ we also find higher log terms,
\begin{align}
{z^{h + \delta h}} = {z^h}\left( {1 + \delta h\log z + \frac{1}{2}\delta {h^2}{{\log }^2}z +  \ldots } \right)~. \label{eq:Expanddh}
\end{align}
The $\log^k z$ terms for an integer $k$ in the generating function are found by inverting an infinite sum of $k$-twist operators~\cite{Fitzpatrick:2015qma,Simmons-Duffin:2016wlq}. Therefore, if we can assume the anomalous dimensions are small, e.g., when studying $s$-channel, double-twist operators with large spin, we can calculate the anomalous dimension by looking at the $\log$ term.

However, this is insufficient if we want to study double-twist operators at finite spin when the anomalous dimension can be large, and the above approximation is no longer valid. In that case, we take a different approach and attempt to fit our approximate generating function from inverting operators of bounded twists to the exact form given in (\ref{anomal}) \cite{Simmons-Duffin:2016wlq,Iliesiu:2018zlz}. For the case of $\<\s\s\s\s\>$ this method is very simple, we take the logarithmic derivative of the generating function and evaluate it at a fixed, small $z$,
\begin{align}
\delta {h_{{{[\sigma \sigma ]}_0}}}(\bar h) =\frac{{\partial C(z,\bar h)}}{{C(z,\bar h)}} - 2{h_\sigma }\bigg|_{z=z_{0}}. \label{eq:dhFixedz}
\end{align}
There is now an arbitrariness in what value to choose for $z_0$. If we choose $z_0$ small we can ignore higher, multi-twist operators in the generating function, e.g., $[\sigma\sigma]_{n=1}(\bar{h})$. On the other hand, if we choose $z_0$ to be {\it too} small, we see that the logs in (\ref{eq:Expanddh}) are becoming large, and we can no longer ignore them, i.e., we are no longer justified in inverting operators of bounded twist.

This happens when
\begin{align}
\left| {\delta h\log {z_0}} \right| \sim 1~,
\end{align}
or equivalently
\begin{align}
z_0 \sim \exp \left(-\frac{1}{{\delta h}}\right)~.
\end{align}
In this work, we make a  choice $z_0=0.1$, the same as the value used in \cite{Simmons-Duffin:2016wlq} in the lightcone bootstrap, and in \cite{Albayrak:2019gnz} when using the inversion formula. In appendix \ref{z0}, we discuss some other choices of $z_0$ and try to make a comparison. We have found that in many situations, the results are very stable against changing the value of $z_0$. 

With these issues in mind, (\ref{eq:dhFixedz}) gives us $h$ as a function $\bar{h}$ but we are specifically interested in operators of integer spin, $\ell$. To determine the physical spectrum, we then need to solve the equation,
\begin{align}
\bar{h}-h(\bar{h})=\ell~.
\end{align}
This is a complicated, transcendental equation, but in practice can be solved iteratively. If $\delta h$ is small we can first plug in the MFT result, $\bar{h}=2h_{\sigma}$ to find an initial approximation $\delta h^{(0)}$.\footnote{Here the superscript denotes the order in the iterative approximation for finding integer spin operators and not the twist of the trajectory.} We then use this as our input $\bar{h}=2h_{\sigma}+\delta h^{(0)}+\ell$ to find another $\delta h^{(1)}$. We can iterate this procedure several times, and the result quickly converges to the actual solution, e.g., we typically need around five iterations. If the anomalous dimension is large, one can also obtain estimates by making a plot in $(h,\bar{h})$ and drawing lines of constant $\ell=\bar{h}-h$.

Finally, we discuss the predictions for OPE coefficients. Once we have the prediction for the anomalous dimensions, we can easily compute the OPE coefficient predictions using
\begin{align}
{\left( { - \frac{1}{2}} \right)^\ell }f_{\sigma \sigma {{[\sigma \sigma ]}_0}}^2(\bar{h}) = 2{\left( {1 - \frac{{\partial \delta {h_{{{[\sigma \sigma ]}_0}}}(\bar h)}}{{\partial \bar h}}} \right)^{ - 1}}\frac{{{C^t}({z_0},\bar h)}}{{z_0^{2{h_\sigma } + \delta {h_{{{[\sigma \sigma ]}_0}}}(\bar h)}}}~.
\end{align}
The derivative term appearing in the above formula is given by a Jacobian since we need to find the residue in $\Delta$ at fixed spin as opposed to the residue in $\bar{h}$ at fixed $h$ \cite{Caron-Huot:2017vep}. The factor of $2$ is because the $t$ and $u$-channels give the same contribution. 
\subsection{The twist Hamiltonian}
\label{sec:twisthamiltonian}
In the previous section, we assumed the leading double-twist operators in the $s$-channel were non-degenerate. In that case, we can use (\ref{eq:dhFixedz}) to find the anomalous dimensions by taking $z_0$ sufficiently small that we can neglect heavier $s$-channel operators. However, this becomes insufficient once there are multiple double-twist operators with commensurate twists. In that case, to find the spectrum at finite spin, we need to introduce the \emph{twist Hamiltonian} \cite{Simmons-Duffin:2016wlq}. Here we describe how to define the twist Hamiltonian in the context of the Lorentzian inversion formula.

We will focus on a practical mixing example that arises in the O(2) model, the mixing of charge 1 operators. We can observe that the charge-1, double twist operators $[\phi t]_{n=0}^1$ and $[\phi s]_{n=0}^1$ have twist which become approximately degenerate for $\bar{h}\gg1$,
\begin{align}
h_{[\phi t]_{0}^1}(\bar{h})&\approx 0.877689~,
\\
h_{[\phi s]_{0}^1}(\bar{h})&\approx 1.01522~.
\end{align}
Recall that we use the notation $[\cO_1\cO_2]_n^q$ described in (\ref{eq:superscriptnotation}), where the subscript $n$ indicates the asymptotic twist of the double-twist family and the superscript $q$ indicates the charge.
Given the exact generating functions, we could of course distinguish the two trajectories. However, as mentioned before we are in practice finding approximations to the generating function and to resolve the mixing need to consider the crossing equations $\left\langle {\phi t\phi t} \right\rangle $, $\left\langle {\phi s\phi s} \right\rangle $, and $\left\langle {\phi s\phi t} \right\rangle $ simultaneously. In particular, we construct the matrix of generating functions,
\begin{align}
{M_1^t}(z,\bar{h}) \equiv \left( {\begin{array}{*{20}{l}}
{C_1^{\phi s s\phi}(z,\bar{h})}&{C_1^{\phi ts \phi}(z,\bar{h})}\\
{C_1^{\phi ts \phi }(z,\bar{h})}&{C_1^{\phi t t \phi}(z,\bar{h})}
\end{array}} \right)~,\label{eq:M1t}
\end{align}
where the subscript $_1$ denotes the charge 1 family.\footnote{To compute these requires group theory factors we have not given yet, but will introduce shortly.} We have also used $C_1^{\f ss\f}$ to denote the $t$-channel contribution to the $\<\f ss \f\>$ generating function. To include $u$-channel contributions, we also need to define:
\begin{align}
{M_1^u}(z,\bar{h}) \equiv \left( {\begin{array}{*{20}{l}}
{C_1^{\phi s \phi s}(z,\bar{h})}&{C_1^{\phi t \phi s}(z,\bar{h})}\\
{C_1^{\phi t \phi s}(z,\bar{h})}&{C_1^{\phi t  \phi \f}(z,\bar{h})}
\end{array}} \right)~, \label{eq:M1u}
\end{align}
so the full matrix is $M_1(z,\bar{h})=M_1^t(z,\bar{h})+(-1)^{\ell}M_1^u(z,\bar{h})$.

Next, we want to fit this matrix to some function. To do this, we define a Hamiltonian of operator twists,
\begin{align}
{H}(\bar h) = \left( {\begin{array}{*{20}{c}}
{{h_{[\phi s]_{0}^1}(\bar{h})}}&0\\
0&{{h_{[\phi t]_{ 0}^1}(\bar{h})}}
\end{array}} \right)~,
\end{align}
and a matrix of OPE coefficients
\begin{align}
\Lambda (\bar h) = \left( {\begin{array}{*{20}{c}}
{{f_{\phi s[\phi s]_{0}^1}(\bar{h})}}&{{f_{\phi s[\phi t]_{ 0}^1}(\bar{h})}}\\
{{f_{\phi t[\phi s]_{0}^1}(\bar{h})}}&{{f_{\phi t[\phi t]_{ 0}^1}(\bar{h})}}
\end{array}} \right)~.
\end{align}
We then have:
\begin{align}
M_1(z,\bar{h})\approx 2^{-\ell}\Lambda (\bar h){z^{H(\bar h)}}\Lambda {(\bar h)^T} =2^{-\ell}\sum\limits_{\cO=[\phi t]_{0}^1,[\phi s]_0^1}
 \left( {\begin{array}{*{20}{c}}
{{f^2_{\phi s\cO}(\bar{h})}}&{{f^2_{\phi s\cO}(\bar{h})}}\\
{{f^2_{\phi t\cO}(\bar{h})}}&{{f^2_{\phi t\cO}(\bar{h})}}
\end{array}} \right)z^{h_{\cO}(\bar{h})}~.
\end{align}
We can then follow a similar strategy as before. To obtain an approximation for the twist Hamiltonian, we diagonalize the following matrix
\begin{align}
H\approx \text{diag}\left(\text{eigenvalues}[{M^{ - 1}}(z,\bar{h})\partial M(z,\bar{h})]\right)\big|_{z=z_0}~.
\end{align}
This gives the twist for ${\left[ {\phi s} \right]_0^1}$ and ${\left[ {\phi t} \right]_0^1}$. As before we have to introduce an arbitrary $z_0$ when evaluating the matrices, which we will take to be $z_0=0.1$.

One can also compute the OPE coefficient matrix using similar ideas. Defining $M'=\partial M$, we can perform the Cholesky decomposition for $M$ and $M'$:
\begin{align}
&M = {U_1}U_1^T~,\nonumber\\
&M' = {U_2}U_2^T~.
\end{align}
Then we have
\begin{align}
&{U_1} = \Lambda_1(\bar{h}){z^{H/2}}Q_1^T~,\nonumber\\ 
&{U_2} = \Lambda_1(\bar{h}){z^{H/2}}{H^{1/2}}Q_2^T~. \label{eq:U1}
\end{align}
To solve for $Q_{1,2}$ we use,
\begin{align}
U_1^{ - 1}{U_2} = {Q_1}{H^{1/2}}Q_2^T~.
\end{align}
One can then compute $Q_1$ and $Q_2$ by taking the singular value decomposition (SVD) of $U_1^{ - 1}{U_2}$. Finally, this allows us to compute $\Lambda_1(\bar{h})$ from (\ref{eq:U1}).

The above treatment can be generalized easily to a general number of mixings involving arbitrary double-twist towers $n$ \cite{Simmons-Duffin:2016wlq}. One can show that for a trivial $1\times 1$ matrix, the method reduces to the case described in the previous section.

The mixing effects computed using the twist Hamiltonian can be significant. In appendix \ref{twistcompare}, we make a direct comparison between the results of using or not using the twist Hamiltonian. This comparison shows that using the twist Hamiltonian is crucial to resolve important mixing effects.

\section{Inversion formula for the O(2) model}
\label{InvForO2}
\subsection{Generating functions}
\label{subsec:O2GenFunc}
Now that we have all the ingredients in place, we can turn to the O(2) model. We work with the following values for dimensions and OPE coefficients of light operators:
\begin{align}
&\Delta_{\phi}=0.519088~, \quad \Delta_{t}=1.23629~, \quad \Delta_{s}=1.51136~,
\nonumber \\
&f_{\phi\phi s}=0.687126~, \quad f_{\phi\phi t}=1.213408~, \quad f_{sss}=0.830914~, \quad f_{tts}=1.25213~,
\nonumber \\
&C_{J}/C_{J}^{\text{free}}=0.904395~, \quad C_{T}/C_{T}^{\text{free}}=0.944056~.
\end{align}
Here $C_{J,T}^{\text{free}}$ are the values of the central charge in the free O(2) model. The couplings $f_{\phi\phi J}$ and $f_{\phi\phi T}$ are then fixed by Ward identities:
\begin{align}
f_{\phi \phi J}=\frac{1}{4\pi \sqrt{C_{J}}}~, \qquad f_{\f\f T}=\frac{3\Delta_{\f}}{8\pi \sqrt{C_{T}}}~,
\end{align}
and then for scalar $O$ we have $f_{OOJ}/f_{\f\f J}=q_{O}$ and $f_{OOT}/f_{\f\f T}=\Delta_{O}/\Delta_{\f}$. Here $C_{J,T}$ are the normalization of $J$ and $T$~\cite{Osborn:1993cr}. We will also need the following data about the lightest charge 3 scalar $\chi$ found using the extremal functional,
\begin{align}
\Delta_{\chi}=2.1086~,\nonumber\\
f_{\phi\chi t}=1.3505~.
\end{align}

When applying the inversion formula, the only new ingredient with the O(2) model is that we now have to include group theory factors, corresponding to $6j$-symbols, when computing the generating function. For example, for the correlator $\<\f\f\f\f\>$ we have:
\begin{align}
C^{\f\f\f\f}_{r_s}(\bar{h})=\sum\limits_{r_t}\sum\limits_{O'\in r_t}\mathcal{M}_{r_sr_t}^{\f\f\f\f}\left(-\frac{1}{2}\right)^{\ell_{O'}}f_{\f\f O'}^{2}C^t_{O'}(z,\bar{h})~,
\label{eq:Cffff}\end{align}
where $C^t_{O}(z,\bar{h})$ is the contribution to the generating function from a single block (\ref{eq:inversionOmega}) and $r_{s,t}$ are the representations appearing in the $s$ and $t$-channel. The explicit form of the O(2) crossing matrices $\mathcal{M}$ are given in appendix \ref{app:Crossing_Matrices}. 

Now the only remaining problem is to determine which operators we should invert on the right-hand side. For $\<\f\f\f\f\>$ we will follow \cite{Albayrak:2019gnz} and only include the light isolated operators, $s$, $t$, $J$, and $T$. From this we can then make predictions for $\delta h_{[\phi\phi]_0^{0^{\pm},2}}$ and $f_{\f\f[\phi\phi]_0^{0^{\pm}2}}$ which can in turn be used in the inversion formula. However, due to the $\sin^2$ factor in (\ref{eq:inversionOmega}) and their small anomalous dimensions we can ignore their effects for this correlator.

The next simplest case to consider is the generating function for charge 3 operators\footnote{In this case, we do not have a matrix of generating functions, but to keep the notation consistent we continue to use $M$.}
\begin{align}
M_3(z,\bar{h})=C^{\f tt \f}(z,\bar{h})+(-1)^{\ell}C^{\f t \f t}(z,\bar{h})~.
\end{align}
To compute $C^{\f tt \f}(z,\bar{h})$ we invert the isolated operators $s$, $J$, and $T$, and also the double-twist operators $[\f\f]_{0}^{0^{\pm}}(\bar{h})$. To compute $C^{\f t \f t}(z,\bar{h})$ we only include $\phi$ which gives the leading effect. The charge 3 scalar, $\chi$, does not contribute because $\mathcal{M}^{\f t\f t}_{3,3}=0$. 

The charge 3 generating function is the simplest case where we must perform an infinite sum over spin. Specifically, we include the sum over the $[\f\f]_{0}^q$ trajectories for $q=0^{\pm}$. First, we determine the couplings $f_{\f\f [\f\f]_{0}^{\pm}}(\bar{h})f_{tt [\f\f]_{0}^{\pm}}(\bar{h})$ for all $\bar{h}$. To do this we study the $t$-channel generating function for $\<\f\f tt\>$. The dominant contribution comes from $\phi$ exchange in the $t$-channel,
\begin{align}
C^{\f\f tt}_{0^{\pm}}(z,\bar{h})=\mathcal{M}^{\f\f tt}_{0^{\pm},1}f_{\phi\phi t}^{2}C^t_\f(z,\bar{h})~.
\end{align}
At small $z$ this generating functions contains the powers $z^{2h_{\f}}$ and $z^{2h_{t}}$ corresponding to computing $f_{\f\f [\f\f]_{0}^{\pm}}(\bar{h})f_{tt [\f\f]_{0}^{\pm}}(\bar{h})$ and $f_{\f\f [tt]_{0}^{\pm}}(\bar{h})f_{tt [tt]_{0}^{\pm}}(\bar{h})$, respectively. We will assume we can determine the couplings for the former by following the na\"ive procedure and looking at the coefficient of $z^{2h_\f}$. This is justified because there are no other light operators that $[\f\f]_{0}^{0^{\pm}}$ mixes with. Later we will compare our prediction for $f_{\f\f T}$ and $f_{\f\f J}$ from analytics with the exact answer from the Ward identity. Given these OPE coefficients, we can perform the large-spin sums using identities given in section \ref{sec:DTI} and appendix \ref{app:SL2Sums}.

The procedure for the charge 1 sector is identical, except as discussed in section \ref{sec:twisthamiltonian} we must study a matrix of generating functions, see (\ref{eq:M1t}) and (\ref{eq:M1u}). To compute the matrix elements, we again include the light operators $s$, $\phi$, $t$, $T$, $J$ and the double-twist operators $[\f\f]_0^{0^{\pm},2}(\bar{h})$ in the $t$-channel. To compute the OPE coefficients $f_{ss[\f\f]_0^{0^{+}}}(\bar{h})$ and $f_{st[\f\f]_0^{2}}(\bar{h})$ we look at the generating functions for $\<\f\f ss\>$ and $\<\f\f st\>$, respectively, and include $\phi$ exchange in the $t$-channel.

To study the charge 4 sector, we must resolve potential mixing between the charge 4 double-twist operators $[tt]_{0}^{4}$ and $[\phi \chi]_0^4$. At asymptotically large spin they have similar twists:
\begin{align}
\lim_{\ell\to\oo} h_{[tt]_0^4}(\bar{h})&= 1.236~,\nonumber\\
\lim_{\ell\to\oo} h_{[\phi\chi]_0^4}(\bar{h})&= 1.314~.
\end{align} 
In order to do this, we study the following matrices of generating functions,
\begin{align}
M^{t}_{4}(z,\bar{h})\equiv \left( {\begin{array}{*{20}{c}}
{{C^{tttt}_{4}(z,\bar{h})}}&{{C^{tt\chi\phi}_{4}(z,\bar{h})}}\\
{{C^{\chi\phi tt}_{4}(z,\bar{h})}}&{{C^{\phi\chi\chi\phi}_{4}(z,\bar{h})}}
\end{array}} \right)~,
\nonumber\\
M^{u}_{4}(z,\bar{h})\equiv \left( {\begin{array}{*{20}{c}}
{{C^{tttt}_{4}(z,\bar{h})}}&{{C^{tt\chi\phi}_{4}(z,\bar{h})}}\\
{{C^{\chi\phi tt}_{4}(z,\bar{h})}}&{{C^{\phi\chi\phi\chi}_{4}(z,\bar{h})}}
\end{array}} \right)~.
\end{align} 
For the even spin trajectories, the full matrix is $M_{4,\text{even}}(z,\bar{h})=M^t_{4}(z,\bar{h})+M^u_{4}(z,\bar{h})$. For simplicity, we will only include the isolated operators $s$, $\phi$, $t$, $J$, and $T$ in the generating function. As we will demonstrate in section \ref{comparing}, this yields accurate results for the even spin charge 4 sector. To further improve the results, we would need to understand how the double-twist operators mix with triple and quadruple twist operators, which lies outside the scope of the current work. 

For the odd spin charge 4 sector the matrix of generating functions collapses to a single function,
\begin{align}
M_{4,\text{odd}}(z,\bar{h})=C^{\phi\chi\chi\phi}_{4}(z,\bar{h})-C^{\phi\chi\phi\chi}_{4}(z,\bar{h})~,
\end{align}
because the $[tt]_0^4(\bar{h})$ trajectory only exists for even spin. For the same reason, we cannot compare our prediction to the current numerical results, which do not include $\chi$ as an external operator. 

For both sectors we must compute $C^{\phi\chi\chi\phi}_{4}(z,\bar{h})$, which involves the OPE coefficients, $f_{\chi\chi O}$ for $O=s,J,T$. The latter two can be fixed by Ward identities, but the OPE coefficient $f_{\chi\chi s}$ is not currently known. While we expect this will give a small contribution in comparison to the exchange of $\phi$ for $C^{\phi\chi\phi\chi}_{4}(z,\bar{h})$, it would still be useful to have an estimate for this coupling. To get an analytic estimate for this OPE coefficient we will argue that $\chi$ sits on the $[\phi t]_{0}^{3}(\bar{h})$ trajectory. Evidence for this conjecture, that in the O(2) model we have analyticity down to spin 0, will be given in section \ref{comparing} where we show the analytic prediction for $\Delta_{\chi}$, obtained by studying $M_3(z,\bar{h})$, is close to the numerical result. Assuming analyticity does hold down to spin 0, we can compute $f_{\chi\chi s}$ by studying the generating function for $\<s\chi\phi t\>$:
\begin{align}
M_{3,\text{off-diag}}(z,\bar{h})=C^{s\chi \phi t}(z,\bar{h})+(-1)^{\ell}C^{s\chi  t\phi}(z,\bar{h})~.
\end{align}
In the small $z$ expansion this generating function contains two powers, $z^{h_s+h_\chi}$ and $z^{h_\phi +h_t}$, which determine the product of couplings $f_{\phi t [\phi t]_0^3}(\bar{h})f_{s\chi [\phi t]_0^3}(\bar{h})$ and $f_{\phi t [s\chi]_0^3}(\bar{h})f_{\s\chi [s\chi]_0^3}(\bar{h})$. When we continue down to spin $0$, the former becomes $f_{\phi t \chi} f_{\chi\chi s}$ and we already know $f_{\phi t \chi}$ from numerics. From this approach we find the estimate 
\begin{align}
f_{\chi\chi s}\approx1.45~.
\end{align}
This is a rough estimate because we are continuing our results down to $\bar{h}_{\chi}\approx 1.05$. For large $\bar{h}$ the contribution of heavy operators $O$ in the inversion formula is suppressed by $\bar{h}^{-h_{O}}$ while around $\bar{h}\sim1$ we lose this suppression. 
It would be interesting to compare this result with future determinations from the numerical bootstrap and/or Monte Carlo simulations.
\subsection{Double-twist sums}
To explain how sums over double-twist operators are performed in practice in the O(2) model, let us study the generating function for charge 3 operators. 
Specifically, we look at:
\begin{align}
C^{\phi tt \phi}_3(z,\bar{h})\supset & \sum\limits_{\ell'=4,6,\cdots}\left(-\frac{1}{2}\right)^{\ell'}f_{\phi \f[\phi \f]_{0}^{0^+}}(\bar{h}_{\ell}')f_{tt[\phi \f]_{0}^{0^+}}(\bar{h}_{\ell}')C^{t}_{[\phi \f]_{0}^{0^+}(\bar{h}_{\ell}')}(z,\bar{h})
\nonumber \\
 +&\sum\limits_{\ell'=3,5,\cdots}\left(-\frac{1}{2}\right)^{\ell'}f_{\phi \f[\phi \f]_{0}^{0^-}}(\bar{h}_{\ell}')f_{tt[\phi \f]_{0}^{0^-}}(\bar{h}_{\ell}')C^{t}_{[\phi \f]_{0}^{0^-}(\bar{h}_{\ell}')}(z,\bar{h})~.\label{eq:Ch3GenFuncLargeSpinPt0}
\end{align}
where we used that $\mathcal{M}^{\f tt \f}_{3,0^{\pm}}=1$. Also, to keep the notation compact, when $\bar{h}_{\ell}'$ appears as an argument, e.g.,~in $f_{tt[\phi \f]_{0}^{0^+}}(\bar{h}_{\ell}')$, we leave it implicitly defined as the solution to $\bar{h}_\ell'=h_{[\f\f]_0^{0^+}}(\bar{h}_\ell')+\ell'$. Next we split each $\ell'$ sum into two pieces, a sum over operators with $\ell'<\ell_*$ and those with $\ell'\geq\ell_*$. When performing the large-spin part of the sum we will set $\bar{h}=2h_{\f}+\ell$. To find the large-spin data, we use results from the lightcone bootstrap \cite{Fitzpatrick:2012yx,Komargodski:2012ek,Simmons-Duffin:2016wlq}:
\begin{align}
\left(-\frac{1}{2}\right)^{\ell'}f_{\phi \f[\phi \f]_{0}^{0^\pm}}(\bar{h}_{\ell}')f_{tt[\phi \f]_{0}^{0^\pm}}(\bar{h}_{\ell}')\sim\pm f_{\f\f t}^{2} \frac{2\Gamma\left(2(h_t-h_\f)\right)\Gamma(2h_\f)}{\Gamma^2(h_t)}S^{0,0}_{-h_t}(\bar{h}_\ell')~.\label{eq:OPECh3LargeSpin}
\end{align}
We also need to expand one of the sin factors which comes from the dDisc to leading order in the anomalous dimension,
\begin{align}
C^{t}_{[\f\f]_0^{0^{\pm}}(\bar{h}_\ell')}(z,\bar{h})\ \approx \ &2\pi\kappa_{2\bar{h}} \delta h_{[\f\f]_0^{0^{\pm}}}(\bar{h}_\ell') \sin\left(2\pi(h_\f-h_t)\right)
\nonumber \\ &\sum\limits_{p=0}^{\infty}\sum\limits_{q=-p}^{p}\widehat{\mathcal{A}}^{0,0}_{p,q}(2h_{\f})\frac{z^{h_\f+h_t}}{(1-z)^{2h_t}}\Omega^{h_\f h_t h_t h_\f}_{\bar{h},2h_\f+p,2h_t}k^{0,0}_{2h_\f+\ell'+q}(1-z)~.\label{eq:GenFuncCh3LargeSpin}
\end{align}
The anomalous dimensions for ${[\f\f]_0^{0^{\pm}}}(\bar{h})$ are in turn computed from the lightcone bootstrap for $\<\f\f\f\f\>$, and get contributions from $s$, $t$, $J$, and $T$ exchange:\footnote{Here we use the notation $O$ to denote the internal operator in $\<\phi\phi\phi\phi\>$.}
\begin{align}
\delta h_{[\f\f]_0^{0^{\pm}}}(\bar{h})\sim  \sum\limits_{O=s,t,T,J}\frac{\cM^{\f\f\f\f}_{0^{\pm}r_O}}{\cM^{\f\f\f\f}_{0^{\pm}0^+}}\left(-\frac{1}{2}\right)^{\ell_O}f_{\f\f O}^{2}\frac{\Gamma(2\bar{h}_O)}{\Gamma^2(\bar{h}_O)}\frac{S^{0,0}_{h_O-2h_\f}(\bar{h})}{S^{0,0}_{-2h_{\f}}(\bar{h})} \label{eq:dhLargespinCh3}~.
\end{align}
To use the identities given in appendix \ref{app:SL2Sums} we need to express all of the $\bar{h}'_\ell$ dependence in terms of $S^{0,0}_{a}(\bar{h}_\ell')$ directly. To do this, we use that in the limit of large $\bar{h}$ we have
\begin{align}
\frac{S_{a}(\bar{h})S_{b}(\bar{h})}{S_{c}(\bar{h})}\approx \frac{\Gamma(-c)^{2}\Gamma(-a-b+c)^{2}}{\Gamma(-a)^{2}\Gamma(-b)^{2}}S_{a+b-c}(\bar{h})~.\label{eq:3Sto1S}
\end{align}
While we can perform the sums over the operators in the $0^+$ and $0^-$ sector individually, in practice, it will be more useful to combine the sums. First, we can note from  (\ref{eq:OPECh3LargeSpin}) that the weighted product of OPE coefficients for these two trajectories has the opposite sign. The sign for the anomalous dimension is determined by a ratio of crossing matrices. Using the results summarized in appendix \ref{app:Crossing_Matrices}, we find at large $\bar{h}$:
\begin{eqnarray}
&\delta h_{[\f\f]_0^{0^+}}(\bar{h})\bigg|_{s,T} &\sim  \delta h_{[\f\f]_0^{0^-}}(\bar{h})\bigg|_{s,T}~,
\\
&\delta h_{[\f\f]_0^{0^+}}(\bar{h})\bigg|_{J} &\sim  \delta h_{[\f\f]_0^{0^-}}(\bar{h})\bigg|_J~,
\\
&\delta h_{[\f\f]_0^{0^+}}(\bar{h})\bigg|_{t} &\sim  -\delta h_{[\f\f]_0^{0^-}}(\bar{h})\bigg|_t~.
\end{eqnarray}
This implies if we include the effects from $s$, $J$, or $T$ on the anomalous dimensions, the sums over the two trajectories add destructively, i.e., we have alternating signs. In this case, we can use equation (\ref{eq:SL2Sumalt}), and the infinite sum does not lead to any new powers of $z$ in the generating function, i.e., there are no new multi-twist operators. By contrast, when we use the correction from $t$, we get a constructive sum and will find higher logs.

When we combine (\ref{eq:Ch3GenFuncLargeSpinPt0}), (\ref{eq:GenFuncCh3LargeSpin}), (\ref{eq:dhLargespinCh3}) for $t$ exchange, and \eqref{eq:3Sto1S} we find the following sum over spin, where we drop some overall coefficients:
\begin{align}
\lim_{\epsilon \rightarrow 0}\sum\limits_{\substack{ \ell'=\ell_*,\ell_*+1,...}}\Gamma(-\epsilon)^{2}S^{0,0}_{\epsilon}(2h_\f+\ell')k_{2h_{\f}+\ell'+q}(1-z)~.
\end{align}
The factor of $\Gamma(-\epsilon)^2$ comes from using $a=-h_t$, $b=h_t-2h_\f$, and $c=2h_\f$ in (\ref{eq:3Sto1S}) and seeing we na\"ively get a divergence from $\Gamma^{2}(0)S_{0}(\bar{h})$. However if we dropped the $\Gamma^2(0)$ and used $S_0(\bar{h})$ in the above sum we would get 0. Being careful with the limits and only taking $\epsilon$ to zero at the end gives the finite result:
\begin{align}
&\lim_{\epsilon \rightarrow 0}\sum\limits_{\substack{ \ell'=\ell_*,\ell_*+1,...}}\Gamma(-\epsilon)^{2}S^{0,0}_{\epsilon}(2h_\f+\ell')k_{2h_{\f}+\ell'+q}(1-z) \approx
\nonumber \\
& \frac{2^{2q}}{6} \left(\frac{3 \left(2 H_{h_0-2}+\log\left(\frac{z}{1-z}\right)\right) \left(2 (h_0-1) H_{h_0-2}+(h_0-1) \log\left(\frac{z}{1-z}\right)+2\right)}{h_0-1}+\pi ^2\right)\bigg|_{h_0=2h_{\f}+\ell_*}~,
\end{align}
where we used (\ref{eq:sum2dBlocks}) and (\ref{eq:shiftS}). Here $H_a$ is the harmonic number. We see that the sum over $\ell'$ has produced a $\log^2(z)$ factor while each individual block contains powers of $z^k$ and $z^k \log(z)$. The $\log^2$ factor is expected if the anomalous dimensions for $[\f t]_0^{3}(\bar{h})$ exponentiate. This also explains why when we included corrections to the $[\f\f]^{0^\pm}_0(\bar{h})$ anomalous dimensions from $s$, $J$, or $T$ we had to find an alternating sum: if the sum were constructive we would find new powers of $z$ corresponding to $[\phi s]_0$, $[\phi J]_0$ or $[\phi T]_0$. However, we are looking at the generating function for charge 3 operators and those double-twist trajectories all have charge 1.  

Finally putting all the ingredients together, we find the large spin contribution to the charge-3 generating function is:
\begin{align}
&C^{\phi tt \phi}_3(z,\bar{h})\bigg|_{\text{large-spin}}\approx \sum\limits_{\ell'=\ell_*,\ell_*+1,...}\left(-\frac{1}{2}\right)^{\ell'}f_{\phi \f[\phi \f]_{0}^{0^+}}(\bar{h}_{\ell}')f_{tt[\phi \f]_{0}^{0^+}}(\bar{h}_{\ell}')C^{t}_{[\phi \f]_{0}^{0^+}(\bar{h}_{\ell}')}(z,\bar{h})
\nonumber \\ \approx &\frac{4 \pi  f_{\f\f t}^4 \kappa_{2 \bar{h}} \Gamma (2 h_\f)^3 \Gamma (2 h_t) \sin (2 \pi  (h_\f-h_t)) \Gamma (2 h_t-2 h_\f) }{3\Gamma (h_t)^6 \Gamma (2 h_\f-h_t)^2} 
\nonumber \\ &\sum\limits_{p=0}^{\infty}\sum\limits_{q=-p}^{p}\widehat{A}^{0,0}_{p,q}(2 h_\f) \Omega^{h_\f,h_t,h_t,h_\f}_{\bar{h},2 h_\f+p,2 h_t}2^{2q} \frac{2^{2q}}{6(h_0-1)} \bigg(3 \left(2 H_{h_0-2}+\log\left(\frac{z}{1-z}\right)\right) 
\nonumber \\ & \left(2 (h_0-1) H_{h_0-2}+(h_0-1) \log\left(\frac{z}{1-z}\right)+2\right)+\pi^{2}(h_0-1)\bigg)\bigg|_{h_0=2h_{\f}+\ell_*}~.
\end{align}

\section{Comparing numerical and analytic predictions}\label{comparing}

In this section, we will present the predictions for the O(2) model using our analytic formalism, and compare them with our 40 extremal spectra from the numerical bootstrap at $\Lambda=35$. The operators we use in the inversion formula are summarized in table \ref{tab:operatorinclude}. 
\begin{table}[H]
\begin{center}
\begin{tabular}{@{}c|c|c|c@{}}
\hline
charge sector & isolated op. & Correlators & DTI \\
\hline
$0^{\pm}$ & $s,t,J,T$ & $\<\f\f\f\f\>$ & No \\
1 & $s,\phi,t,J,T$ & $\left( {\begin{array}{*{20}{c}}
{\<\phi ss\phi\> }&{\<\phi st \phi \>}\\
{\<\phi t s\phi \>}&{\<\phi tt\phi\>}
\end{array}} \right)$ & $[\phi\phi]_{0}^{0^{\pm},2}$ \\
2 even &  $s,t,J,T$  & $\<\f\f\f\f\>$ & No   \\
3 & $\phi,s,J,T$ & $\<\f tt \f\>$ & $[\phi\phi]_{0}^{0^{\pm}}$  \\
4 even & $s,\phi,J,T$ & $\left( {\begin{array}{*{20}{c}}
{\<tttt\>}&{\<tt\chi \phi\> }\\
{\<\phi \chi  tt\>}&{\< \phi \chi  \chi \phi \>}
\end{array}} \right)$ & No \\
4 odd &  $s,t,J,T$  & $\<\f\chi\chi\phi\>$  & No \\
\hline
\end{tabular}
\end{center}
\caption{\label{tab:operatorinclude} We compare analytic and numerical predictions of the conformal data in the 3d O(2) CFT. For analytics, the table summarizes different operators in the cross channels and different metrologies we use.}  
\end{table}

\subsection{Charge $0^{\pm},1,2,3$}
\begin{figure}
  \centering	
  \includegraphics[width=0.8\textwidth]{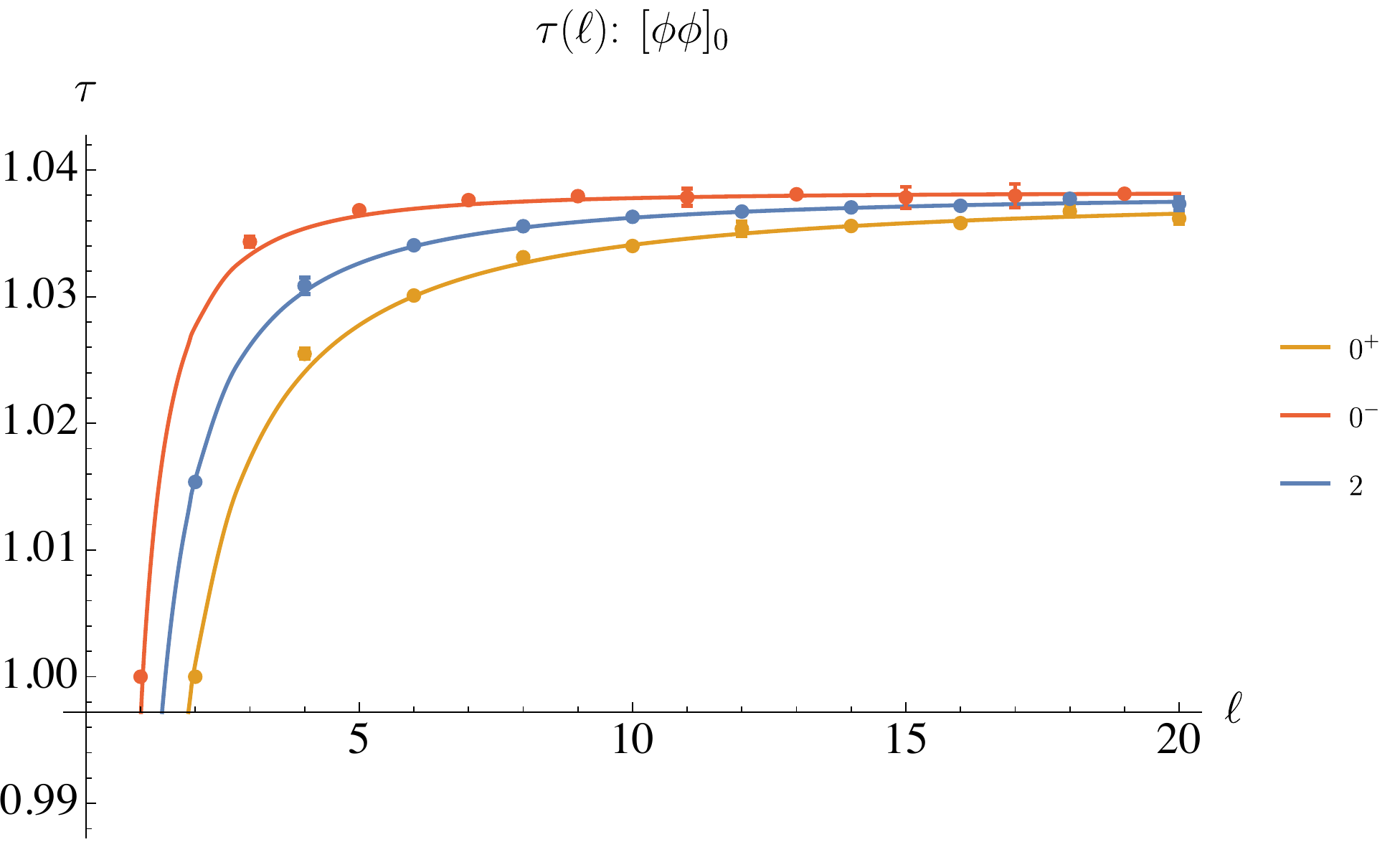}
  \caption{\label{phiphidimOPE} Analytical and numerical predictions for the twists $\tau=\De-\ell$ of the leading-twist Regge trajectories $[\phi\phi]^{0^{\pm},2}_{0}$ in the charge sectors $0^{\pm},2$ as a function of spin $\ell$. The lowest trajectory corresponds to charge $0^+$ (the trivial representation of O(2)). Note that both the spin-2 operator in the charge $0^+$ sector and the spin-1 operator in the charge $0^-$ sector have twist 1, corresponding to the stress tensor and O(2) current, respectively.}
\end{figure}

\begin{figure}
  \centering	
  \includegraphics[width=0.8\textwidth]{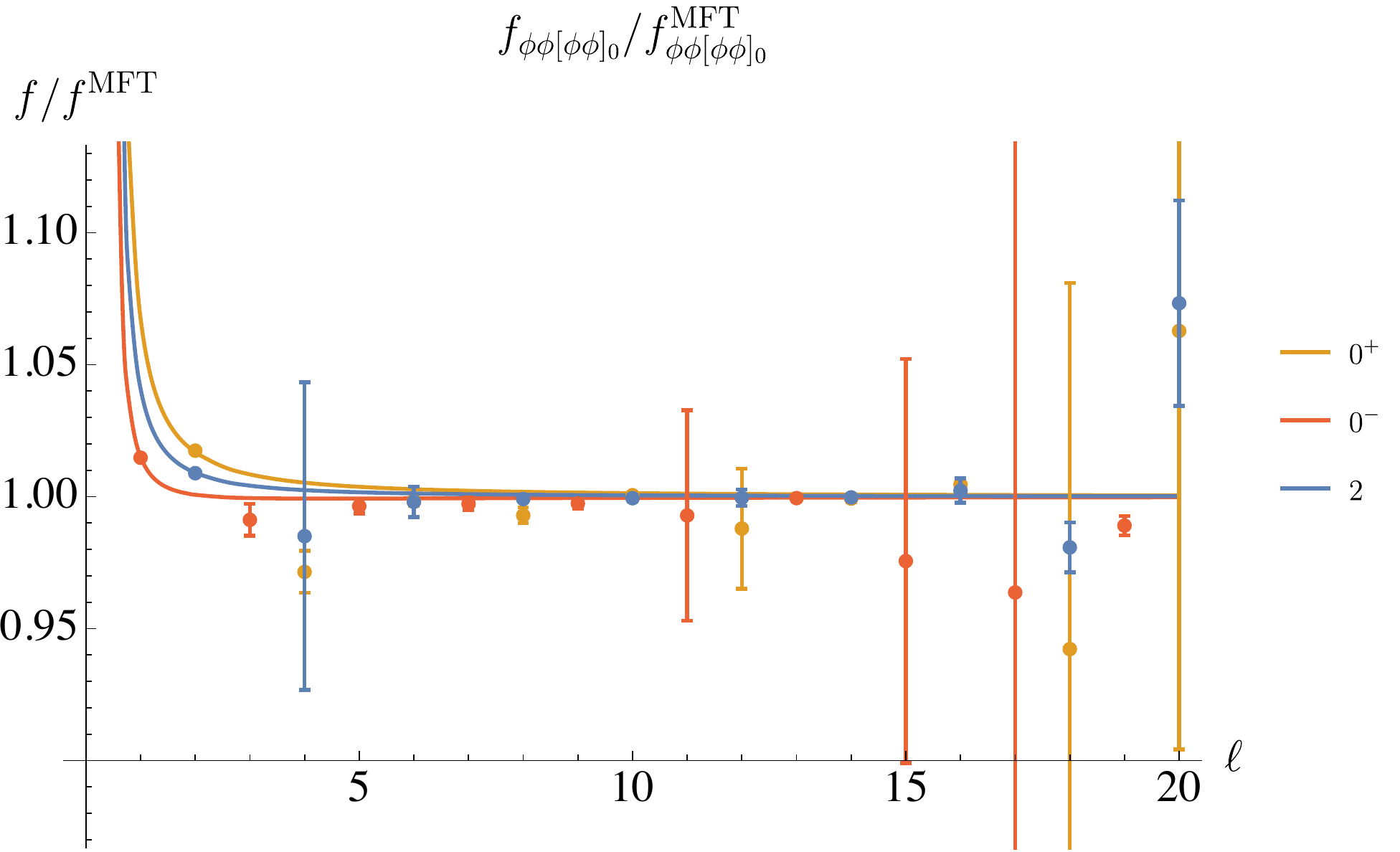}
  \caption{\label{phiphidimOPE2} Analytical and numerical predictions for the OPE coefficients of the leading-twist Regge trajectories $[\phi\phi]^{0^{\pm},2}_{0}$ in the charge sectors $0^{\pm},2$, as a function of spin $\ell$. We divide each coefficient by the corresponding coefficient in Mean Field Theory (MFT). }
\end{figure}

\begin{figure}
  \centering
  \includegraphics[width=0.8\textwidth]{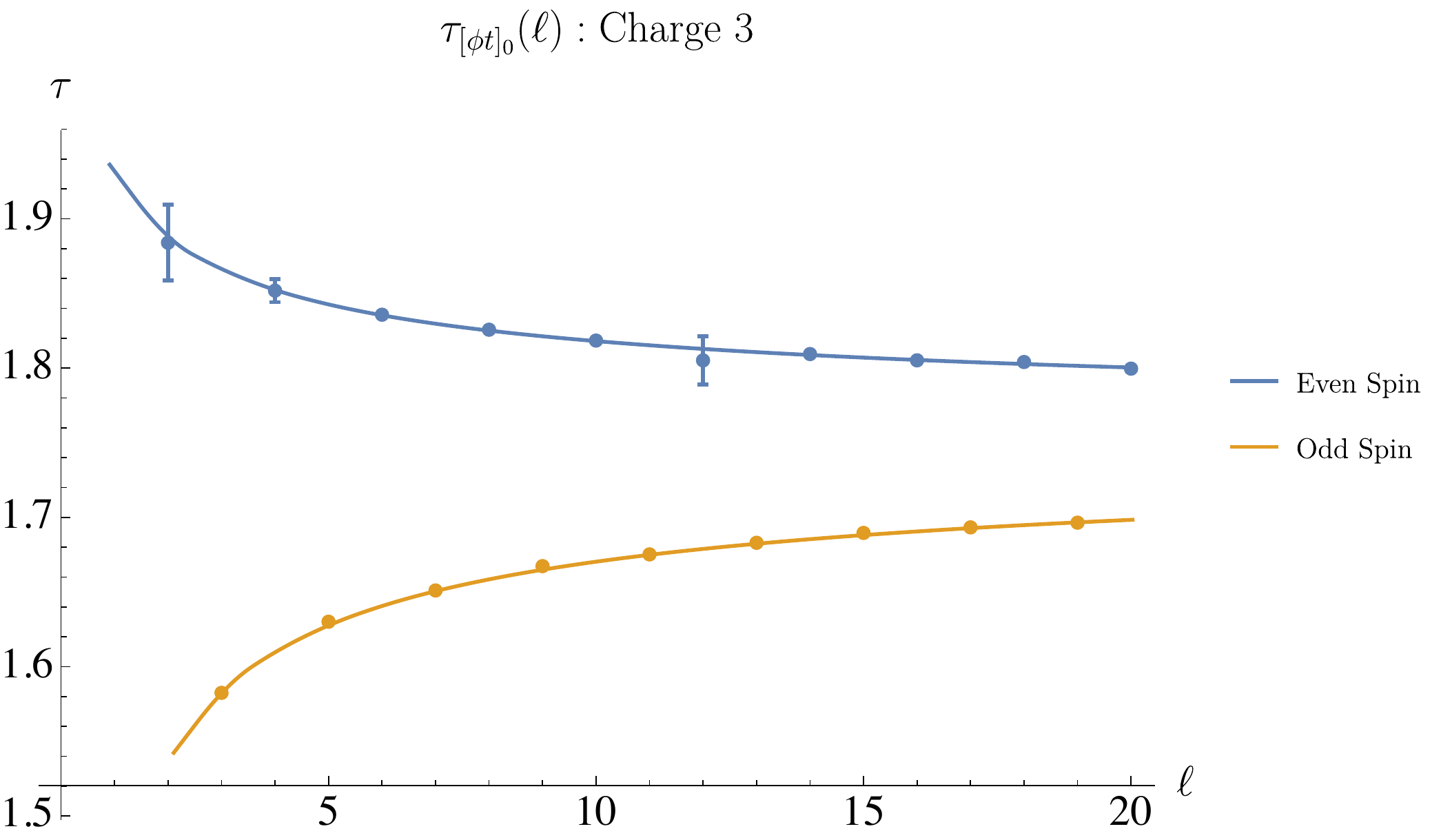}
  \caption{\label{3dim}Analytic and numerical predictions for the spectrum of the $[\phi t]^{3}_{0}$ Regge trajectories. The upper (blue) curve represents even spin operators, while the lower (orange) curve represents odd-spin operators.}
\end{figure}

\begin{figure}
  \centering
   \includegraphics[width=0.8\textwidth]{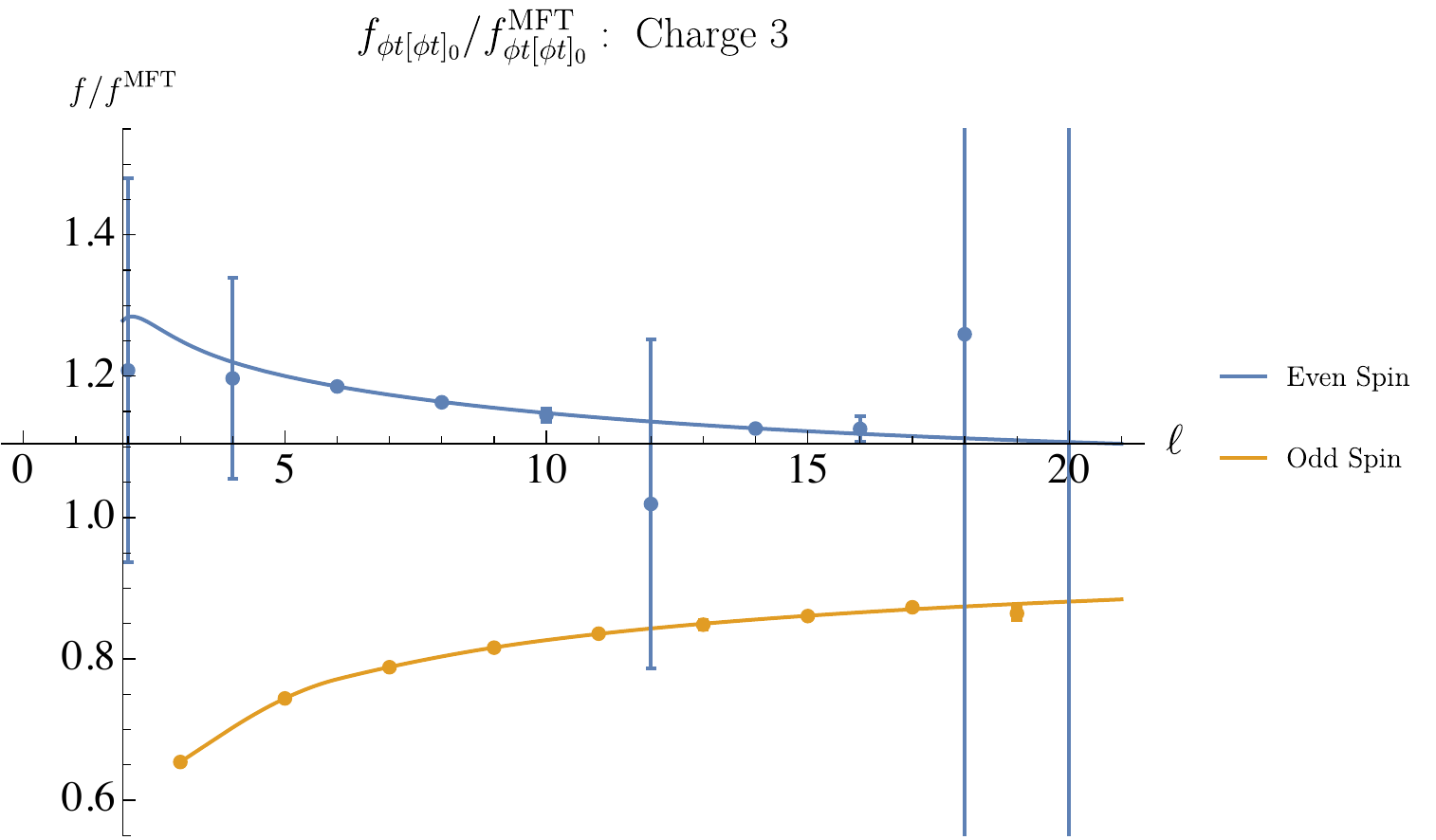}
  \caption{\label{3dim2} Analytic and numerical predictions for three-point coefficients of the $[\phi t]^{3}_{0}$ Regge trajectories. The upper (blue) curve represents even spin operators, while the lower (orange) curve represents odd-spin operators.  We divide each coefficient by the corresponding coefficient in Mean Field Theory (MFT).}
\end{figure}

\begin{figure}
  \centering
  \includegraphics[width=0.8\textwidth]{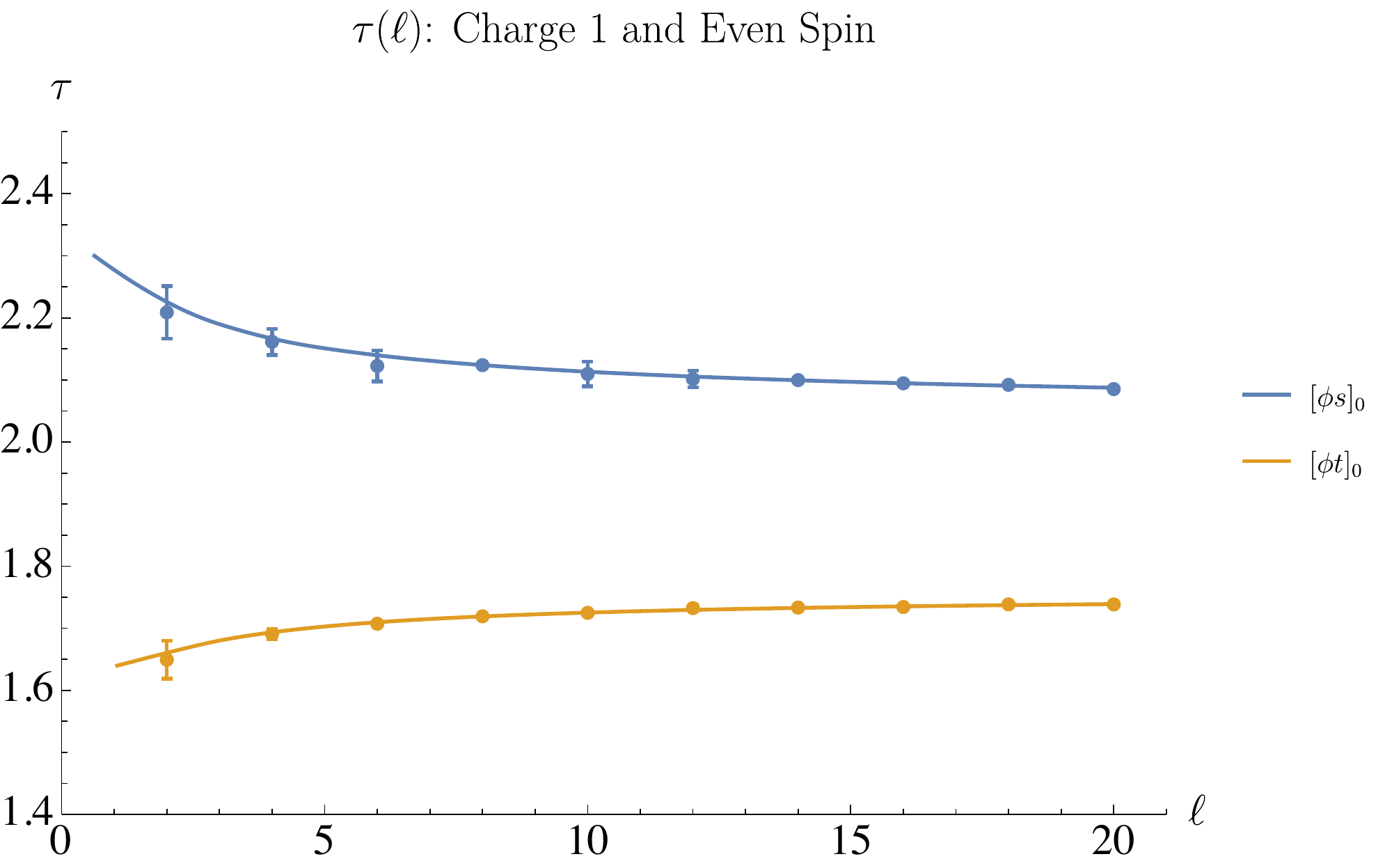}
  \caption{\label{1dim} Analytic and numerical predictions for twists of the double-twist trajectories $[\phi s]_0^1, [\phi t]_0^1$ with charge 1 and even spin. The blue curve represents the $[\phi s]_0^1$ trajectory, while the orange curve represents the $[\phi t]_0^1$ trajectory.}
\end{figure}

\begin{figure}
  \centering
  \includegraphics[width=0.8\textwidth]{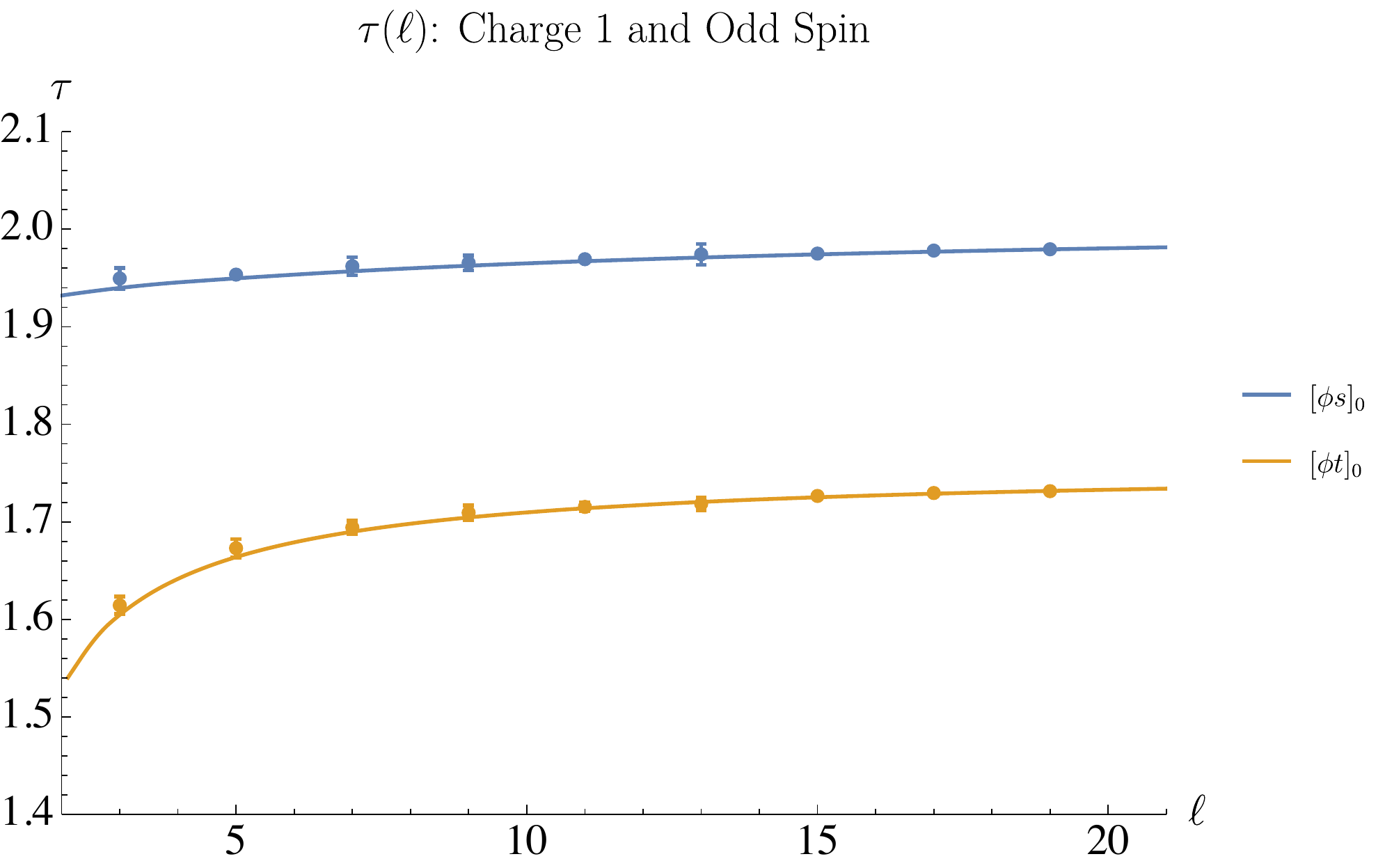}
  \caption{\label{1dim2} Analytic and numerical predictions for twists of the double-twist trajectories $[\phi s]_0^1, [\phi t]_0^1$ with charge 1 and odd spin. The blue curve represents the $[\phi s]_0^1$ trajectory, while the orange curve represents the $[\phi t]_0^1$ trajectory.}
\end{figure}

\begin{figure}
  \centering
  \includegraphics[width=0.8\textwidth]{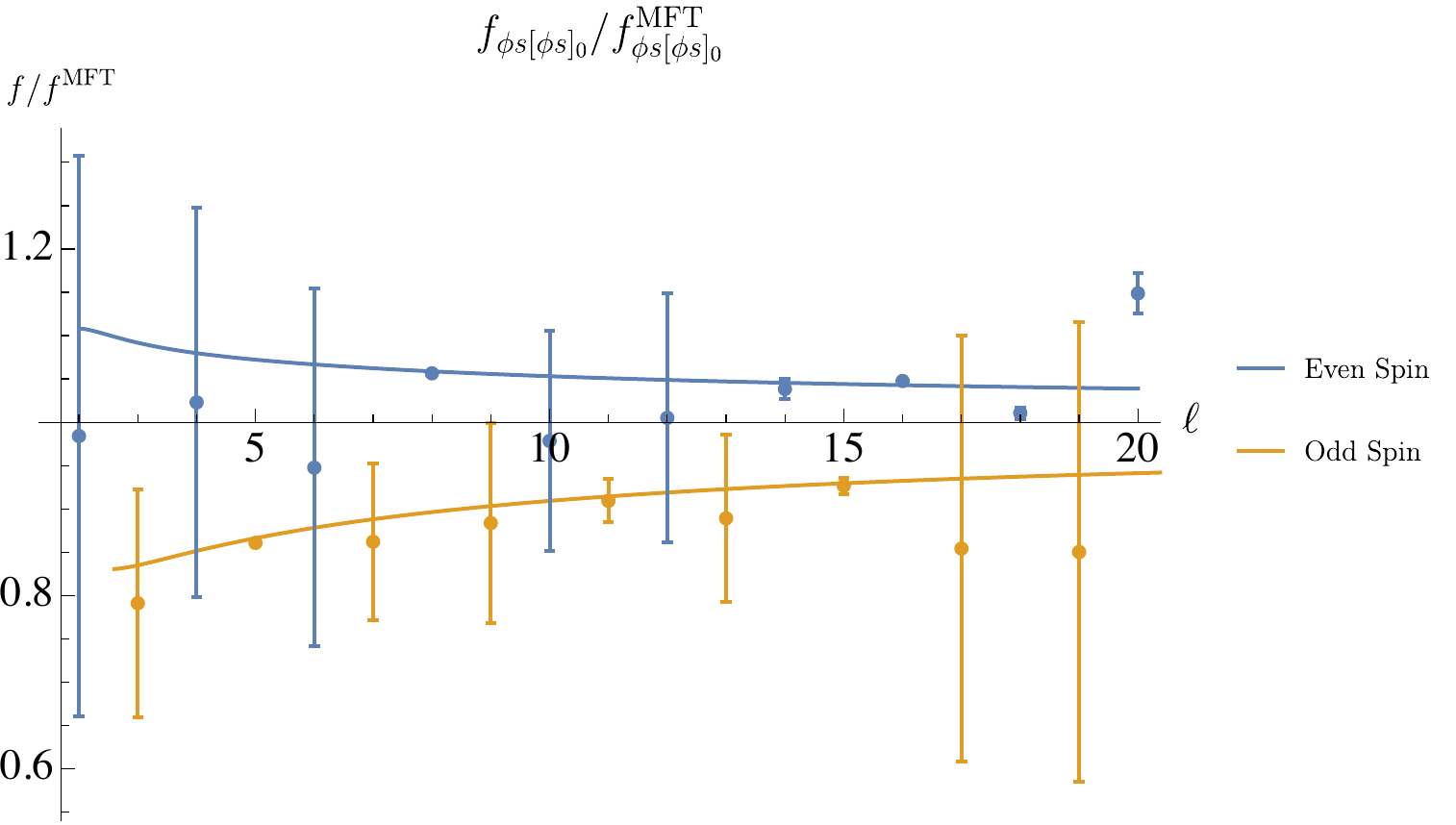}
  \caption{\label{1OPE} Analytic and numerical predictions for ``diagonal" three-point coefficients between external operators $\phi$, $s$ and operators on the double-twist trajectories $[\phi s]_0^1$ with charge $1$ and both even and odd spin. The blue curve represents the even-spin trajectory while the orange curve represents the odd-spin trajectory.  We divide each coefficient by the corresponding coefficient in Mean Field Theory (MFT).
  }
  \end{figure}

\begin{figure}
  \centering
  \includegraphics[width=0.8\textwidth]{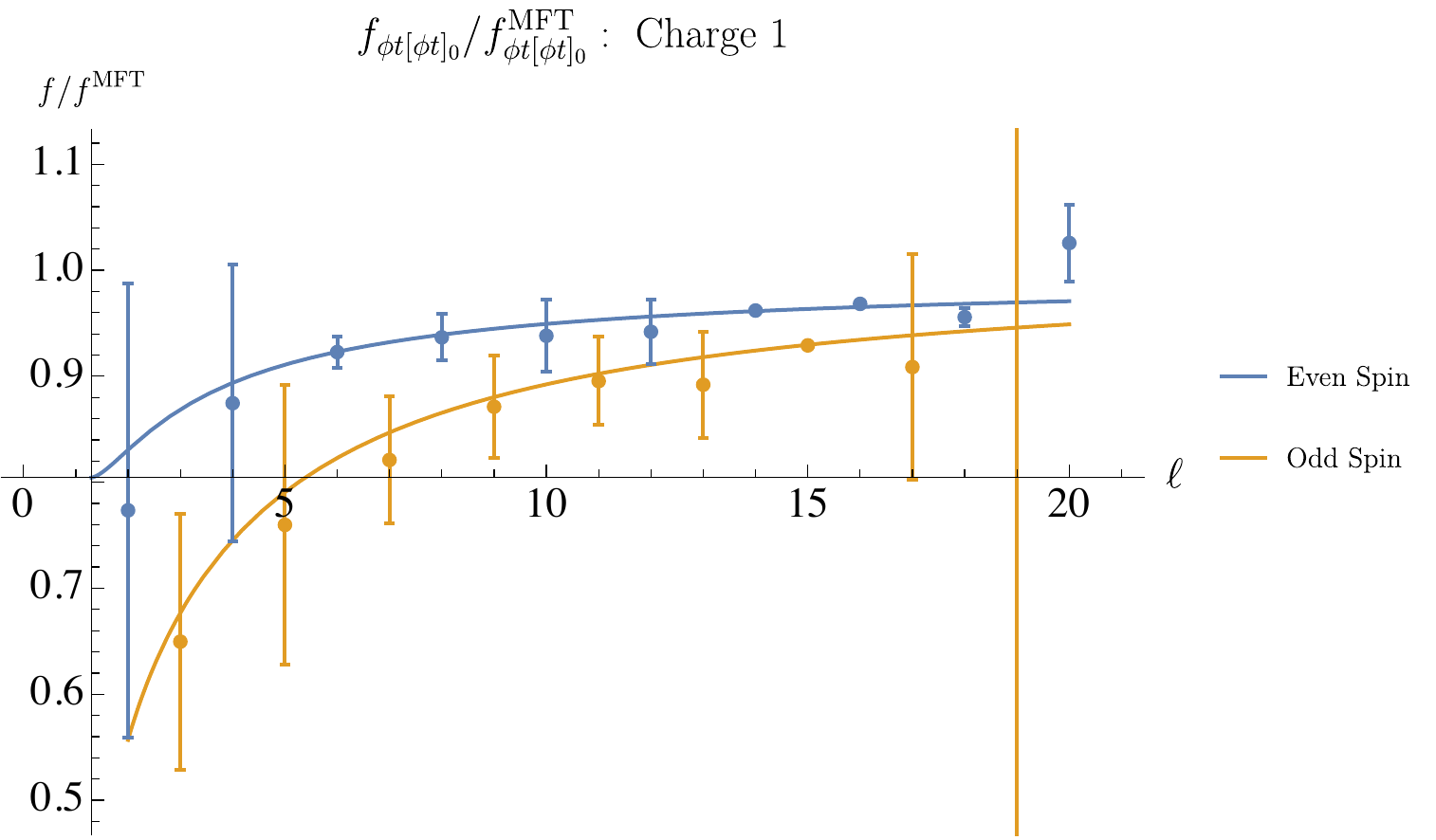}
  \caption{\label{1OPE2} Analytic and numerical predictions for ``diagonal" three-point coefficients between external operators $\phi$, $t$ and operators on the double-twist trajectories $[\phi t]_0^1$ with charge $1$ and both even and odd spin. The blue curve represents the even-spin trajectory while the orange curve represents the odd-spin trajectory.  We divide each coefficient by the corresponding coefficient in Mean Field Theory (MFT). }
\end{figure}

\begin{figure}
  \centering
  \includegraphics[width=0.8\textwidth]{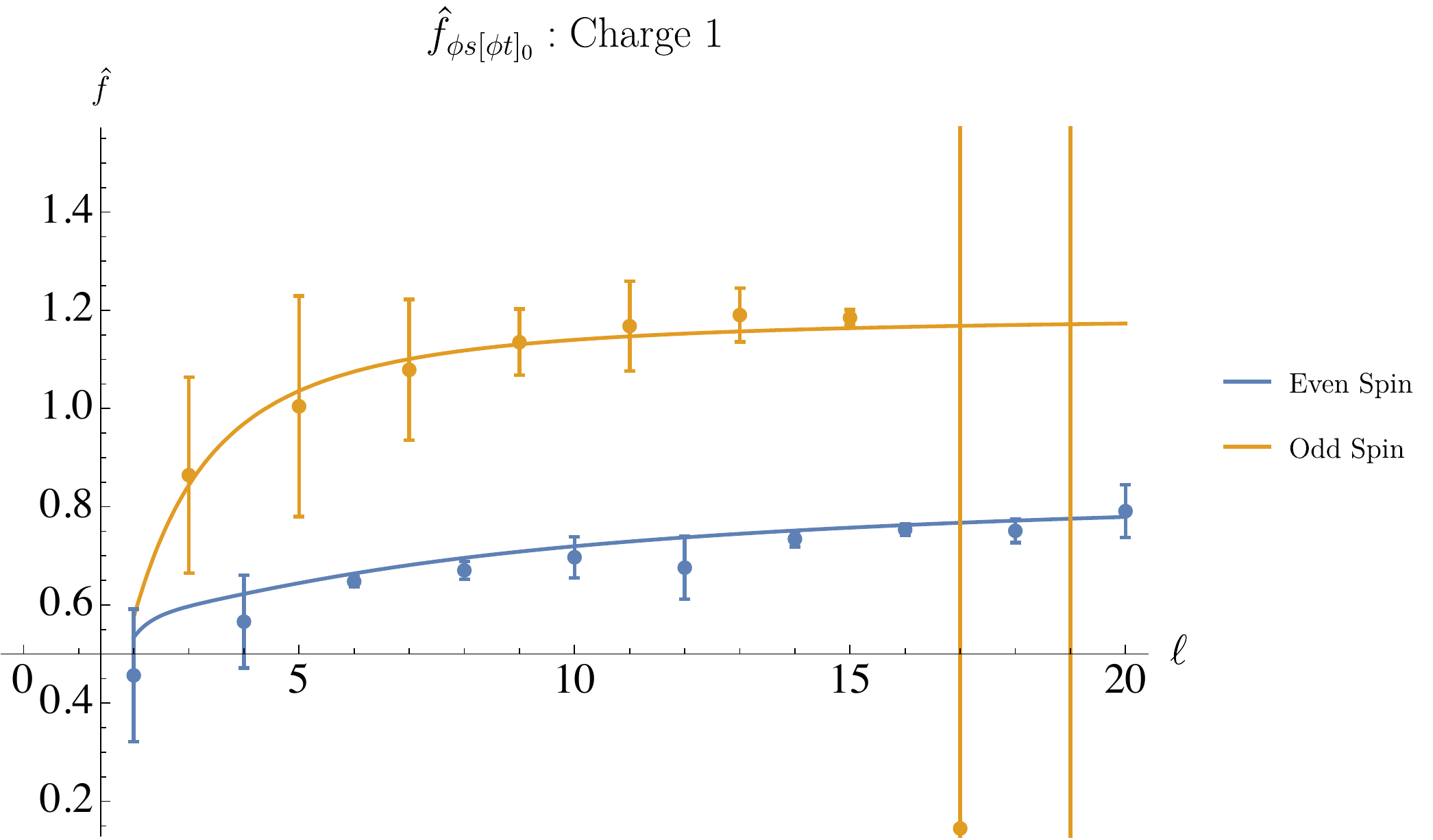}
  \caption{\label{1OPEoff} 
  Analytic and numerical predictions for ``off-diagonal" three-point coefficients between external operators $\phi$, $s$ and operators on the double-twist trajectories $[\phi t]_0^1$ with charge $1$ and both even and odd spin. The blue curve represents the even-spin trajectory while the orange curve represents the odd-spin trajectory. We normalize the OPEs such that the asymptotic values are one.}
\end{figure}

\begin{figure}
  \centering
  \includegraphics[width=0.8\textwidth]{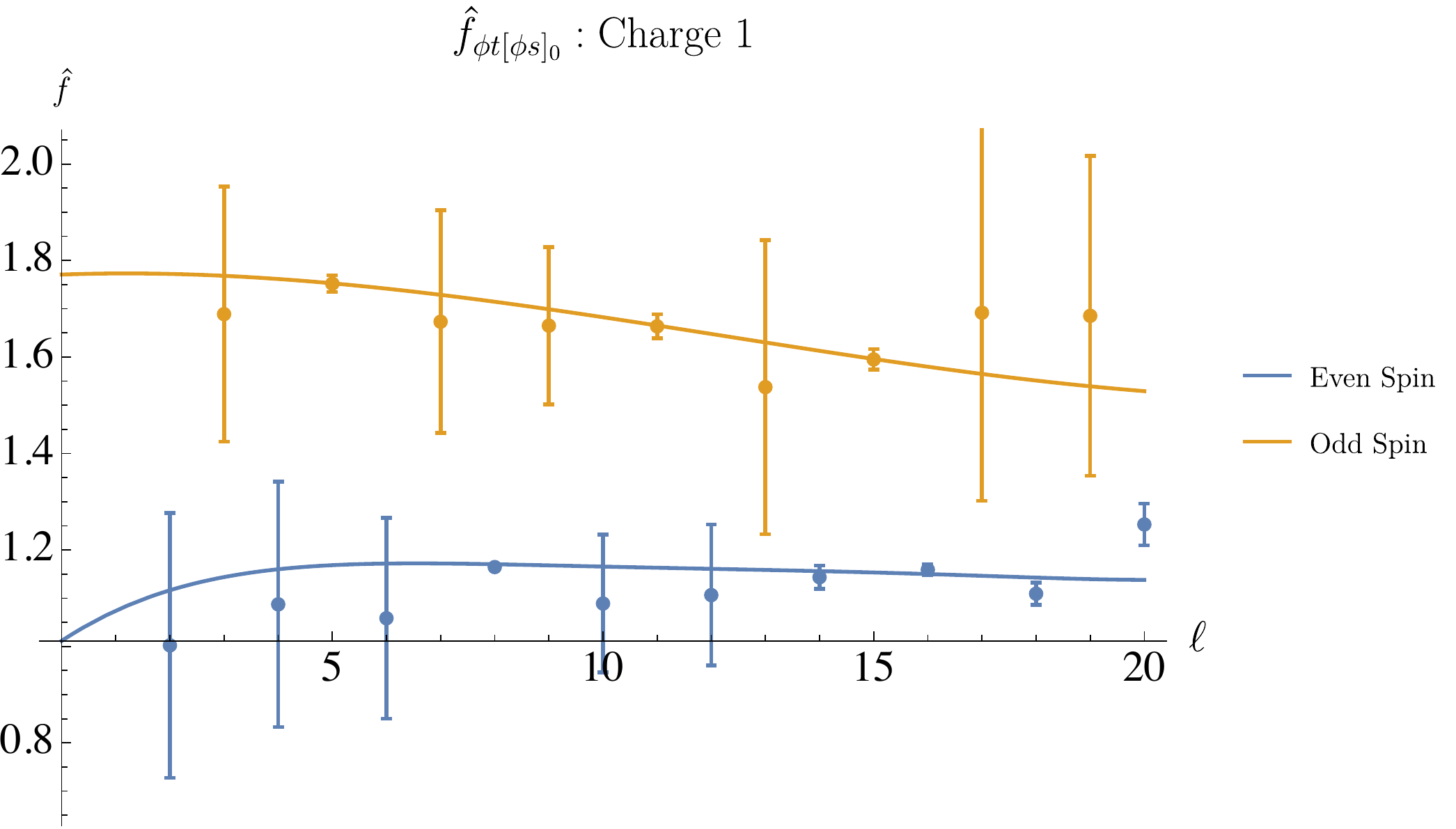}
  \caption{\label{1OPEoff2} 
   Analytic and numerical predictions for ``off-diagonal" three-point coefficients between external operators $\phi$, $t$ and operators on the double-twist trajectories $[\phi s]_0^1$ with charge $1$ and both even and odd spin. The blue curve represents the even-spin trajectory while the orange curve represents the odd-spin trajectory.
  We normalize the OPEs such that the asymptotic values are one.}
\end{figure}

First, we present plots for the low-lying operators and their OPE coefficients in the charge $0^{\pm},1,2,3$ sectors, and compare our analytic and numerical results. We make plots for twists and OPE coefficients in figures \ref{phiphidimOPE}, \ref{phiphidimOPE2}, \ref{3dim}, \ref{3dim2}, \ref{1dim}, \ref{1dim2}, \ref{1OPE}, \ref{1OPE2}, \ref{1OPEoff} and \ref{1OPEoff2} for different spins up to $\ell=20$. The OPE coefficients are normalized by the MFT coefficients when those are non-zero. The ratio of OPE coefficients then approaches 1 at large spins.  

For all plots, the curves are found via the inversion formula, and the dots show our numerical results found via the extremal functional method. In the inversion formula calculations, it is sufficient to work to $1^{\text{st}}$ order in dimensional reduction, i.e., $p_{\max}=1$, when comparing with the numerical predictions for operators with spin $\ell\geq2$. Working to higher orders in $p_{\text{max}}$ gives a negligible contribution which is not visible on the plot. The exceptions are figures \ref{phiphidimOPE} and \ref{phiphidimOPE2} where we work to $5^{\text{th}}$ order since we are extrapolating down to $\bar{h}=3/2$ to reach the spin-1 current $J$ and dimensional reduction converges slower for smaller $\bar{h}$. For the double-twist operators with spin $\ell_*\leq 20$ we will invert them individually using OPE data found from the inversion formula. For the operators with spins $\ell\geq 21$ we approximate their sum using their large-spin asymptotics.

In figure \ref{phiphidimOPE} and \ref{phiphidimOPE2}, we see that for the leading towers $[\phi\phi]^{0^{\pm}, 2}_0$  the analytic results for the twist $\tau$ agree with the numerical data to high precision. (Recall that we use the notation (\ref{eq:superscriptnotation}) for double-twist families.) Similar analytic results for charge $0^{\pm}$ were presented previously in \cite{Albayrak:2019gnz}. The new results in this figure are the analytic curve for charge $2$ and the numerical data for all three charged trajectories up to spin 20. To find these results we inverted the light operators, $s$, $t$, $J$, and $T$ as described in table~\ref{tab:operatorinclude}. While the prediction for the OPE coefficients from analytics agrees well with numerics at low spin, e.g., $\ell=1, 2$, there are large numerical errors and disagreement as we go to higher spins. It would be interesting to understand how to reduce these errors to better understand the validity of the analytic predictions. 

In figure \ref{3dim}, we see that the analytic predictions for the charge 3 spectrum also agree with the numerical values down to low spin. We see a similar agreement with the OPE coefficients in figure \ref{3dim2}, with the analytic curve passing through the numerical data points.  

In figure \ref{1dim} and \ref{1dim2}, we once again see agreement between the analytic and numerical methods for computing the charge 1 spectrum. This is also the first example where the twist Hamiltonian is used. As we demonstrate in appendix \ref{twistcompare}, it is crucial that we use it to obtain this agreement. In other words, the minimal twist, charge 1 double-twist operators in the O(2) model do have sizable mixing at low spin.

In figure \ref{1OPE} and \ref{1OPE2} we see that the numerics and analytics also agree for the charge 1 OPE data, although here the numerical errors are larger in comparison to the results for the spectrum. Taking these errors into account, we see the analytic curve is consistent with the prediction from the extremal functional.

Finally, by studying the charge-1 twist Hamiltonian we also have access to the off-diagonal OPE coefficients, $f_{\phi s[\phi t]_0^1}(\bar{h})$ and $f_{\phi t[\phi s]_0^1}(\bar{h})$ which are given in figure \ref{1OPEoff} and \ref{1OPEoff2}. These OPE coefficients very quickly go to zero at large $\bar{h}$, so in the above plots, we have normalized them by their asymptotic large-spin value. For both OPE coefficients, this can be found by studying $\<\phi s \phi t\>$ and inverting the contribution of $\phi$ in the $t$-channel. In all cases, the analytic results are in agreement with predictions from the extremal functional approach.

We summarize the lessons we have learned from this calculation:
\begin{itemize}
\item Generically, the numerics and analytics agree pretty well below spins $\ell\le 20$, which means that the analytic methods we have established seem to be effective across different double-twist sectors. Examples include the charge $0^\pm$ and charge 2 trajectories $[\phi\phi]_0^{0^{\pm}, 2}$, double-twist towers $[\phi t]_0^{1,3}$ built out of non-identical operators in the charge 1 and 3 sectors, and the importance of mixing with $[\phi s]_0^1$ in the charge 1 sector in order to obtain accurate predictions.
\item The twist Hamiltonian approach is seen to be very powerful in the charge 1 sector. We make a more detailed comparison about predictions with or without the twist Hamiltonian in appendix \ref{twistcompare}.
\item Generically, we have noticed that the OPE coefficients behave worse in the numerical calculations. This might occur due to some limitations of the extremal functional method. This might also be related to the sharing effect we have discussed previously: the predictions for OPE coefficients of the low-lying operators could be affected by contributions of fake operators at the gap we impose in the semidefinite program.
\end{itemize}

\subsection{Charge 4 predictions} 
\label{sec:ch4predictions}
Next, we discuss the analytic predictions for the charge 4 sector. As discussed earlier, in order to do this, we need to include $\chi$ as an external operator in order to resolve the mixing effects between $[tt]_0^4$ and $[\phi \chi]_0^4$.

\begin{figure}[H]
  \centering
  \includegraphics[width=0.8\textwidth]{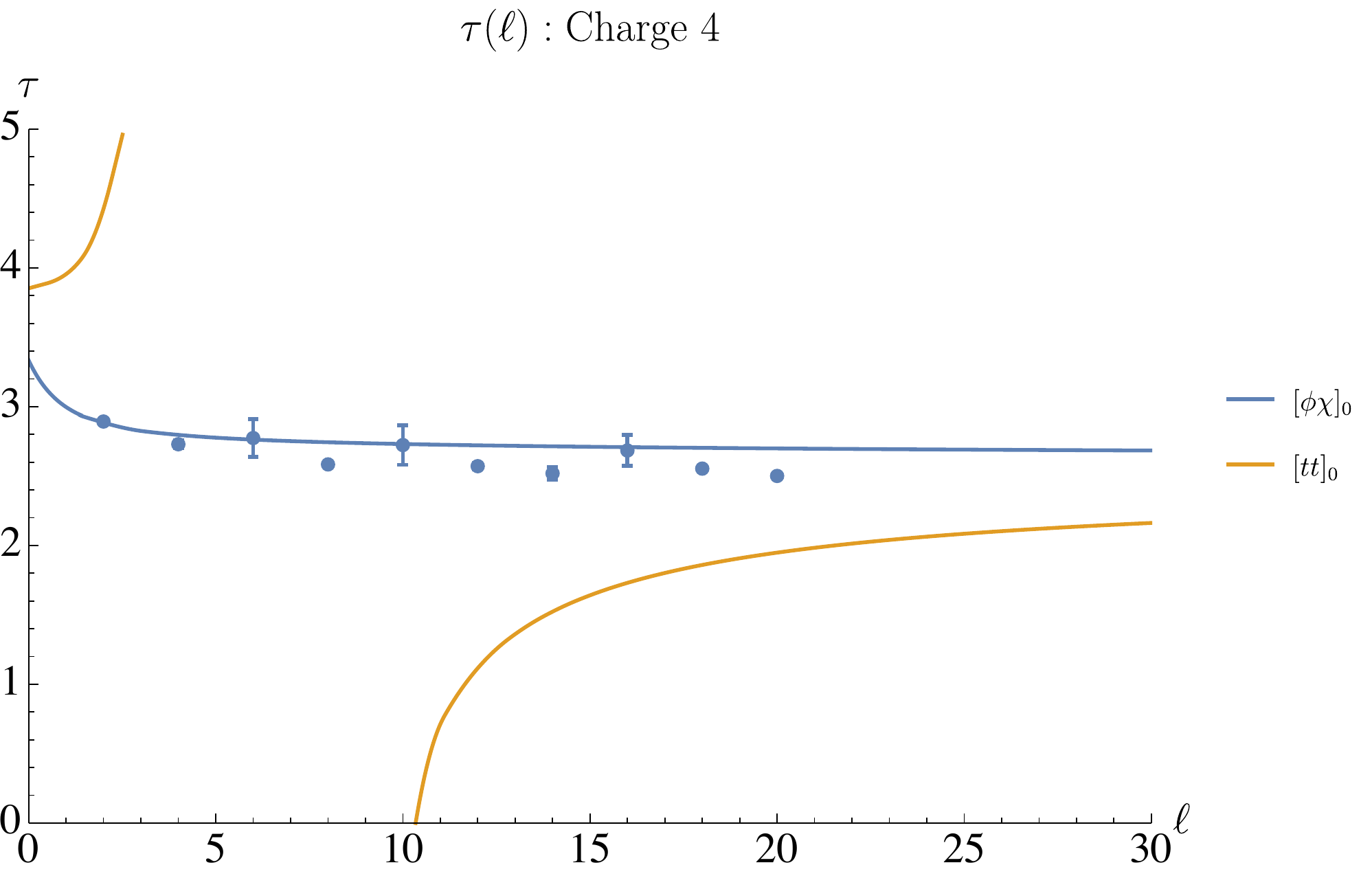}
  \caption{\label{4dimeven} Analytic predictions for charge 4 even spin anomalous dimensions, and their comparison with numerics. In order of increasing spins, the combined numerical data from 40 extremal spectra contains $\{40, 40, 10, 40, 14, 40, 38, 17, 38, 39\}$ operators around this twist range for spins 2-20. Despite the limited data at some spins we still see decent agreement with the $[\phi\chi]_0$ trajectory.}
\end{figure}

We will start by focusing on operators of even spin since we can then make a comparison to the numerical bootstrap.\footnote{As a caveat, the numerics has relatively lower accuracy here because we do not have as many operators in this sector.} Using the OPE data known from numerics along with the analytic estimates described in section~\ref{subsec:O2GenFunc}, in figure \ref{4dimeven}, we show our predictions for low-lying charge 4 dimensions for even spins. We see that despite the fact we have less numerical data and had to use an estimate for $f_{\chi\chi s}$ by assuming analyticity down to spin 0, the analytics and the numerics are consistent. In particular, we see the analytic prediction for one of the curves passes almost exactly through the spin-2 numerical point.

However, there is a funny feature about this plot. If we continue the curves to asymptotically large spin, we see the orange curve corresponds to the $[tt]_{0}^4(\bar{h})$ trajectory while the blue curve corresponds to the ${[\phi \chi ]_{0}^4(\bar{h})}$ trajectory. While the blue curve is well-behaved down to spin 0, the orange curve displays a pole at relatively high spin. This is problematic because this curve is going below the unitarity bound $\tau=1$ while the O(2) CFT obeys all the usual unitarity bounds.

The likely resolution is that our methods cannot be trusted for this curve because we are ignoring other charge 4 operators with lower twists. For example we could consider the triple-twist operators $[\phi\phi t]_{0}^4(\bar{h})$ and $[\phi\phi\phi\phi]_{0}^4$. At large spin their twists asymptotically approach:
\begin{align}
h_{[\phi\phi t]_0}^4(\bar{h})\rightarrow&~1.14~,\nonumber\\
h_{[\f\f\f\f]_{0}^4}(\bar{h})\rightarrow&~1.04~.
\end{align}
In the exact plot, we would, therefore, expect that the $[tt]_{0}^4$ trajectory will intersect the $[\phi\phi t]_0^4$ trajectory and there will be large mixing which prevents the curve from diverging downwards. In order to make this more concrete, we need analytic tools to resolve mixing for these multi-twist operators, either by considering double-twists as external operators or by studying higher-point functions of $\f$ and $t$.

\begin{figure}[H]
  \centering
  \includegraphics[width=0.8\textwidth]{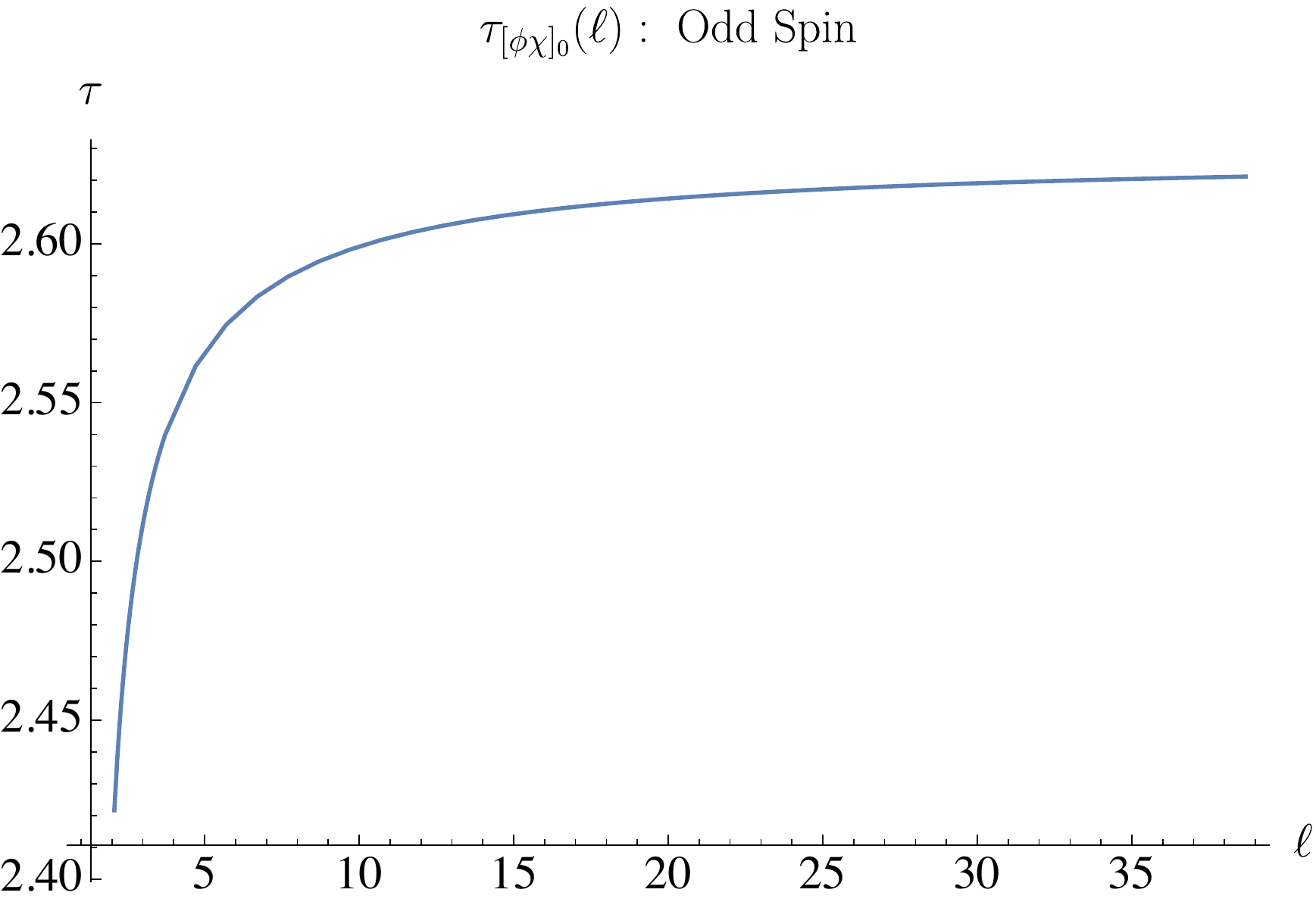}
  \caption{\label{4dimodd} Analytic predictions for twists of operators on the Regge trajectory $[\f \chi]^4_0$ with charge $4$ and odd spin.}
\end{figure}

By contrast, the odd-spin charge 4 sector is simpler to study because we only have the $[\phi \chi]_{0}^{4}$ trajectory, so mixing effects will be suppressed. However as mentioned before, here we do not have any numerical data to compare with, and in figure \ref{4dimodd} we just present the analytic results. It would be interesting to compare with future numerical work for $\left\langle {\chi \phi \chi \phi } \right\rangle $, which would give better access to the charge 4 sector.

\subsection{Ward identity checks}
As discussed in section \ref{subsec:O2GenFunc}, to determine the OPE coefficients $f_{tt[\f\f]_0^{0^{\pm}}}(\bar{h})$, $f_{ss[\f\f]_0^{0^{+}}}(\bar{h})$, and $f_{ts[\f\f]_0^{2}}(\bar{h})$ we use the inversion formula for $\<\f\f tt\>$, $\<\f\f ss\>$, and $\<\f\f ts\>$, respectively, where in all cases the leading contribution comes from $\f$ exchange. To get an additional test of the validity of this method, we can use the fact that the conserved operators $T$ and $J$ sit on the $[\f\f]_{0}^{0^{\pm}}$ trajectories~\cite{Albayrak:2019gnz}. Therefore, we can compute the OPE coefficients $f_{tt T}$, $f_{ttJ}$, and $f_{ssT}$ from the inversion formula and compare their values to the Ward identity constraints, which relate them to $f_{\f\f T}$ and $f_{\f\f J}$. Using the numerical bootstrap values of the latter, we find:
\begin{eqnarray}
&f_{ttT,\text{inv}}/f_{\f\f T,\text{num}}=2.428~, \qquad &f_{ttT,\text{Ward}}/f_{\f\f T,\text{num}}=2.382~,
\nonumber\\
&f_{ssT,\text{inv}}/f_{\f\f T,\text{num}}=3.050~, \qquad &f_{ssT,\text{Ward}}/f_{\f\f T,\text{num}}=2.912~,
\nonumber\\
&f_{ttJ,\text{inv}}/f_{\f\f J,\text{num}}=2.32~, \qquad &f_{ttJ,\text{Ward}}/f_{\f\f J,\text{num}}=2~,
\end{eqnarray}
which correspond to errors of $\sim 1.9\%$, $4.7$\%, and $16\%$. To obtain the analytic estimates involving $T$ and $J$ we worked to $5^{\text{th}}$ and $15^{\text{th}}$ order in dimensional reduction, respectively, which ensures that the numerical values shown above are stable to increasing $p_{\text{max}}$. We see that while the analytic predictions work well for the stress tensor, they start to deviate significantly for the current. Since we will only use inversion formula results for  $\ell \geq 2$ for charge 2, $\ell \geq 3$ for charge $0^-$, and $\ell \geq 4$ for charge $0^+$, we expect the corresponding errors are small. To improve the result for $f_{ttJ}$ we will likely need to include more operators in the inversion formula.

We can also repeat the same exercise for $\<\f\f \chi \chi\>$. In that case, the dominant contribution comes inverting $t$ exchange, and we find:
\begin{align}
&f_{\chi\chi T,\text{inv}}/f_{\f\f T,\text{num}}=4.049~, ~~~~ f_{\chi\chi T ,\text{Ward}}/f_{\f\f T,\text{num}}=4.062~,
\nonumber\\
&f_{\chi\chi J,\text{inv}}/f_{\f\f J,\text{num}}=3.31~, ~~~~~~ f_{\chi\chi J,\text{Ward}}/f_{\f\f J,\text{num}}=3~,
\end{align}
which corresponds to errors of $\sim$ $0.3\%$ and $10.3\%$ respectively. So we see that this simple approximation works very well for studying spin-2 operators, but getting accurate predictions at lower-spin likely requires inverting more operators.

\subsection{Leading scalar predictions}

Given that the inversion formula works well down to spin 2 and perhaps down to spin 1, it is natural to ask if we can push it further down to spin 0.\footnote{An explanation that scalar operators and their shadows should appear on Regge trajectories continued to $\ell=0$ was given in \cite{Caron-Huot-Talk}. This idea has been explored in perturbation theory in \cite{Alday:2017zzv}.} In previous sections, we have already used this idea to get an estimate for $f_{\chi\chi s}$ and we will now make a comparison between analytic and known numerical results for scalars. In table \ref{tab:finaltable}, we compare results for the dimensions of scalars from the inversion formula with known numerical values.
\begin{table}
\begin{center}
\begin{tabular}{@{}c|c|c|c|c@{}}
\hline
Operator & Inversion & Numerics & Relative error & Monte Carlo \\
\hline
Charge 1, shadow of $\phi$ & 2.37 & $2.480912(\bf{22})$ & 4\% & 2.480950(40) \cite{hasenbusch2019monte} \\
Charge 3, $\chi$ & 1.99 & $2.1086(3^*)$ & 6\%  & 2.1085(20) \cite{Hasenbusch_2011}\\
Charge 4, $\tau$ & 3.35 & $3.11535(73^*)$ & 7.5\% & 3.108(6) \cite{Hasenbusch_2011}\\
\hline
\end{tabular}
\end{center}
\caption{\label{tab:finaltable} Predictions from the inversion formula versus numerical bootstrap values for the leading scalar operator dimensions. For the relative error between the inversion formula and numerical bootstrap methods, we use the formula $\left| {\frac{{{\text{numerics}} - {\text{inversion}}}}{{{\text{numerics}}}}} \right|$. The charge 1, 3, and 4 inversion formula predictions come from evaluating the $[\f s]_0^1$, $[\f t]_0^3$, and $[\f \chi]_0^4$ trajectories at spin 0. We also include a comparison to Monte Carlo results.}

\end{table}
Before explaining how we get these results, we should note the large-spin expansion should actually be thought of as a $1/\bar{h}$ expansion. This is especially relevant for scalars since we will be setting $\ell=0$, but from experience with the lightcone bootstrap, we should expect the inversion formula will give more accurate results for heavier operators. In practice, this means dimensional reduction will converge faster when we are studying heavier scalars.

Starting with the charge 1 trajectory, we claim that the shadow of $\phi$ sits on the $[\phi s]_0^1$ trajectory.\footnote{As a reminder, the OPE function $c(\Delta,J)$ is shadow symmetric, a pole at $\Delta_{*}$ also implies the existence of a pole at $d-\Delta_{*}$. As explained in \cite{Kravchuk:2018htv}, the poles in $c(\Delta,J)$ represent light-ray operators with quantum numbers $(\Delta^L,J^L)=(1-J,1-\Delta)$. The integer spin points represent light-transforms of local operators $\mathbf{L}[\cO]$. The shadow operation for light-ray operators is the ``spin shadow" $\mathbf{S}_J:(\Delta^L,J^L)\to (\Delta^L,2-d-J^L)=(1-J,\De-d+1)$. Thus, when we refer to the shadow $\tilde\cO$ of a local operator $\cO$, we are really discussing the spin-shadow of its light-transform $\mathbf{S}_J [\mathbf{L} [\cO]]$.} 

To test our hypothesis, we must evaluate $\widetilde \Delta_\f^\mathrm{Regge}= \tau_{[\phi s]_0^1}(0)$, where $\tau_{[\phi s]_0^1}(\ell)$ is the upper curve in figure~\ref{1dim}.  To do so, we computed $M_1(z,\bar{h})$ to $20^{\text{th}}$ and $15^{\text{th}}$ order in dimensional reduction for the isolated and double-twist operators $[\phi\phi]_{0}$, respectively. For the double-twist operators we also inverted the first 50 operators exactly, i.e. we chose $\ell_{*}=50$ and performed the sum for $\ell=51,...,\infty$ using the identities given in section \ref{sec:DTI}. At this order we find $\widetilde\Delta^\mathrm{Regge}_{\phi}\approx 2.37$, which is close to the expected result $\widetilde \Delta_\phi = 2.48091(22)$.

\begin{figure}
  \begin{center}
  \includegraphics[width=0.8\textwidth]{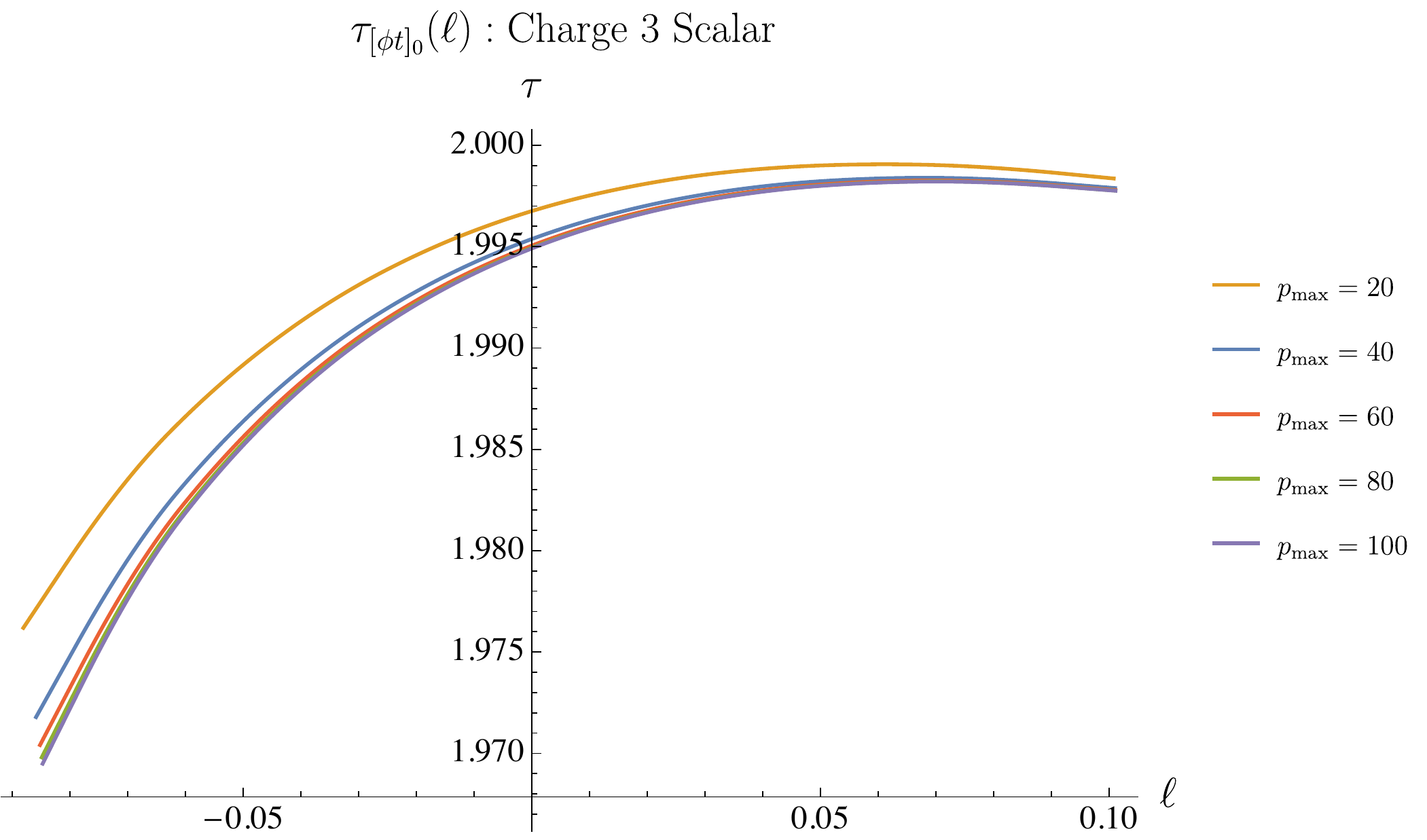}
  \end{center}
  \caption{\label{3scalar} The analytic prediction for the twist of the $[\phi t]_0^3$ Regge trajectory near $\ell=0$, for various orders in dimensional reduction for isolated operators in the inversion formula. }
  \end{figure}

For the charge 3 scalar, we can see in figure \ref{3dim} that the even-spin trajectory is increasing as we go down to spin 0. To make the structure clearer, in figure \ref{3scalar}, we zoom in on the region around spin 0 and plot the charge 3 scalar dimension at various orders in the dimensional reduction for the isolated operators. For the double-twist operators, we work to $15^{\text{th}}$ order in dimensional reduction, treat operators with spin $\ell\leq 50$ exactly, and approximate the remaining operators using the large-spin asymptotics.
We see that for $p_{\text{max}}=100$ the curve intersects around $\Delta_\chi^{\mathrm{Regge}}\approx 1.995$. This is to be compared with the expected value $\Delta_\chi\approx 2.1086(3)$ in table~\ref{tab:resultsefm}. Note that $\bar{h}_{\chi}\approx 1$, so we are clearly no longer in the regime of large-spin perturbation theory, but the inversion formula still works well.

Next, we can study the charge 4 scalar $\tau$. Here we have a tradeoff, while $\bar{h}_{\tau}$ is larger than $\bar{h}_{\chi}$ or $\bar{h}_{\widetilde{\phi}}$, we know there are potentially large mixing effects to consider when studying $M_4(z,\bar{h})$. The former means dimensional reduction converges very quickly and we can work to $1^{\text{st}}$ order to find the converged value $\Delta_{\tau}^\mathrm{Regge}\approx 3.35$, which can be compared to the expected result $\Delta_\tau=3.11535(73)$ in table~\ref{tab:resultsefm}. This can also be seen from the figure \ref{4dimeven} by looking at the $[\phi \chi]_{0}^{4}$ trajectory. It would be interesting to understand how to improve this result and especially how to properly include mixing with higher multi-twist operators.

\subsection{Regge intercepts}
\label{sec:ReggeInt}
In this work, we have mostly focused on using the inversion formula to make predictions for the spectrum and couplings of local operators. However, another interesting observable when studying CFTs in Lorentzian signature are the Regge intercepts for the minimal-twist trajectories. The Regge intercept of a given trajectory, $\ell_{\text{int}}$, is defined by analytically continuing the dimensions as a function of $\ell$, $\Delta(\ell)$ and finding the spin such that $\Delta(\ell_{\text{int}})=d/2$.\footnote{There will exactly be two solutions to this equation and we will take the larger one. This is the solution that determines the leading behavior of the CFT correlator in the Regge limit.} The Regge intercept of the leading trajectory, $\Delta_{\text{min}}(\ell_{\text{int}})=d/2$, determines the behavior of the correlator in the Regge limit  \cite{Costa:2012cb}.

\begin{figure}
  \begin{center}
     \includegraphics[width=0.8\textwidth]{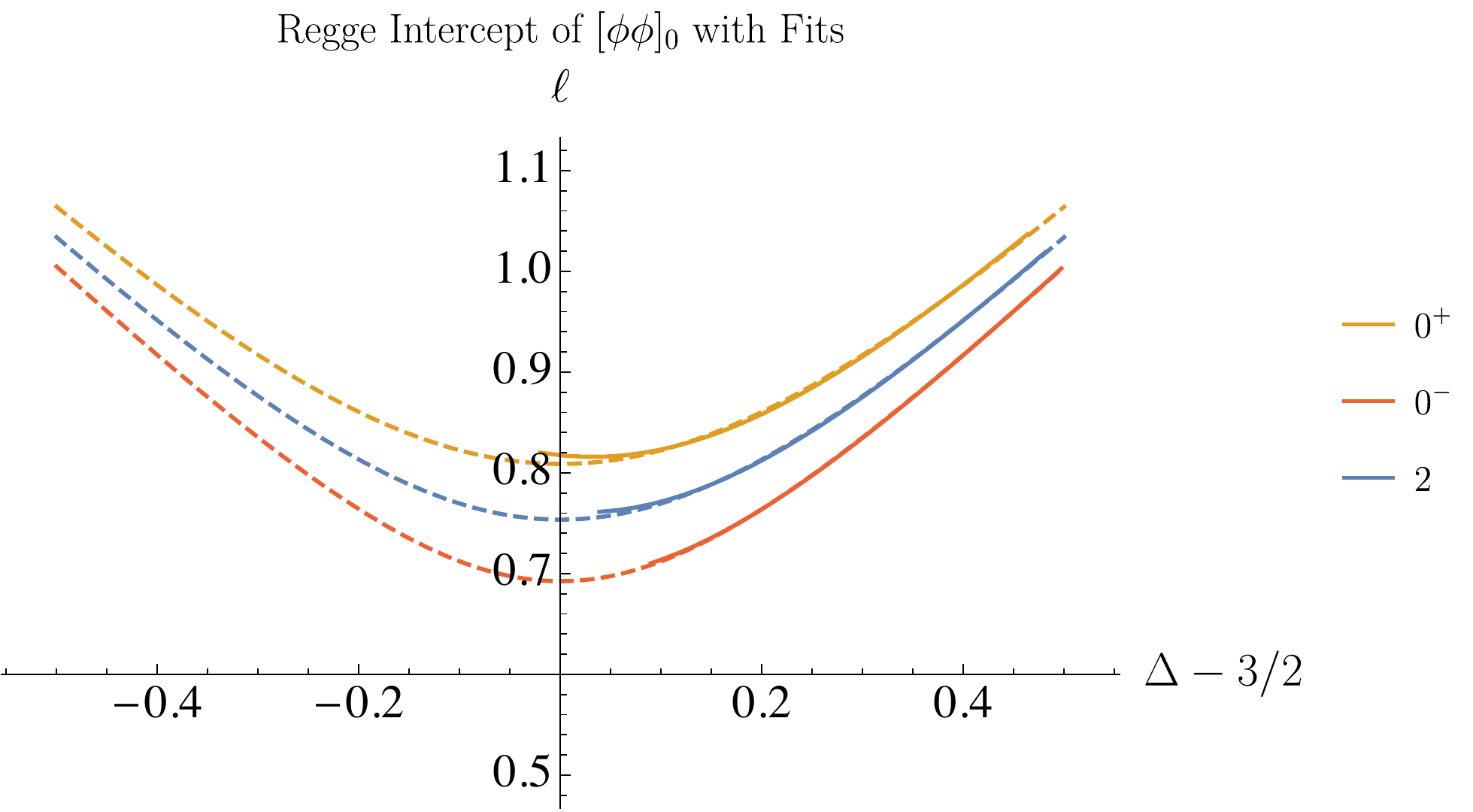}
  \end{center}
  \caption{\label{reggeintercept} The $[\phi\phi]_0^{0^\pm,2}$ trajectories continued down to the intercept point $\Delta=3/2$. The undotted lines give the prediction from the inversion formula. The curves are truncated at $\bar{h}=1.15$ as below this value the dimensional reduction expansion converges too slowly to make reliable predictions. The dashed lines correspond to a simple shadow-symmetric quadratic fit.
}
  \end{figure}

In the $\text{O}(2)$ model the leading trajectory corresponds to $[\phi\phi]_0^{0^+}$, which can be seen directly from figure \ref{phiphidimOPE}.\footnote{One can also show on general grounds that the leading trajectory has to be a singlet under global symmetries \cite{Meltzer:2018tnm}.} To find its Regge intercept in practice, we make a $(\Delta,\ell)$ plot for this trajectory and find the $\Delta=3/2$ point. In figure \ref{reggeintercept} we plot this trajectory as well as the trajectories for $[\phi\phi]_0^{0^-,2}$ as we approach this point. To make this plot we worked to $p_{\text{max}}=200$ in dimensional reduction when inverting the light, isolated operators, and then made a linear fit to extrapolate to $p_{\text{max}}=\infty$. The plot is made by evaluating the generating functions from $\bar{h}=1.15$ to $1.50$ in steps of $0.01$. For lower $\bar{h}$ dimensional reduction converges very slowly and we cannot make reliable predictions.

\begin{figure}[h]
  \begin{center}
     \includegraphics[width=0.8\textwidth]{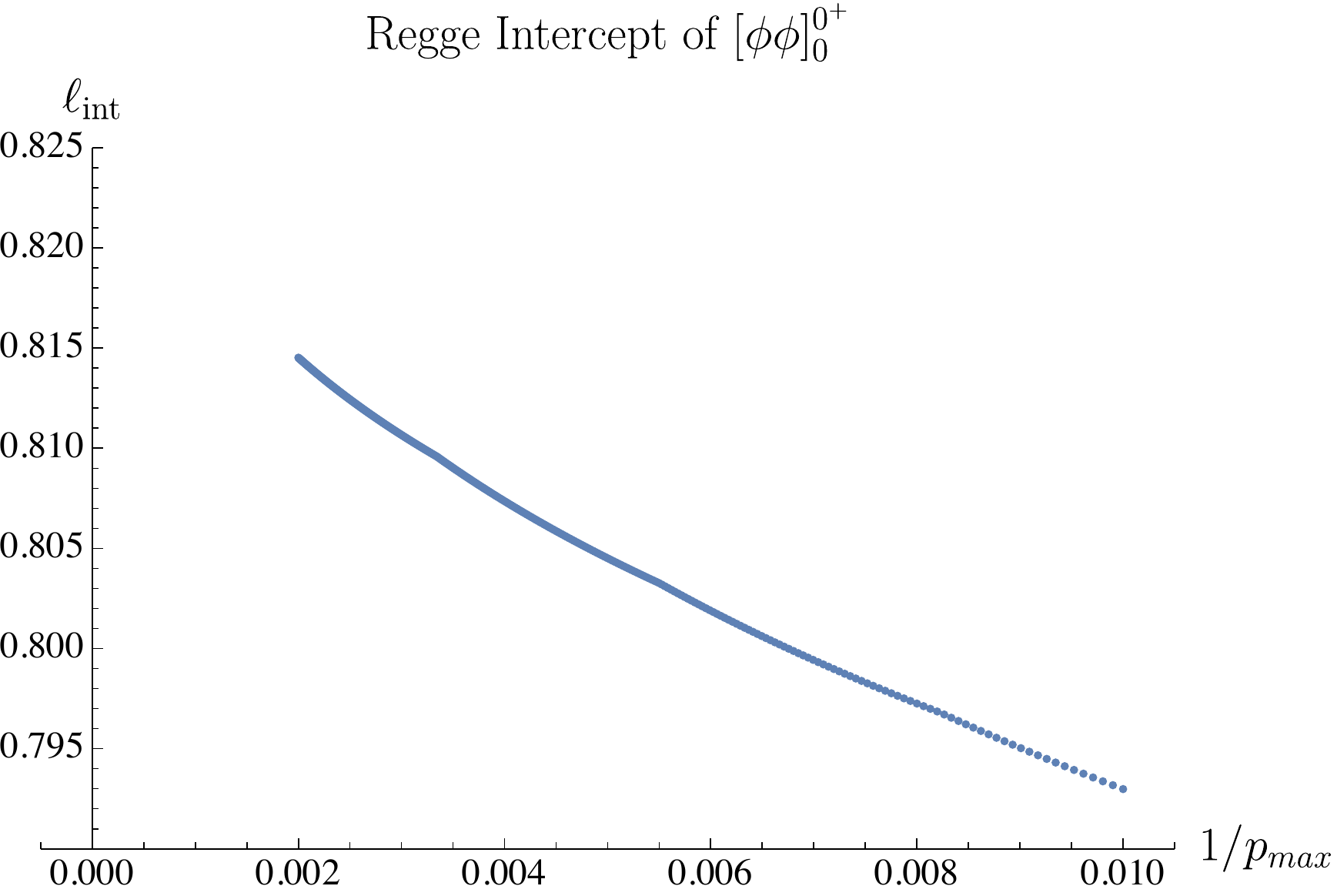}
  \end{center}
  \caption{\label{reggeinterceptvspmax} The $[\f\f]_0^{0^+}$ Regge intercept as a function of $1/p_{\text{max}}$. 
}
  \end{figure}

From this plot, we see that the Regge intercept is slightly below $0.82$. Note that there is an ambiguity, from figure \ref{reggeintercept} we see that our function $\ell^{0^+}(\Delta)$ is almost, but not quite shadow symmetric, i.e. invariant under $\Delta\rightarrow 3-\Delta$. In the exact theory, the minimum of $\ell(\Delta)$ will be exactly at $\Delta=3/2$, while here we see it is slightly to the right. The lack of exact shadow symmetry is not surprising and will likely only emerge when we know the exact correlator. For the sake of making an approximation, we define the Regge intercept as corresponding to the minimum of $\ell^{0^+}(\Delta)$, which here is slightly off-center. To obtain a better estimate, in figure \ref{reggeinterceptvspmax} we plot the Regge intercept as a function of $1/p_{\text{max}}$ from $p_{\text{max}}=100$ to $500$. Making a linear fit we see that the extrapolation points to the value $\ell^{0^+}_{\text{int}}\approx .82$. 

\begin{table}[t]
\begin{center}
\begin{tabular}{@{}c|c@{}}
\hline
Trajectory & $\ell_{\text{int}}$\\
\hline
$[\f\f]_0^{0^+}$ & 0.82 \\
$[\f\f]_0^{0^-}$ & 0.75 \\
$[\f\f]_0^{2}\,\,\,$ & 0.69 \\
\hline
\end{tabular}
\end{center}
\caption{\label{tab:intercepts} Analytical estimates of the Regge intercepts $\ell_{\text{int}}$ of the leading $[\f\f]_0^{0^{\pm},2}$ trajectories.}
\end{table}

To obtain estimates for the Regge intercepts of the other trajectories, we need to make an ansatz for their form. One option is to assume shadow symmetry for $\ell(\Delta)$ and make a quadratic fit based on the results of the inversion formula. Here we will take a related approach, but choose an ansatz which is motivated by the generalized free field spectrum. If we have a generalized free field $\f$ then we can construct the $n=0$, double-twist operators $[\f\f]_{0,\ell}$ with dimension $\Delta_{0,\ell}=2\Delta_\f+\ell$. By shadow symmetry we also have the trajectory with $\widetilde{\Delta}=d-\Delta_{0,\ell}$. These two trajectories can be found by solving
\begin{align}
(\Delta-\ell-2\Delta_\f) (d-\Delta -\ell-2\Delta_\f)=0~.
\end{align}
We can deform this result by introducing two new parameters $a$ and $b$:
\begin{align}
(\Delta-\ell-a) (d-\Delta -\ell-a)=b~. \label{eq:ansatzRegge}
\end{align}
This ansatz makes shadow symmetry, $\Delta\rightarrow d-\Delta$, manifest. Solving for $\ell$ yields two trajectories:
\begin{align}
\ell=d/2-a\pm\frac{1}{2}\sqrt{4b+(d-2\Delta)^{2}}~.\label{eq:two_trajectories}
\end{align}
Choosing the plus sign gives both the physical trajectory and its shadow. Choosing the minus sign gives a trajectory whose spin decreases as we increase the scaling dimension. 

Setting $d=3$, taking the data in figure \ref{reggeintercept} from $\bar{h}=1.15, ..., 1.50$ and making a fit for the $0^+$ trajectory yields:
\begin{align}
a_{0^+}=1.05~,
\\
b_{0^+}=0.13~.
\end{align}
Setting $\Delta=3/2$ yields the Regge intercept $\ell^{0^+}_{\text{int}} \approx .81$ which is consistent with (but slightly differs from) our previous result based on extrapolation. 

For the charge $2$ trajectory we find:
\begin{align}
a_{2}=1.05~,
\\
b_{2}=0.09~.
\end{align}
Plugging this into the ansatz yields a Regge intercept of $\ell^{2}_{\text{int}} \approx 0.75$. 

Finally, for the charge $0^-$ trajectory we find:
\begin{align}
a_{0^-}=1.05~,
\\
b_{0^-}=0.06~.
\end{align}
This gives a Regge intercept of $\ell^{0^-}_{\text{int}} \approx 0.69$.

We can note that for all three trajectories we have $a_{q}\approx 1.05$. If we take (\ref{eq:two_trajectories}) and expand it at large $\Delta$ we see $a_q$ is the twist of each trajectory. It is then not surprising $a_q$ does not significantly vary between trajectories as they all have twist $\tau\approx 2\Delta_{\f}$. The parameter $b_q$ is effectively measuring the deviation of the exact trajectory from the generalized free field result. For all three trajectories we see $b_q$ is relatively small and that the $0^+$ trajectory has the largest deviation from generalized free field theory. 

Finally, an interesting feature of the ansatz (\ref{eq:ansatzRegge}) is that it also produces a lower trajectory which can intersect the $\ell=0$ line for real $\Delta$. Specifically, we find for the $0^+$ trajectory a scalar with dimension $1.76$ and for the charge $2$ trajectory a scalar with dimension $\Delta=1.17$. We see that the charge 2 result is fairly close to the numerical prediction $\Delta_t = 1.23629(\bf{11})$, with a relative error around $5.5\%$. However, the charge $0^+$ result differs more significantly from the numerical result $\Delta_s = 1.51136(\bf{22})$, with a relative error around $16.5\%$. The likely resolution is that our simple ansatz (\ref{eq:ansatzRegge}) is really only a good local fit to the upper branch for $\Delta\in(3/2,2)$, and does not correctly describe the lower branch. The fact that we can see physical, scalar operators on the lower branch is encouraging, but to make precise predictions we likely need a better ansatz and/or method for inverting 3d blocks.

\subsection{Crossing symmetry and the dDisc}
In this section, we will briefly return to the problem of studying crossing symmetry for the full correlation function. In the numerical bootstrap, bounds are derived by imposing crossing symmetry and unitarity on the full correlation function. However, when studying the Lorentzian inversion formula, we start by making an ansatz for the double-discontinuity and use this to find the OPE data in a given channel.\footnote{In this discussion, we will assume the analyticity in spin holds to $\ell=0$ so we can ignore any ambiguities at finite spin.} In principle, given the full double-discontinuity, the output of the inversion formula is a solution to crossing symmetry. In practice, we only have an approximation for the double-discontinuity and can only make reliable predictions for low-twist operators. Here we will study to what extent the CFT data we have saturates the crossing symmetry condition and, in the process, understand where our ansatz for the double-discontinuity is reliable.

\begin{figure}
  \centering	
  \includegraphics[width=0.8\textwidth]{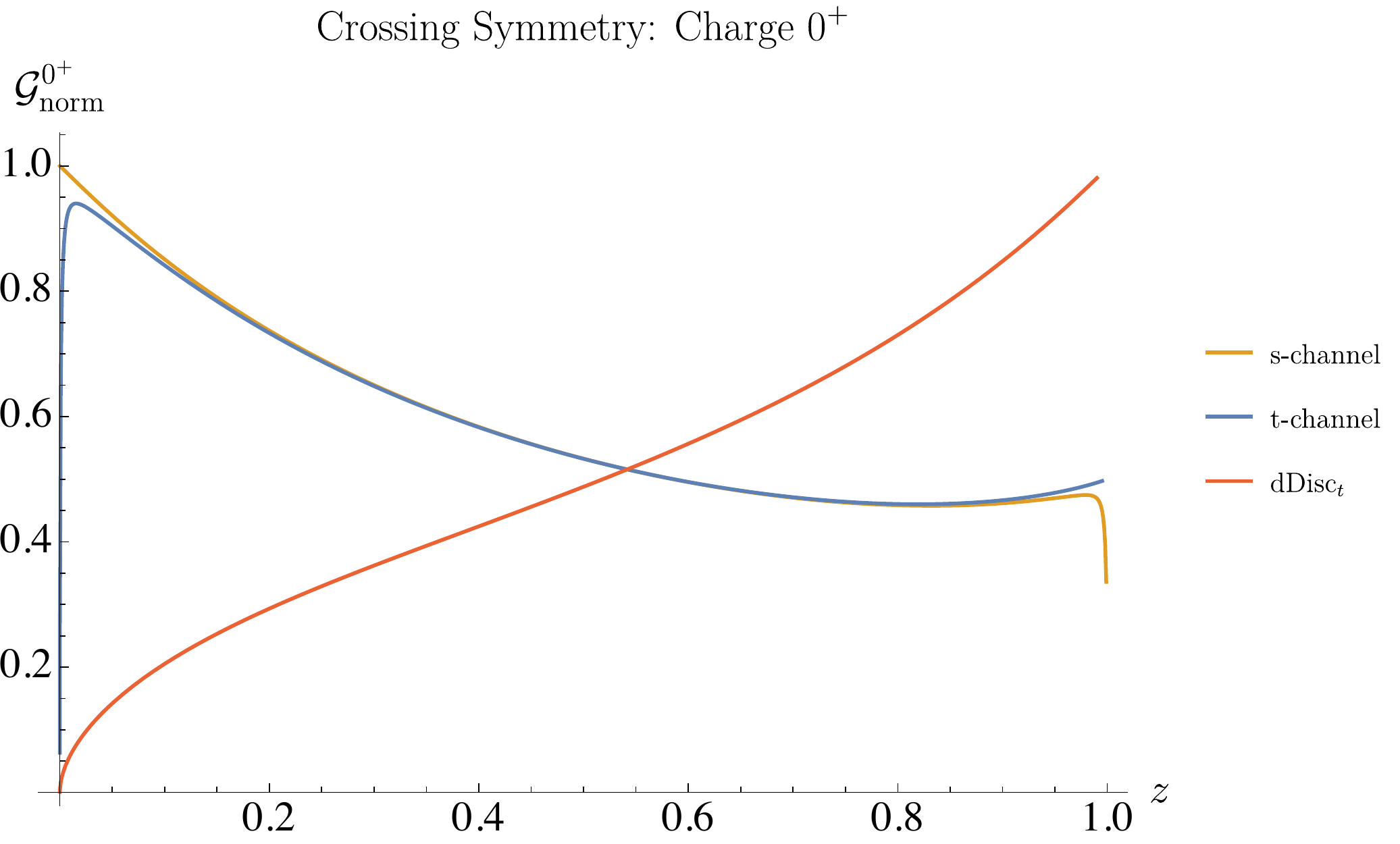}
  \caption{\label{0pCrossing} We study the correlator $\<\phi \phi \phi \phi \>$ and project onto $0^+$ exchange in the $s$-channel. We plot the prediction for $\mathcal{G}^{0^+}_{\text{norm}}(z,z)$ using the $s$ and $t$-channel OPEs. We also plot its dDisc$_t$, which we compute by expanding in the $t$-channel OPE and then taking the double-discontinuity. We normalize all three functions by an MFT four-point function. Because dDisc$_t$ is computed using the $t$-channel OPE, it is reliable where the $t$-channel OPE is reliable. This includes the region $z>1/2$ and also somewhat smaller $z \gtrsim 0.1$ where crossing symmetry holds to high precision.}
\end{figure}

Specifically, we will study the correlator $\<\phi \phi \phi \phi \>$ and use data from the numerical bootstrap \cite{Chester:2019ifh} for the light, isolated operators $s$, $t$, $J$, and $T$, and use data from the inversion formula for the double-twist operators $[\f\f]_0^q(\ell)$ up to spin 20. Equivalently, one can determine the double-twist couplings and spectrum via the extremal functional method and the exact parameters can be found in the attached \texttt{Mathematica} file. As an example, in figure \ref{0pCrossing} we project onto charge $0^+$ exchange in the $s$-channel and compare the predictions from the $s$ and $t$-channel OPEs. Recall $\mathcal{G}(z,\bar{z})$ is defined in (\ref{eq:mGdef}) by factoring out an $s$-channel prefactor. To make the plots clearer, we specialize to the diagonal $z=\bar{z}$ and define the normalized correlator as:
\begin{align}
\mathcal{G}_{\text{norm}}^{q}(z,z)=\frac{\mathcal{G}^q(z,z)}{1+z^{2\Delta_\f}+\left(\frac{z}{1-z}\right)^{2\Delta_\f}}~.
\end{align}
That is, we are dividing by the MFT four-point function for a fictitious, uncharged scalar with dimension $\Delta_\f$. To compute the blocks, we also use dimensional reduction up to $p_{\text{max}}=30$.

From figure \ref{0pCrossing}, we see that crossing symmetry is obeyed in a large neighborhood around $z=1/2$. As we go to smaller $z$, for example $z\sim 1/10$, we see significant deviations between the $s$ and $t$-channel OPEs. This is important because we compute the double-discontinuity, dDisc$_t$, by expanding in the $t$-channel and then taking the discontinuities. Recall that computing dDisc$_t$ requires analytically continuing outside the regime where the $s$-channel OPE converges,\footnote{Relatedly, dDisc$_t$ annihilates individual $s$-channel conformal blocks \cite{Caron-Huot:2017vep}.} so we cannot use a finite number of $s$-channel blocks to approximate  dDisc$_t$. For $z\gtrsim 9/10$, we also see that crossing symmetry no longer holds, but in this region, the $t$-channel OPE is valid. 

\begin{figure}
  \centering	
  \includegraphics[width=0.8\textwidth]{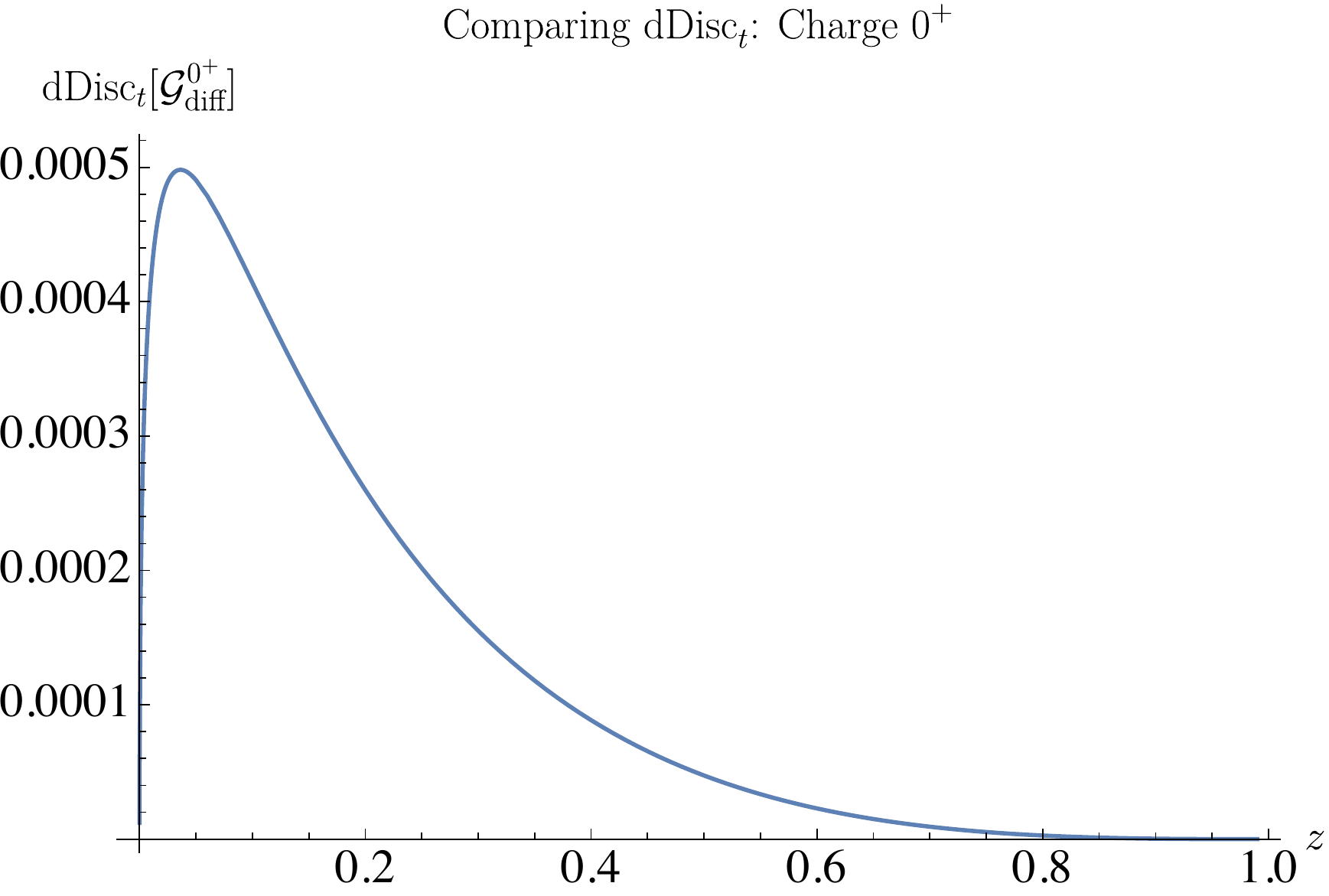}
  \caption{\label{compdDisc} We study the effect of the double-twist operators on dDisc$_t[\mathcal{G}_{\text{norm}}^{0^+}(z,z)]$ and see that the difference between including or not including them is small.
  }
\end{figure}

To make figure \ref{0pCrossing}, we included the double-twist operators $[\f\f]_0^{0^+}(\ell)$, up to spin 20, for all three functions. It is important to include these operators in the OPE to see crossing symmetry emerge for the correlator, $\mathcal{G}(z,z)$. We also included these operators when plotting its double-discontinuity, but here they play a minor role. In figure \ref{compdDisc} we plot the difference dDisc$_t$$[\mathcal{G}^{q}_{\text{diff}}(z,z)]$ for $\mathcal{G}^{q}_{\text{diff}} = \mathcal{G}^q_{\text{norm}}\big|_{\text{DT}} - \mathcal{G}^q_{\text{norm}}\big|_{\text{no DT}}$ , again specialized to the diagonal $z=\bar{z}$, when we include or exclude the double-twist operators. We see that including the double-twists gives a small effect, which is of order $\sim0.0005$. Therefore, while the double-twist operators have a large effect on the full correlator, the double-discontinuity itself is relatively stable against further corrections. This gives more evidence that just including the light operators of low-spin gives a good approximation to the full double-discontinuity.

Using the same data, we can also make a plot for $\mathcal{G}^{0^+}(z,\bar{z})$ in Euclidean configurations, as shown in figure \ref{fig:Euc4pt0p}. Here we restricted to the configuration $\bar{z}=z^*$, worked to $5^{\text{th}}$ order in dimensional reduction, and normalized the correlator by its value in MFT. To make the plot we defined two regions:
\begin{align}
R_{s}&=\{z\in \mathbb{C} \ | \ |z|< 1, \ \text{Re}[z]\leq 1/2 \}\,,
\\
R_{t}&=\{z\in \mathbb{C} \ | \ |z-1|< 1, \ \text{Re}[z]\geq 1/2 \}\,,
\end{align}
and used the $s$ and $t$-channel to compute the correlator in $R_s$ and $R_t$ respectively. We apply the transformation $z \rightarrow \frac{z}{z-1}$ to the $t$-channel data to obtain the correlator outside of these regions (corresponding to the $u$-channel region). We see that there are two peaks, centered around $z=0$ and $z=1$, corresponding to identity exchange in each channel. The correlator decreases away from these peaks and the minimum of $\mathcal{G}^{0^+}/\mathcal{G}^{0^+}_{\text{MFT}}$ in these two regions is approximately $0.704$. This tells us that the O(2) CFT, like the critical 3d Ising model \cite{Rychkov:2016mrc}, does exhibit large deviations from the MFT prediction. 

\begin{figure}
  \centering	
  \includegraphics[width=0.8\textwidth]{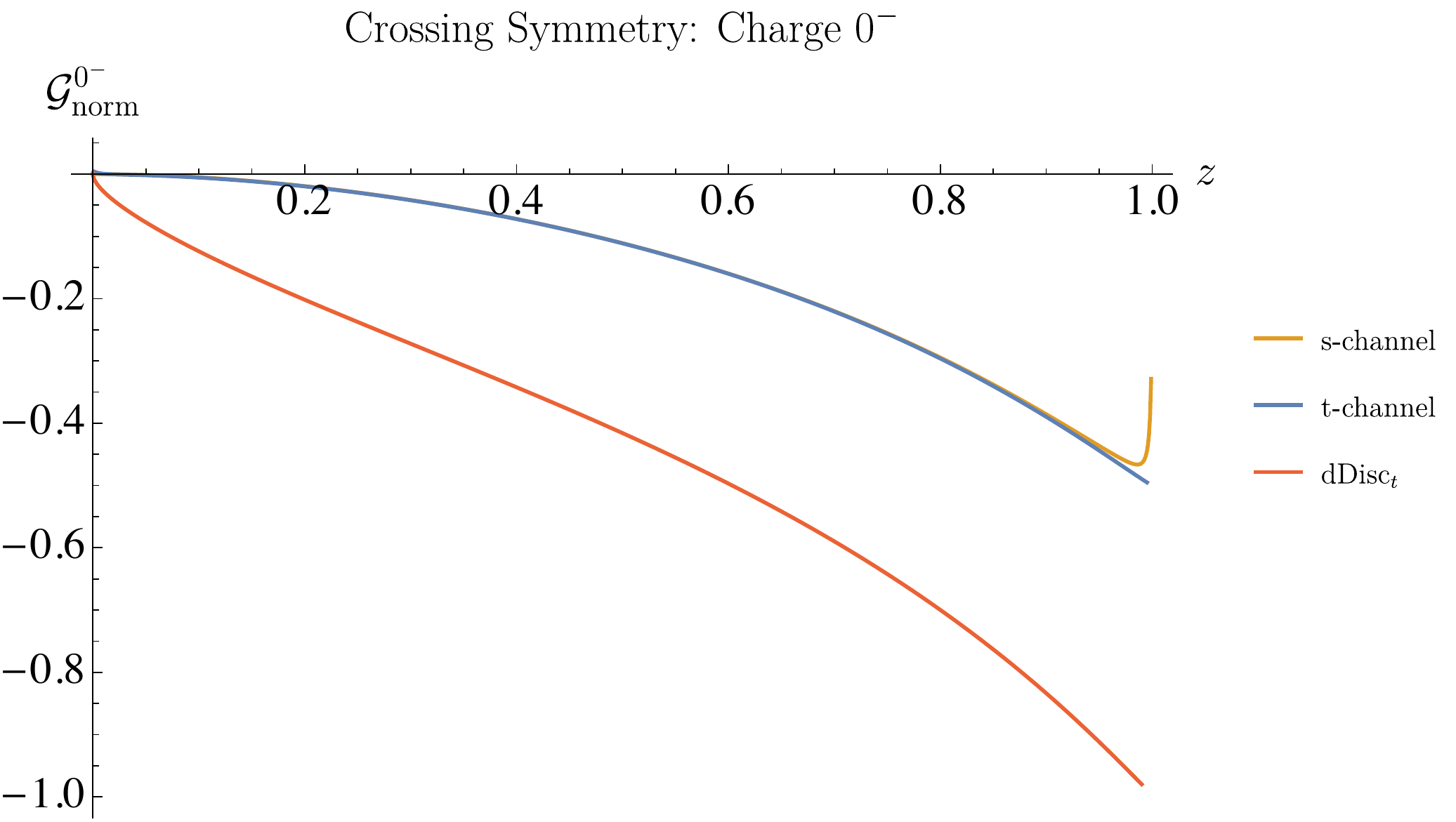}
  \includegraphics[width=0.8\textwidth]{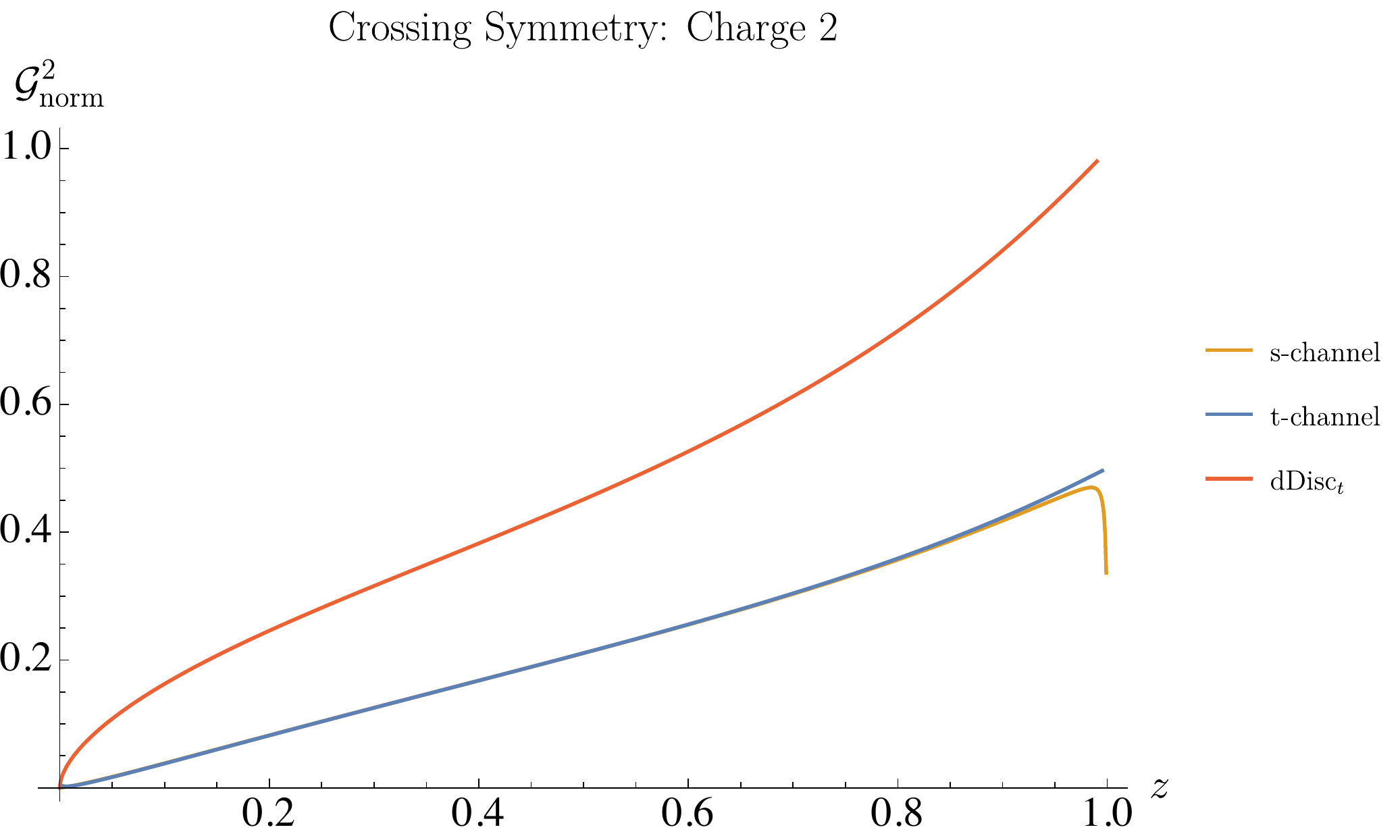}
  \caption{\label{0m2Crossing} We study $\<\phi \phi \phi \phi \>$ and project onto charge $0^-$ and charge 2 exchange in the $s$-channel. We compare the prediction for the full normalized correlator on the diagonal, $\mathcal{G}_{\text{norm}}^{0^-,2}(z,z)$, using the $s$ and $t$-channel OPEs. We also plot its dDisc$_t$, which is computed using the $t$-channel OPE.}
\end{figure}

We can also repeat this analysis when we project onto charge $0^-$ and charge-2 exchange in the $s$-channel, as shown in figure \ref{0m2Crossing}. For both cases, we see that crossing symmetry is obeyed in a large region around $z=1/2$. Here deviations between the $s$ and $t$-channel OPEs are not visible around $z=0$ because to define $\mathcal{G}^q$ we pulled out an overall $s$-channel prefactor. This has the effect of weighting each correlator by a factor of $(z\bar{z})^{\Delta_{\f}}$.

As a final application, one could also try to use the above approximation for the dDisc$_t$ to estimate the Regge growth of the correlator. In section \ref{sec:ReggeInt} we approximated the Regge intercepts by continuing the physical trajectories down to $\Delta=3/2$. Alternatively, one could directly study $\text{dDisc}_t[\mathcal{G}(z,\bar{z})]$ and take the limit $z,\bar{z}\rightarrow 0$ with the ratio $z/\bar{z}$ held fixed. The current difficulty with this approach is that we need to study the double-discontinuity around $z=0$, while the expansion we have is around $z=1$. 
Indeed, if we na\"ively fit the dDisc$_t$ in figure~\ref{0pCrossing} to a power law at small $z$, we obtain a Regge intercept of $3/2-2\Delta_{\f}\approx 0.46$. This is a kinematic effect, coming from the fact that the dDisc$_t$ in figure~\ref{0pCrossing} is dominated by a finite number of $t$-channel blocks. Each individual $t$-channel block has Regge spin $3/2-2\Delta_{\f}$.\footnote{One way to see this is to note that $t$-channel blocks have an expansion in $s$-channel double-twists, which have $\Delta=2\Delta_{\f}+\ell$. Setting $\De=\frac d 2$ gives $\ell_\mathrm{int}=d/2-2\De_\f$. More directly, one can take an individual $t$-channel block, $g_{\Delta,\ell}(1-z,1-\bar{z})$, and take the limit $z,\bar{z}\rightarrow 0$ with $z/\bar{z}$ held fixed. In this limit, each block scales like $z^{1-\frac{d}{2}}$, which can be verified by solving the Casimir equation or using known expressions for blocks on the diagonal $z=\bar{z}$ \cite{Hogervorst:2013kva,KravchukUnpublished}. Multiplying by the factor $(z\bar z)^{\De_\f}/((1-z)(1-\bar z))^{\De_\f}$ coming from crossing symmetry, we obtain $z^{1-\p{\frac{d}{2}-2\De_\f}}$, which corresponds to spin $d/2-2\De_\f$ growth in the Regge limit.}

To find the genuine Regge intercept, we must study a sum over an infinite family of operators in the $t$-channel.\footnote{For example, in large $N$ theories the non-trivial Regge growth comes from performing infinite sums of the double-twist operators over both $n$ and $\ell$ \cite{Cornalba:2006xm,Cornalba:2007zb,Li:2017lmh,Meltzer:2019pyl}.} Alternatively, in section \ref{sec:ReggeInt} we found a Regge intercept of $\ell_{\text{int}}^{0^+}\approx 0.82$ by also inverting a few light operators. However there we assumed the $\log$s in the generating function exponentiate, following the discussion in section \ref{subsec:ExactvsApprox}. Understanding a similar approximation for the full double-discontinuity could be a useful alternative to performing infinite sums over 3d blocks.

\section{Future directions}\label{future}
\subsection{Numerical bootstrap}
Recent improvements to numerical bootstrap algorithms~\cite{Landry:2019qug,Chester:2019ifh} allow us to solve problems with significantly higher numerical precision and at a much larger scale. In particular, we can get a much clearer picture of the full spectrum for a variety of different CFTs that can be solved using bootstrap techniques. Some immediate candidates are to study in detail the full spectrum of the 3d Ising CFT, the full set of O($N$) CFTs, Gross-Neveu-Yukawa models, and related theories (e.g., various supersymmetric extensions) using the methods developed in~\cite{Chester:2019ifh} and in this paper. We also hope that targets like 3d QED and 4d QCD in their conformal windows will soon become accessible using large-scale bootstrap methods~\cite{Chester:2016wrc,Li:2018lyb,Li:2020bnb}. 

For the O(2) model itself, we see several ways to improve the numerics used in our work. For instance, we have seen that overcoming the ``sharing effect", which arises due to spurious operators at the gap threshold sharing OPE coefficients with physical operators, will be important for obtaining accurate numerical conformal data. Thus, to overcome the sharing effect and mitigate the error, one could impose more carefully chosen gaps in different sectors. It will be interesting to explore which precise gaps could be implemented and how to control the error induced by the sharing effect. In addition, it might be helpful to study mixed correlators involving other relevant operators, such as the leading charge 3 scalar and conserved O(2) current (building on~\cite{Reehorst:2019pzi}). Pursuing these directions may, for example, be helpful for obtaining precise numerical data for the higher-twist families in the O(2) model.

It will also be interesting to explore methods for spectrum extraction beyond the current numerical computational framework. The standard extremal functional method has some drawbacks. For instance, one needs to average over several spectra to stabilize the operators, and there is no known way to make rigorous estimates. It is worth thinking about how to overcome this problem and finding an improved version of the extremal functional method in which errors are under better control. Other possible numerical improvements include finding a more efficient dimension and OPE scan method, a better basis of functionals, or numerical bootstrap algorithms beyond the usual semidefinite programming interior point method. We believe that those directions will be helpful for obtaining precise conformal data for specific CFTs.  

\subsection{Analytic bootstrap}
For the analytic bootstrap, an immediate direction is to apply the new techniques developed in this work to other CFTs where very precise low-lying data is known from the numerical bootstrap, such as the 3d Ising model~\cite{Kos:2016ysd}, its supersymmetric extension~\cite{Rong:2018okz,Atanasov:2018kqw}, or other O($N$) models~\cite{Kos:2015mba,Kos:2016ysd}. Using the recent results~\cite{Albayrak:2020rxh}, they can also now be readily applied to 3d CFTs with fermions such as the Gross-Neveu-Yukawa models~\cite{Iliesiu:2015qra,Iliesiu:2017nrv}. These models will be important testing grounds for a better understanding of which contributions are needed in order to make precise predictions for higher-twist trajectories. 

It will also be helpful to further refine the $z=z_0$ method formalized in this paper. Can we find a more systematic scheme to determine the optimal values of $z_0$ in different models, charge sectors, and spins? Is there a better way to move into the proper small $z$ regime while suppressing the logarithmic divergence? Along these lines, it will be important to improve methods for the twist Hamiltonian and resummations of double-twist operators. 

A related problem is how to properly study triple- and higher-twist operators. As we study double-twist trajectories with larger twists, we need to understand how to resolve mixing with the higher-twist trajectories. Double-twist operators are fairly simple. The trajectory is non-degenerate, i.e., there is a single operator with a given twist and spin, and we have a single accumulation point in twist space. On the other hand, triple-twist operators are composed of double-twists, e.g., $[\f\f\f]=[[\f\f]\f]$, so their degeneracy grows with spin, and we find accumulation points of accumulation points in twist space. Therefore, resolving their mixing and understanding their large-spin structure is a significantly more complicated problem and a proper understanding may require studying higher-point functions, e.g., $\<\f\f\f\f s\>$ in the O(2) model.

It would also be useful to understand how to more effectively streamline these computations. At our current stage, we have an iterative procedure which involves computing the contributions of operators to a matrix of generating functions, extracting the physical data by diagonalizing the twist Hamiltonian, and then plugging the results back into the inversion formula. While the procedure is straightforward, in practice, it would be useful to develop tools that avoid intermediate steps, e.g., diagonalizing the twist Hamiltonian. 

Finally, our work is mostly based on the Lorentzian inversion formula, and it would be interesting to understand its connection to other analytic methods. For example, we can compare to more traditional lightcone bootstrap \cite{Simmons-Duffin:2016wlq,Albayrak:2019gnz}, the CFT dispersion relation \cite{Carmi:2019cub}, or analytic functional methods \cite{Mazac:2019shk}. Eventually, it would be interesting to find a way to incorporate the precise analytical solutions computed via these methods back into the numerical bootstrap to find a robust iterative scheme for efficient bootstrap computations. In order to do this, it is important to understand how to rigorously bound errors when using the inversion formula.

\subsection{Additional applications}

Our preliminary result for the Regge intercept of the O(2) model, $\ell_\mathrm{int} \approx 0.82 < 1$ implies that charge-charge event shapes are well-defined in this theory \cite{Hofman:2008ar,Kologlu:2019bco,Kologlu:2019mfz}. Charge-charge correlators are natural observables in quantum-critical realizations of the O(2) model: one could potentially measure them by exciting a critical sample in the center and measuring the distribution of charge excitations at the boundary. It would be very interesting to compute these correlators using a combination of analytical and numerical bootstrap techniques. A natural approach would be to compute the data appearing in current-current OPEs using the numerical bootstrap, analytically continue the results in spin, and use the light-ray OPE of \cite{Kologlu:2019mfz} to construct the event shape in an expansion in the angle between charge detectors. A comparison between theory and experiment for these observables would constitute a nontrivial check of nonperturbative CFT techniques in intrinsically Lorentzian regimes.

Now that we are obtaining more complete pictures of the full spectrum of CFTs, it is important to think about the broader applications of this data. One possible application is to apply this data to the Hamiltonian or conformal truncation program (recent work includes~\cite{Hogervorst:2014rta,Rychkov:2015vap,Katz:2016hxp,Anand:2017yij,Elias-Miro:2017xxf,Elias-Miro:2017tup,Fitzpatrick:2018ttk,Fitzpatrick:2018xlz,Hogervorst:2018otc,Delacretaz:2018xbn,Anand:2019lkt,EliasMiro:2020uvk}). The conformal truncation method starts from fixed points described by CFTs and probes non-conformal physics (e.g., the mass spectrum) using the UV conformal basis. Thus, to obtain precise predictions, it is important to obtain precise sets of conformal data. 

It is also interesting to study the physics of condensed matter or other experimental systems away from their critical points. For example, transport properties or thermal coefficients of quantum critical systems at finite temperature can be computed using CFT data~\cite{Katz:2014rla,Witczak-Krempa:2015pia,Lucas:2016fju,Lucas:2017dqa,Iliesiu:2018fao,Iliesiu:2018zlz}. These computations can now be pursued with much higher precision for O(2) quantum critical points, and one can try to make direct connections with experimental or quantum Monte Carlo data.

\section*{Acknowledgments}
We thank Soner Albayrak, Simon Caron-Huot, Shai Chester, Rajeev Erramilli, Walter Landry, Zhijin Li, Junchen Rong, Slava Rychkov, Ning Su, Alessandro Vichi, and Zahra Zahraee for discussions. We particularly thank Walter Landry and Ning Su for their valuable assistance with computing and software. DSD and JL are supported by Simons Foundation grant 488657 (Simons Collaboration on the Nonperturbative Bootstrap). DSD and JL are also supported by a Sloan Research Fellowship, and a DOE Early Career Award under grant no. DE-SC0019085. The research of DM is supported by the Walter Burke Institute for Theoretical Physics and the Sherman Fairchild Foundation. DP is supported by Simons Foundation grant 488651 (Simons Collaboration on the Nonperturbative Bootstrap) and DOE grant no.~DE-SC0020318. This work uses the Caltech High-Performance Cluster, partially supported by a grant from the Gordon and Betty Moore Foundation, and the Grace computing cluster, supported by the facilities and staff of the Yale University Faculty of Sciences High-Performance Computing Center.

\appendix

\section{Summary of notation}\label{tablenota}
We summarize some of the notation used in this paper in table \ref{tab:notations}.
\begin{table}
\centering
\begin{tabular}{|c|c|c|c|c|c}
\hline
Notation & Description \\
\hline
$s,\phi,t,\chi,\tau$ & Leading scalar operators with O(2) charge $0^+$,1,2,3,4. \\
\hline
$J,T$ & Current and stress tensor. \\
\hline
$f_{O_1O_2O_3}$ & The OPE coefficient with respect to operators $O_1,O_2,O_3$.\\
\hline
${[{{O}_1}{{O}_2}]_{n,\ell }}$ \text{or} ${[{{O}_1}{{O}_2}]_{n,\ell }^q}$ & The $n$-th double-twist family $[{{O}_1}{{O}_2}]$ with spin $\ell$ and charge $q$.\\
\hline
$g_{h,\bar{h}}^{h_{12},h_{34}} (z,\bar{z})$ & Conformal block in three dimensions using the \\
or $g_{O}^{h_{12},h_{34}} (z,\bar{z})$ & convention \cite{Simmons-Duffin:2016wlq} for external scalar operators 1234.\\
\hline
$\Delta,\ell$ & Conformal dimension and spin in general,\\
or $h,\bar{h}$ & or conformal dimension and spin in the $s$-channel.\\
\hline
$\Delta',\ell',h',\bar{h}'$& Mostly conformal dimension and spin in the $t$ or $u$-channels.\\
\hline
$c({h,\bar{h}})$ & Coefficients in the conformal partial wave\\
or $c^{\text{channel}}({h,\bar{h}})$ & decomposition of four-point functions.\\
\hline
$r_O$& The O(2) representation for operator $O$.\\
\hline
$\mathcal{G}$& Conformal blocks factoring out kinematic factors.\\
\hline
$C(z,\bar{h})$, $C^\text{channel}(z,\bar{h})$, & Generating functions.\\
or $C^{abcd}(z,\bar{h})$& \\
\hline
$k_{h}^{r,s}(z)$ &  The $\text{SL}_2$ block $k_{h}^{r,s}(z)\equiv {z^h}\times {_2}{F_1}(h - r,h + s,2h,z).$\\
\hline
$A$, &  Three-dimensional conformal block expansion coefficients\\
$\mathcal{A}$, & in the context of $\text{SL}_2$, dimensional reduction, \\
$\mathcal{C}$ & and Weyl-reflected blocks.\\
\hline
$M$ &  Matrices of correlators for the twist Hamiltonian.\\
\hline
$\mathcal{M}$ &  Crossing matrices in the O(2) model. \\
\hline
\end{tabular}
\caption{\label{tab:notations}Some of the notation used in this paper.}
\end{table}

\section{Integrals of hypergeometric functions}\label{integral}
We define $\Omega$ as
\begin{align}
\Omega _{{h_5},{h_6},p}^{{h_1}{h_2}{h_3}{h_4}} = \int_0^1 {\frac{{dz}}{{{z^2}}}} {\left( {\frac{z}{{1 - z}}} \right)^p}{z^{{h_1} - {h_3}}}k_{{h_5}}^{\tilde 1\tilde 2\tilde 3\tilde 4}(z)k_{{h_6}}^{3214}(1 - z)~.
\end{align}
Here, $\tilde{i}$ means the shadow transform in $d=2$, which maps $h_i$ to $1-h_i$ and $\bar{h}_i$ to $1-\bar{h}_i$, and we define the $\SL _2$ blocks as
\begin{align}
k_{h}^{1234}(z) \equiv k_{h}^{r,s}(z) \equiv {\bar z^{\bar h}}{  _2}{F_1}(h - r,h + s,2h,z)~.
\end{align}
The result of the $\Omega$ integral is given by \cite{Liu:2018jhs}
\begin{align}
&\Omega _{{h_5},{h_6},p}^{1234} =\nonumber\\
& \frac{{\Gamma \left( {2{h_6}} \right)\Gamma \left( {{h_1} - {h_3} + {h_5} + p - 1} \right)\Gamma \left( { - {h_2} + {h_4} + {h_5} + p - 1} \right)\Gamma \left( { - {h_1} + {h_2} - {h_5} + {h_6} - p + 1} \right)}}{{\Gamma \left( { - {h_1} + {h_4} + {h_6}} \right)\Gamma \left( {{h_2} - {h_3} + {h_6}} \right)\Gamma \left( {{h_1} - {h_2} + {h_5} + {h_6} + p - 1} \right)}}\nonumber\\
&{ \times _4}{F_3}\left( {\begin{array}{*{20}{c}}
{{h_1} - {h_2} + {h_5},{h_3} - {h_4} + {h_5},{h_1} - {h_3} + {h_5} + p - 1, - {h_2} + {h_4} + {h_5} + p - 1}\\
{2{h_5},{h_1} - {h_2} + {h_5} - {h_6} + p,{h_1} - {h_2} + {h_5} + {h_6} + p - 1}
\end{array}} \right)\nonumber\\
&+ \frac{{\Gamma \left( {2{h_5}} \right)\Gamma \left( {{h_6} - p + 1} \right)\Gamma \left( {{h_1} - {h_2} + {h_5} - {h_6} + p - 1} \right)\Gamma \left( { - {h_1} + {h_2} + {h_3} - {h_4} + {h_6} - p + 1} \right)}}{{\Gamma \left( {{h_1} - {h_2} + {h_5}} \right)\Gamma \left( {{h_3} - {h_4} + {h_5}} \right)\Gamma \left( { - {h_1} + {h_2} + {h_5} + {h_6} - p + 1} \right)}}\nonumber\\
&{ \times _4}{F_3}\left( {\begin{array}{*{20}{c}}
{{h_2} - {h_3} + {h_6}, - {h_1} + {h_4} + {h_6},{h_6} - p + 1, - {h_1} + {h_2} + {h_3} - {h_4} + {h_6} - p + 1}\\
{2{h_6}, - {h_1} + {h_2} - {h_5} + {h_6} - p + 2, - {h_1} + {h_2} + {h_5} + {h_6} - p + 1}
\end{array};1} \right)~.
\end{align}

We will also define an $R$ symbol,
\begin{align}
R_{\alpha,\beta,\gamma}^{\rho,\sigma}=\int_0^1 {{x^\rho }} {(1 - x)^\sigma }_2F_1(\alpha ,\beta ;\gamma ;x)dx = \frac{{\Gamma (\rho  + 1)\Gamma (\sigma  + 1)}}{{\Gamma (\rho  + \sigma  + 2)}}{}_3{F_2}(\alpha ,\beta ,\rho  + 1;\gamma ,\rho  + \sigma  + 2;1)~.
\end{align}

When inverting individual blocks using the $\text{SL}_2$ expansion, we have to be a little careful about the analytic continuation of the function $_3 F_2$, defined as 
\begin{align}
_3{F_2}({a_1},{a_2},{a_3};{b_1},{b_2};z) \equiv \sum\limits_{k = 0}^\infty  {\frac{{{{\left( {{a_1}} \right)}_k}{{\left( {{a_2}} \right)}_k}{{\left( {{a_3}} \right)}_k}}}{{{{\left( {{b_1}} \right)}_k}{{\left( {{b_2}} \right)}_k}}}} \frac{{{z^k}}}{{k!}}~.
\end{align}
On the circle $\left| z \right| = 1$, the function itself is absolute convergent when 
\begin{align}
{\mathop{\text{Re}}\nolimits}[{\gamma _{a,b}}] = {b_1} + {b_2} - {a_1} - {a_2} - {a_3}>0~.
\end{align}
It is convergent except at $z=1$ when 
\begin{align}
- 1 < {\mathop{\text{Re}}\nolimits} [{\gamma _{a,b}}] \le 0~.
\end{align}
On the other hand, it is divergent for
\begin{align}
{\mathop{\text{Re}}\nolimits} [{\gamma _{a,b}}] \le  - 1~.
\end{align}
Now, one might worry about the divergence at $z=1$ when $\gamma_{a,b}\le 0$ in our inversion formula. However, we wish to point out that a hypergeometric transformation can solve the problem. In our inversion formula, the $R$ symbol appears as
\begin{align}
R_{\bar h - {h_{12}},\bar h + {h_{34}},2\bar h}^{\bar h + {h_1} + {h_2} - 2,{h_O} + p - {h_1} - {h_4}}~.
\end{align}
It has the following symmetry when we swap the operators $1 \leftrightarrow 2,3 \leftrightarrow 4$:
\begin{align}\label{Requ}
R_{\bar h - {h_{12}},\bar h + {h_{34}},2\bar h}^{\bar h + {h_1} + {h_2} - 2,{h_O} + p - {h_1} - {h_4}} = R_{\bar h - {h_{21}},\bar h + {h_{43}},2\bar h}^{\bar h + {h_1} + {h_2} - 2,{h_O} + p - {h_2} - {h_3}}~.
\end{align}
This formula holds when both sides are well-defined, and could be proven easily using a hypergeometric transformation. Thus, when one side is not well-defined, we can use the other side to define an analytic continuation. The following is a specific example.

Consider including the isolated operators $T$ and $J$ in the $t$-channel of $\left\langle {\phi ss\phi } \right\rangle $. We see that for the left-hand side of (\ref{Requ}):
\begin{align}
{\gamma _{a,b}} = {h_O} + p - {h_2} - {h_3} + 1 = \frac{3}{2} + p - 2{h_s} = p - {\text{0.02}}~,
\end{align}
which is smaller than 0 for $p=0$. To resolve the divergence, we swap $1 \leftrightarrow 2,3 \leftrightarrow 4$ and we obtain a formula for $\gamma_{a,b}$ using the right-hand side of (\ref{Requ}):
\begin{align}
{\gamma _{a,b}} = {h_O} + p - {h_1} - {h_4} + 1 = \frac{3}{2} + p - 2{h_\phi } = p + 0.98~,
\end{align} 
which is always in the convergence regime of the $_3F_2$. We find that the above treatment can resolve all the non-convergence issues in our analytic bootstrap in the O(2) model. Moreover, the non-convergence situation does not appear in the dimensional reduction version, which makes use of $_4F_3$ hypergeometric functions. It might be interesting to ask for a deeper interpretation of the above phenomenon.

\section{Crossing equations in the O(2) model}\label{cross}
In this paper, we mainly study crossing equations for the external operators $\phi, s, t$. The crossing equations and their derivation are given in more detail in \cite{Chester:2019ifh}. Here, we will review some basics of their derivation, and present crossing equations in conventions consistent with the main text of this paper. Moreover, we give some explicit Mean Field Theory (MFT) coefficients suitable for our analytic bootstrap analysis in the O(2) system. We also derive some crossing equations involving $\chi$ used in the analytic bootstrap, which is new compared to \cite{Chester:2019ifh}.

\subsection{Index-free notation and practical implementation}
We first give a brief introduction to the index-free notation established in \cite{Chester:2019ifh}. For a charge-$q$ operator $O_{i_1i_2\cdots i_q}$ containing $q$ O$(N)$ indices, we define the index-free notation
\begin{align}
O(v) = {O_{{i_1}{i_2} \cdots {i_q}}}{v^{i_1}v^{i_2} \cdots v^{i_q}}~,
\end{align}
where $v$ is a null O$(N)$ vector
\begin{align}
{v^2} = 0~.
\end{align}
Since we are taking $N=2$, we can make further simplifications. We introduce the basis vectors
\begin{align}
e = \left( \begin{array}{l}
1\\
i
\end{array} \right)~,~~~~~\bar e = \left( \begin{array}{l}
1\\
 - i
\end{array} \right)~.
\end{align}
So for degree-$n$ tensors represented in the index-free notation, $f(v)$ and $g(v)$, we have the expansions
\begin{align}
&f(v) = f(e){\bigg(\frac{{v \cdot \bar e}}{2}\bigg)^n} + f(\bar e){\bigg(\frac{{v \cdot e}}{2}\bigg)^n}~,\nonumber\\
&g(v) = g(e){\bigg(\frac{{v \cdot \bar e}}{2}\bigg)^n} + g(\bar e){\bigg(\frac{{v \cdot e}}{2}\bigg)^n}~.
\end{align}
This expansion is consistent with both the degree in $v$ and the result when $v$ is taken to be $e$ or $\bar{e}$. Thus, it is uniquely fixed. When contracting two tensors, $f$ and $g$ with degree $n$ in the index-free notation, we compare the degree for common vectors and remove the coefficients, so we define the inner product
\begin{align}
&(f,g) = \left(f(e){\left(\frac{{v \cdot \bar e}}{2}\right)^n} + f(\bar e){\left(\frac{{v \cdot e}}{2}\right)^n},g(e){\left(\frac{{v \cdot \bar e}}{2}\right)^n} + g(\bar e){\left(\frac{{v \cdot e}}{2}\right)^n}\right)\nonumber\\
&= f(\bar e)g(e){\left(\frac{{e \cdot \bar e}}{4}\right)^n} + f(e)g(\bar e){\left(\frac{{e \cdot \bar e}}{4}\right)^n}\nonumber\\
&= \frac{1}{2^{n}}\left(f(e)g(\bar e) + f(\bar e)g(e)\right)~.
\end{align}
One can use the formula to compute some examples. For instance, if we wish to compute
\begin{align}
\left( {{{({v_1} \cdot {v_0})}^2}{{({v_2} \cdot {v_0})}^2},{{({v_3} \cdot {v_0})}^2}{{({v_4} \cdot {v_0})}^2}} \right)~,
\end{align}
the result is 
\begin{align}
\frac{1}{{{2^4}}}\left({({v_1} \cdot e)^2}{({v_2} \cdot e)^2}{({v_3} \cdot \bar e)^2}{({v_4} \cdot \bar e)^2} + {({v_1} \cdot \bar e)^2}{({v_2} \cdot \bar e)^2}{({v_3} \cdot e)^2}{({v_4} \cdot e)^2} \right)~.
\end{align}
Furthermore, we introduce
\begin{align}
&{v_i} \cdot e = {w_i}~,\nonumber\\
&{v_i} \cdot \bar e = {{\bar w}_i}~,
\end{align}
where
\begin{align}
&v_i^2 = {w_i}{{\bar w}_i}~,\nonumber\\
&{v_i} \cdot {v_j} = \frac{1}{2}({w_i}{{\bar w}_j} + {{\bar w}_i}{w_j})~,\nonumber\\
&{v_i} \cdot {{\tilde v}_j} = \frac{i}{2}({w_i}{{\bar w}_j} - {{\bar w}_i}{w_j})~,\nonumber\\
&\tilde{v}\equiv \epsilon^{ij} v_j~,
\end{align}
and $\epsilon$ is the Levi-Civita symbol. Then we can rewrite the above example as
\begin{align}
\frac{1}{2^4}\left(w_1^2w_2^2\bar w_3^2\bar w_4^2 + \bar w_1^2\bar w_2^2w_3^2w_4^2\right)~.
\end{align}
Practically, we could use either $v$ or $w$ variables to compute the O(2) crossing equations. One important thing to notice is that we should keep $v_i$ to be null. For instance, if we write crossing equations in terms of the $w$ variables, because $v_i$ is null, we know that
\begin{align}
w_i=0 \text{ or }\bar{w}_i=0~.
\end{align}
So when we derive crossing equations for different charge sectors, we can always impose this condition. 
\subsection{General setup}
Next we compute crossing equations and MFT coefficients using the index-free notation. 

We will completely follow the convention of \cite{Simmons-Duffin:2016wlq}. We define the $s$, $t$ and $u$-channel expansion as the following:
\begin{align}
&\left\langle {{\phi _1}({x_1}){\phi _2}({x_1}){\phi _3}({x_3}){\phi _4}({x_4})} \right\rangle  \nonumber\\
&=\frac{1}{{x_{12}^{{\Delta _1} + {\Delta _2}}x_{34}^{{\Delta _3} + {\Delta _4}}}}\frac{{x_{14}^{{\Delta _{34}}}x_{24}^{{\Delta _{12}}}}}{{x_{13}^{{\Delta _{34}}}x_{14}^{{\Delta _{12}}}}}\sum\limits_O {{{( - \frac{1}{2})}^\ell }{f_{12O}}{f_{34O}}g_{\Delta ,\ell }^{{\Delta _{12}},{\Delta _{34}}}(z,\bar z)} \nonumber\\
&= \frac{1}{{x_{32}^{{\Delta _3} + {\Delta _2}}x_{14}^{{\Delta _1} + {\Delta _4}}}}\frac{{x_{34}^{{\Delta _{14}}}x_{24}^{{\Delta _{32}}}}}{{x_{31}^{{\Delta _{14}}}x_{34}^{{\Delta _{32}}}}}\sum\limits_O {{{( - \frac{1}{2})}^{\ell'} }{f_{32O}}{f_{14O}}g_{\Delta' ,\ell' }^{{\Delta _{32}},{\Delta _{14}}}(1 - z,1 - \bar z)}  \nonumber\\
&= \frac{1}{{x_{42}^{{\Delta _4} + {\Delta _2}}x_{13}^{{\Delta _1} + {\Delta _3}}}}\frac{{x_{43}^{{\Delta _{13}}}x_{23}^{{\Delta _{42}}}}}{{x_{41}^{{\Delta _{13}}}x_{43}^{{\Delta _{42}}}}}\sum\limits_O {{{( - \frac{1}{2})}^{\ell'} }{f_{42O}}{f_{13O}}g_{\Delta' ,\ell' }^{{\Delta _{42}},{\Delta _{13}}}(1/z,1/\bar z)} ~,
\end{align}
where we use $\Delta_i$ for the dimensions of the external scalars $\phi_i$, the internal operator $O$ has dimension $\Delta$ (or $\Delta'$) and spin $\ell$ (or $\ell')$. In general, we use $\Delta'$, $\ell'$ for $t-$ and $u$-channel operators, and $\Delta$, $\ell$ for $s$-channel operators. We define $\Delta_{ij}=\Delta_i-\Delta_j$ and $x_{ij}=|x_i-x_j|$. The cross ratios are defined by
\begin{align}
u = \frac{{x_{12}^2x_{34}^2}}{{x_{13}^2x_{24}^2}} = z\bar z\quad ~,~~~~v = \frac{{x_{23}^2x_{14}^2}}{{x_{13}^2x_{24}^2}} = (1 - z)(1 - \bar z)~.
\end{align}
Sometimes, we will exchange ${{f_{12O}}{f_{34O}}} = {{f_{12O}}{f_{43O}}}(-1)^\ell$ to cancel the $(-1)^\ell$ factors appearing in the OPEs. From now on, we will write the prefactors as
\begin{align}
&{\mathbb{S}_{}} = \frac{1}{{x_{12}^{{\Delta _1} + {\Delta _2}}x_{34}^{{\Delta _3} + {\Delta _4}}}}\frac{{x_{14}^{{\Delta _{34}}}x_{24}^{{\Delta _{12}}}}}{{x_{13}^{{\Delta _{34}}}x_{14}^{{\Delta _{12}}}}}~,\nonumber\\
&{\mathbb{T}_{}} = \frac{1}{{x_{32}^{{\Delta _3} + {\Delta _2}}x_{14}^{{\Delta _1} + {\Delta _4}}}}\frac{{x_{34}^{{\Delta _{14}}}x_{24}^{{\Delta _{32}}}}}{{x_{31}^{{\Delta _{14}}}x_{34}^{{\Delta _{32}}}}}~,\nonumber\\
&{\mathbb{U}_{}} = \frac{1}{{x_{42}^{{\Delta _4} + {\Delta _2}}x_{13}^{{\Delta _1} + {\Delta _3}}}}\frac{{x_{43}^{{\Delta _{13}}}x_{23}^{{\Delta _{42}}}}}{{x_{41}^{{\Delta _{13}}}x_{43}^{{\Delta _{42}}}}}~.
\end{align}
Now we set up the conventions for MFT coefficients. We have
\begin{align}
&\left\langle {{\phi _1}\left( {{x_1}} \right){\phi _2}\left( {{x_2}} \right){\phi _1}\left( {{x_3}} \right){\phi _2}\left( {{x_4}} \right)} \right\rangle \nonumber\\
&= \frac{1}{{x_{13}^{2{\Delta _1}}x_{24}^{2{\Delta _2}}}} = \mathbb{S}{z^{\frac{{{\Delta _1} + {\Delta _2}}}{2}}}{z^{\frac{{{\Delta _1} + {\Delta _2}}}{2}}} = \mathbb{S} \sum\limits_{n,\ell  = 0}^\infty  {C_{n,\ell }^{{\text{MFT}}}} \left( {{\Delta _1},{\Delta _2}} \right){( - 1)^\ell }g_{{{[\phi_1\phi_2]}_{n,\ell }}}^{\frac{{{\Delta _{12}}}}{2},\frac{{{\Delta _{12}}}}{2}}(z,\bar z)~,\nonumber\\
&\left\langle {{\phi _1}\left( {{x_1}} \right){\phi _2}\left( {{x_2}} \right){\phi _2}\left( {{x_3}} \right){\phi _1}\left( {{x_4}} \right)} \right\rangle \nonumber\\
&= \frac{1}{{x_{14}^{2{\Delta _1}}x_{23}^{2{\Delta _2}}}} = \mathbb{S} {y^{\frac{{{\Delta _1} + {\Delta _2}}}{2}}}{{\bar y}^{\frac{{{\Delta _1} + {\Delta _2}}}{2}}}{v^r} = \mathbb{S} \sum\limits_{n,\ell  = 0}^\infty  {C_{n,\ell }^{{\text{MFT}}}} \left( {{\Delta _1},{\Delta _2}} \right)g_{[\phi_1\phi_2]_{n,\ell}}^{\frac{{{\Delta _{12}}}}{2},\frac{{{\Delta _{12}}}}{2}}(z,\bar z)~,
\end{align}
where we define
\begin{align}
&y = \frac{z}{{1 - z}}~~~,~~~\bar y = \frac{{\bar z}}{{1 - \bar z}}~,\nonumber\\
&r = \frac{{{\Delta _{12}}}}{2}~~~,~~~s = \frac{{{\Delta _{34}}}}{2}~.
\end{align}
The above expansions define the $t$ and $u$-channel expansions of double-twist operators ${\left[ {{\phi _1}{\phi _2}} \right]_{n,\ell }}$ with spin $\ell$. Here we use the convention of the MFT coefficients in \cite{Simmons-Duffin:2016wlq}
\begin{align}
&C_{n,\ell }^{{\text{MFT}}}\left( {{\Delta _1},{\Delta _2}} \right) = \\
&\frac{{{{\left( {{\Delta _1} - 1/2} \right)}_n}{{\left( {{\Delta _2} - 1/2} \right)}_n}{{\left( {{\Delta _1}} \right)}_{\ell  + n}}{{\left( {{\Delta _2}} \right)}_{\ell  + n}}}}{{\ell !n!{{(\ell  + 3/2)}_n}{{\left( {{\Delta _1} + {\Delta _2} + n - 2} \right)}_n}{{\left( {{\Delta _1} + {\Delta _2} + 2n + \ell  - 1} \right)}_\ell }{{\left( {{\Delta _1} + {\Delta _2} + n + \ell  - 3/2} \right)}_n}}}~,\nonumber
\end{align}
where we use the Pochhammer symbol $(a)_n\equiv \Gamma(a+n)/\Gamma(a)$. 

Now we write down crossing equations for identical operators, involving $s$, $\phi$, and $t$ as external operators. To keep equations simpler we also define
\begin{align}
{\left( - \frac{1}{2}\right)^{\ell} }{f_{abO}}{f_{cdO}} \equiv {\hat f_{abO}}{\hat f_{cdO}}~,
\end{align}
as a short-hand notation for the products of OPEs.

\subsection{Crossing equations and MFT coefficients for identical operators}
We start with correlators for identical operators:
\begin{itemize}
\item $\left\langle {ssss} \right\rangle$: We have the crossing equation
\begin{align}
&{\mathbb{S}_{ssss}}\sum\limits_{O = {0^ + }} {} {\hat{f}_{ssO}}{\hat{f}_{ssO}}g_{O}^{ss,ss}(z,\bar z)\nonumber\\
&= {\mathbb{T}_{ssss}}\sum\limits_{O' = {0^ + }} {} {\hat{f}_{ssO'}}{\hat{f}_{ssO'}}g_{O'}^{ss,ss}(1-z,1- \bar{z})\nonumber\\
&= {\mathbb{U}_{ssss}}\sum\limits_{O' = {0^ + }} {} {\hat{f}_{ssO'}}{\hat{f}_{ssO'}}g_{O'}^{ss,ss}(1/z,1/\bar z)~,
\end{align}
and for MFT we have
\begin{align}
\left\langle {ssss} \right\rangle  = \frac{1}{{x_{12}^{2{\Delta _s}}x_{34}^{2{\Delta _s}}}} + \frac{1}{{x_{13}^{2{\Delta _s}}x_{24}^{2{\Delta _s}}}} + \frac{1}{{x_{14}^{2{\Delta _s}}x_{23}^{2{\Delta _s}}}} = \frac{1}{{x_{12}^{2{\Delta _s}}x_{34}^{2{\Delta _s}}}}\left( {1 + {u^{{\Delta _s}}} + {y^{{\Delta _s}}}{{\bar y}^{{\Delta _s}}}} \right)~.
\end{align}
\item $\left\langle { \phi  \phi \phi \phi} \right\rangle$: We have the crossing equation 
\begin{align}
&{\mathbb{S}_{\phi \phi \phi \phi }}\sum\limits_{O = {0^ + }} {} {\hat{f}_{\phi \phi O}}{\hat{f}_{\phi \phi O}}g_{O}^{\phi \phi ,\phi \phi }(z,\bar z)\nonumber\\
&= {\mathbb{T}_{\phi \phi \phi \phi }}\left( \begin{array}{l}
\frac{1}{2}\sum\limits_{O' = {0^ + }} {} {\hat{f}_{\phi \phi O'}}{\hat{f}_{\phi \phi O'}}g_{O'}^{\phi \phi ,\phi \phi }(1 - z,1 - \bar z)\\
+\frac{1}{2}\sum\limits_{O' = 2} {} {\hat{f}_{\phi \phi O'}}{\hat{f}_{\phi \phi O'}}g_{O'}^{\phi \phi ,\phi \phi }(1 - z,1 - \bar z)\\
 - \frac{1}{2}\sum\limits_{O' = {0^ - }} {} {\hat{f}_{\phi \phi O'}}{\hat{f}_{\phi \phi O'}}g_{O'}^{\phi \phi ,\phi \phi }(1 - z,1 - \bar z)
\end{array} \right)\nonumber\\
&= {\mathbb{U}_{\phi \phi \phi \phi }}\left( \begin{array}{l}
\frac{1}{2}\sum\limits_{O' = {0^ + }} {} {\hat{f}_{\phi \phi O'}}{\hat{f}_{\phi \phi O'}}g_{O'}^{\phi \phi ,\phi \phi }(1/z,1/\bar z)\\
+\frac{1}{2}\sum\limits_{O' = 2} {} {\hat{f}_{\phi \phi O'}}{\hat{f}_{\phi \phi O'}}g_{O'}^{\phi \phi ,\phi \phi }(1/z,1/\bar z)\\
 - \frac{1}{2}\sum\limits_{O' = {0^ - }} {} {\hat{f}_{\phi \phi O'}}{\hat{f}_{\phi \phi O'}}g_{O'}^{\phi \phi ,\phi \phi }(1/z,1/\bar z)
\end{array} \right)~,
\end{align}
\begin{align}
&{\mathbb{S}_{\phi \phi \phi \phi }}\sum\limits_{O = {0^ - }} {} {\hat{f}_{\phi \phi O}}{\hat{f}_{\phi \phi O}}g_{O}^{\phi \phi ,\phi \phi }(z,\bar z)\nonumber\\
&= {\mathbb{T}_{\phi \phi \phi \phi }}\left( \begin{array}{l}
 - \frac{1}{2}\sum\limits_{O' = {0^ + }} {} {\hat{f}_{\phi \phi O'}}{\hat{f}_{\phi \phi O'}}g_{O'}^{\phi \phi ,\phi \phi }(1 - z,1 - \bar z)\\
+\frac{1}{2}\sum\limits_{O' = 2} {} {\hat{f}_{\phi \phi O'}}{\hat{f}_{\phi \phi O'}}g_{O'}^{\phi \phi ,\phi \phi }(1 - z,1 - \bar z)\\
 + \frac{1}{2}\sum\limits_{O' = {0^ - }} {} {\hat{f}_{\phi \phi O'}}{\hat{f}_{\phi \phi O'}}g_{O'}^{\phi \phi ,\phi \phi }(1 - z,1 - \bar z)
\end{array} \right)\nonumber\\
&= {\mathbb{U}_{\phi \phi \phi \phi }}\left( \begin{array}{l}
\frac{1}{2}\sum\limits_{O' = {0^ + }} {} {\hat{f}_{\phi \phi O'}}{\hat{f}_{\phi \phi O'}}g_{O'}^{\phi \phi ,\phi \phi }(1/z,1/\bar z)\\
 - \frac{1}{2}\sum\limits_{O' = 2} {} {\hat{f}_{\phi \phi O'}}{\hat{f}_{\phi \phi O'}}g_{O'}^{\phi \phi ,\phi \phi }(1/z,1/\bar z)\\
 - \frac{1}{2}\sum\limits_{O' = {0^ - }} {} {\hat{f}_{\phi \phi O'}}{\hat{f}_{\phi \phi O'}}g_{O'}^{\phi \phi ,\phi \phi }(1/z,1/\bar z)
\end{array} \right)~,
\end{align}
\begin{align}
&{\mathbb{S}_{\phi \phi \phi \phi }}\sum\limits_{O = 2} {} {\hat{f}_{\phi \phi O}}{\hat{f}_{\phi \phi O}}g_{O}^{\phi \phi ,\phi \phi }(z,\bar z)\nonumber\\
&= {\mathbb{T}_{\phi \phi \phi \phi }}\left( \begin{array}{l}
\sum\limits_{O' = {0^ + }} {} {\hat{f}_{\phi \phi O'}}{\hat{f}_{\phi \phi O'}}g_{O'}^{\phi \phi ,\phi \phi }(1 - z,1 - \bar z)\\
 + \sum\limits_{O' = {0^ - }} {} {\hat{f}_{\phi \phi O'}}{\hat{f}_{\phi \phi O'}}g_{O'}^{\phi \phi ,\phi \phi }(1 - z,1 - \bar z)
\end{array} \right)\nonumber\\
&= {\mathbb{U}_{\phi \phi \phi \phi }}\left( \begin{array}{l}
\frac{1}{2}\sum\limits_{O' = {0^ + }} {} {\hat{f}_{\phi \phi O'}}{\hat{f}_{\phi \phi O'}}g_{O'}^{\phi \phi ,\phi \phi }(1/z,1/\bar z)\\
 + \frac{1}{2}\sum\limits_{O' = {0^ - }} {} {\hat{f}_{\phi \phi O'}}{\hat{f}_{\phi \phi O'}}g_{O'}^{\phi \phi ,\phi \phi }(1/z,1/\bar z)
\end{array} \right)~,
\end{align}
and for MFT we have
\begin{align}
&\left\langle {\phi \phi \phi \phi } \right\rangle  = \frac{{\left( {{v_1} \cdot {v_2}} \right)\left( {{v_3} \cdot {v_4}} \right)}}{{x_{12}^{2{\Delta _\phi }}x_{34}^{2{\Delta _\phi }}}} + \frac{{\left( {{v_1} \cdot {v_3}} \right)\left( {{v_2} \cdot {v_4}} \right)}}{{x_{13}^{2{\Delta _\phi }}x_{24}^{2{\Delta _\phi }}}} + \frac{{\left( {{v_1} \cdot {v_4}} \right)\left( {{v_2} \cdot {v_3}} \right)}}{{x_{14}^{2{\Delta _\phi }}x_{23}^{2{\Delta _\phi }}}}\nonumber\\
&= \frac{1}{{x_{12}^{2{\Delta _\phi }}x_{34}^{2{\Delta _\phi }}}}\left( {\begin{array}{*{20}{l}}
{\left( {{v_1} \cdot {v_2}} \right)\left( {{v_3} \cdot {v_4}} \right)\left( {1 + \frac{1}{2}{u^{{\Delta _\phi }}} + \frac{1}{2}{y^{{\Delta _\phi }}}{{\bar y}^{{\Delta _\phi }}}} \right)}\\
{ + \frac{1}{4}\left( {{w_1}{w_2}{{\bar w}_3}{{\bar w}_4} + {{\bar w}_1}{{\bar w}_2}{w_3}{w_4}} \right)\left( {{u^{{\Delta _\phi }}} + {y^{{\Delta _\phi }}}{{\bar y}^{{\Delta _\phi }}}} \right)}\\
{ + \left( {{v_1} \cdot {{\tilde v}_2}} \right)\left( {{v_3} \cdot {{\tilde v}_4}} \right)\left( {\frac{1}{2}{u^{{\Delta _\phi }}} - \frac{1}{2}{y^{{\Delta _\phi }}}{{\bar y}^{{\Delta _\phi }}}} \right)}
\end{array}} \right)~.
\end{align}
\item $\left\langle {tttt} \right\rangle$: We have the crossing equation
\begin{align}
&{\mathbb{S}_{t t t t }}\sum\limits_{O = {0^ + }} {} {\hat{f}_{t t O}}{\hat{f}_{t t O}}g_{O}^{t t ,t t }(z,\bar z)\nonumber\\
&= {\mathbb{T}_{t t t t }}\left( \begin{array}{l}
\frac{1}{2}\sum\limits_{O' = {0^ + }} {} {\hat{f}_{t t O'}}{\hat{f}_{t t O'}}g_{O'}^{t t ,t t }(1 - z,1 - \bar z)\\
+\frac{1}{2}\sum\limits_{O' = 4} {} {\hat{f}_{t t O'}}{\hat{f}_{t t O'}}g_{O'}^{t t ,t t }(1 - z,1 - \bar z)\\
 - \frac{1}{2}\sum\limits_{O' = {0^ - }} {} {\hat{f}_{t t O'}}{\hat{f}_{t t O'}}g_{O'}^{t t ,t t }(1 - z,1 - \bar z)
\end{array} \right)\nonumber\\
&= {\mathbb{U}_{t t t t }}\left( \begin{array}{l}
\frac{1}{2}\sum\limits_{O'= {0^ + }} {} {\hat{f}_{t t O'}}{\hat{f}_{t t O'}}g_{O'}^{t t ,t t }(1/z,1/\bar z)\\
+\frac{1}{2}\sum\limits_{O' = 4} {} {\hat{f}_{t t O'}}{\hat{f}_{t t O'}}g_{O'}^{t t ,t t }(1/z,1/\bar z)\\
 - \frac{1}{2}\sum\limits_{O' = {0^ - }} {} {\hat{f}_{t t O'}}{\hat{f}_{t t O'}}g_{O'}^{t t ,t t }(1/z,1/\bar z)
\end{array} \right)~,
\end{align}
\begin{align}
&{\mathbb{S}_{t t t t }}\sum\limits_{O = {0^ - }} {} {\hat{f}_{t t O}}{\hat{f}_{t t O}}g_{O}^{t t ,t t }(z,\bar z)\nonumber\\
&= {\mathbb{T}_{t t t t }}\left( \begin{array}{l}
 - \frac{1}{2}\sum\limits_{O' = {0^ + }} {} {\hat{f}_{t t O'}}{\hat{f}_{t t O'}}g_{O'}^{t t ,t t }(1 - z,1 - \bar z)\\
+\frac{1}{2}\sum\limits_{O' = 4} {} {\hat{f}_{t t O'}}{\hat{f}_{t t O'}}g_{O'}^{t t ,t t }(1 - z,1 - \bar z)\\
 + \frac{1}{2}\sum\limits_{O' = {0^ - }} {} {\hat{f}_{t t O'}}{\hat{f}_{t t O'}}g_{O'}^{t t ,t t }(1 - z,1 - \bar z)
\end{array} \right)\nonumber\\
&= {\mathbb{U}_{t t t t }}\left( \begin{array}{l}
\frac{1}{2}\sum\limits_{O' = {0^ + }} {} {\hat{f}_{t t O'}}{\hat{f}_{t t O'}}g_{O'}^{t t ,t t }(1/z,1/\bar z)\\
 - \frac{1}{2}\sum\limits_{O' = 4} {} {\hat{f}_{t t O'}}{\hat{f}_{t t O'}}g_{O'}^{t t ,t t }(1/z,1/\bar z)\\
 - \frac{1}{2}\sum\limits_{O' = {0^ - }} {} {\hat{f}_{t t O'}}{\hat{f}_{t t O'}}g_{O'}^{t t ,t t }(1/z,1/\bar z)
\end{array} \right)~,
\end{align}
\begin{align}
&{\mathbb{S}_{t t t t }}\sum\limits_{O = 4} {} {\hat{f}_{t t O}}{\hat{f}_{t t O}}g_{O}^{t t ,t t }(z,\bar z)\nonumber\\
&= {\mathbb{T}_{t t t t }}\left( \begin{array}{l}
\sum\limits_{O' = {0^ + }} {} {\hat{f}_{t t O'}}{\hat{f}_{t t O'}}g_{O'}^{t t ,t t }(1 - z,1 - \bar z)\\
 + \sum\limits_{O' = {0^ - }} {} {\hat{f}_{t t O'}}{\hat{f}_{t t O'}}g_{O'}^{t t ,t t }(1 - z,1 - \bar z)
\end{array} \right)\nonumber\\
&= {\mathbb{U}_{t t t t }}\left( \begin{array}{l}
\frac{1}{2}\sum\limits_{O' = {0^ + }} {} {\hat{f}_{t t O'}}{\hat{f}_{t t O'}}g_{O'}^{t t ,t t }(1/z,1/\bar z)\\
 + \frac{1}{2}\sum\limits_{O' = {0^ - }} {} {\hat{f}_{t t O'}}{\hat{f}_{t t O'}}g_{O'}^{t t ,t t }(1/z,1/\bar z)
\end{array} \right)~,
\end{align}
and for MFT we have
\begin{align}
&\left\langle {t t t t } \right\rangle  = \frac{{\left( {{v_1} \cdot {v_2}} \right)\left( {{v_3} \cdot {v_4}} \right)}}{{x_{12}^{2{\Delta _t }}x_{34}^{2{\Delta _t }}}} + \frac{{\left( {{v_1} \cdot {v_3}} \right)\left( {{v_2} \cdot {v_4}} \right)}}{{x_{13}^{2{\Delta _t }}x_{24}^{2{\Delta _t }}}} + \frac{{\left( {{v_1} \cdot {v_4}} \right)\left( {{v_2} \cdot {v_3}} \right)}}{{x_{14}^{2{\Delta _t }}x_{23}^{2{\Delta _t }}}}\nonumber\\
&= \frac{1}{{x_{12}^{2{\Delta _t }}x_{34}^{2{\Delta _t }}}}\left( {\begin{array}{*{20}{l}}
{\left( {{v_1} \cdot {v_2}} \right)\left( {{v_3} \cdot {v_4}} \right)\left( {1 + \frac{1}{2}{u^{{\Delta _t }}} + \frac{1}{2}{y^{{\Delta _t }}}{{\bar y}^{{\Delta _t }}}} \right)}\\
{ + \frac{1}{4}\left( {{w_1}{w_2}{{\bar w}_3}{{\bar w}_4} + {{\bar w}_1}{{\bar w}_2}{w_3}{w_4}} \right)\left( {{u^{{\Delta _t }}} + {y^{{\Delta _t }}}{{\bar y}^{{\Delta _t }}}} \right)}\\
{ + \left( {{v_1} \cdot {{\tilde v}_2}} \right)\left( {{v_3} \cdot {{\tilde v}_4}} \right)\left( {\frac{1}{2}{u^{{\Delta _t }}} - \frac{1}{2}{y^{{\Delta _t }}}{{\bar y}^{{\Delta _t }}}} \right)}
\end{array}} \right)~.
\end{align}
\end{itemize}

\subsection{Crossing equations and MFT coefficients for mixed operators}
Here we list the crossing equations for mixed operators we will use in this paper.
\begin{itemize}
\item $\left\langle { \phi s \phi s} \right\rangle$: We have the crossing equation
\begin{align}
&{\mathbb{S}_{\phi s\phi s}}\sum\limits_{O = 1} {} {\hat{f}_{\phi sO}}{\hat{f}_{\phi sO}}g_{O}^{\phi s,\phi s}(z,\bar z)\nonumber\\
&= {\mathbb{T}_{\phi s\phi s}}\sum\limits_{O' = 1} {} {\hat{f}_{\phi sO'}}{\hat{f}_{\phi sO'}}g_{O'}^{\phi s,\phi s}(1 - z,1 - \bar z)\nonumber\\
&= {\mathbb{U}_{\phi s\phi s}}\sum\limits_{O' = {0^ + }} {} {\hat{f}_{ssO'}}{\hat{f}_{\phi \phi O'}}g_{O'}^{ss,\phi \phi }(1/z,1/\bar z)~,
\end{align}
and for MFT we have
\begin{align}
\left\langle {\phi s\phi s} \right\rangle  = \frac{{({v_1} \cdot {v_3})}}{{x_{13}^{2{\Delta _\phi }}x_{24}^{2{\Delta _s}}}} = \frac{1}{{x_{12}^{{\Delta _\phi } + {\Delta _s}}x_{34}^{{\Delta _\phi } + {\Delta _s}}}}\frac{{x_{14}^{{\Delta _\phi } - {\Delta _s}}x_{24}^{{\Delta _\phi } - {\Delta _s}}}}{{x_{13}^{{\Delta _\phi } - {\Delta _s}}x_{14}^{{\Delta _\phi } - {\Delta _s}}}}({v_1} \cdot {v_3}){z^{\frac{{{\Delta _\phi } + {\Delta _s}}}{2}}}{{\bar z}^{\frac{{{\Delta _\phi } + {\Delta _s}}}{2}}}~.
\end{align}
\item $\left\langle { \phi t \phi t} \right\rangle$: We have the crossing equation
\begin{align}
&{\mathbb{S}_{\phi t\phi t}}\sum\limits_{O = 3} {} {\hat{f}_{t\phi O}}{\hat{f}_{t\phi O}}g_{O}^{\phi t,\phi t}(z,\bar z)\nonumber\\
&= {\mathbb{T}_{\phi t\phi t}}\sum\limits_{O' = 1} {} {\hat{f}_{t\phi O'}}{\hat{f}_{t\phi O'}}g_{O'}^{\phi t,\phi t}(1 - z,1 - \bar z)\nonumber\\
&= {\mathbb{U}_{\phi t\phi t}}\left( \begin{array}{l}
\sum\limits_{O' = {0^ + }} {} {\hat{f}_{\phi \phi O'}}{\hat{f}_{ttO'}}g_{O'}^{tt,\phi \phi }(1/z,1/\bar z)\\
 + \sum\limits_{O' = {0^ - }} {} {\hat{f}_{\phi \phi O'}}{\hat{f}_{ttO'}}g_{O'}^{tt,\phi \phi }(1/z,1/\bar z)
\end{array} \right)~,
\end{align}
\begin{align}
&{\mathbb{S}_{\phi t\phi t}}\sum\limits_{O = 1} {} {\hat{f}_{t\phi O}}{\hat{f}_{t\phi O}}g_{O}^{\phi t,\phi t}(z,\bar z)\nonumber\\
&= {\mathbb{T}_{\phi t\phi t}}\sum\limits_{O' = 3} {} {\hat{f}_{t\phi O'}}{\hat{f}_{t\phi O'}}g_{O'}^{\phi t,\phi t}(1 - z,1 - \bar z)\nonumber\\
&= {\mathbb{U}_{\phi t\phi t}}\left( \begin{array}{l}
\sum\limits_{O' = {0^ + }} {} {\hat{f}_{\phi \phi O'}}{\hat{f}_{ttO'}}g_{O'}^{tt,\phi \phi }(1/z,1/\bar z)\\
 - \sum\limits_{O' = {0^ - }} {} {\hat{f}_{\phi \phi O'}}{\hat{f}_{ttO'}}g_{O'}^{tt,\phi \phi }(1/z,1/\bar z)
\end{array} \right)~,
\end{align}
and for MFT we have,
\begin{align}
&\langle \phi t\phi t\rangle  = \frac{{\left( {{v_1} \cdot {v_3}} \right){{\left( {{v_2} \cdot {v_4}} \right)}^2}}}{{x_{13}^{2{\Delta _\phi }}x_{24}^{2{\Delta _t}}}} = \frac{{\left( {{v_1} \cdot {v_3}} \right){{\left( {{v_2} \cdot {v_4}} \right)}^2}}}{{x_{12}^{{\Delta _\phi } + {\Delta _t}}x_{34}^{{\Delta _\phi } + {\Delta _t}}}}\frac{{x_{14}^{{\Delta _t} - {\Delta _\phi }}x_{24}^{{\Delta _t} - {\Delta _\phi }}}}{{x_{13}^{{\Delta _t} - {\Delta _\phi }}x_{14}^{{\Delta _t} - {\Delta _\phi }}}}{z^{\frac{{{\Delta _\phi } + {\Delta _t}}}{2}}}{{\bar z}^{\frac{{{\Delta _\phi } + {\Delta _t}}}{2}}}\nonumber\\
&= \frac{1}{{x_{12}^{{\Delta _\phi } + {\Delta _t}}x_{34}^{{\Delta _\phi } + {\Delta _t}}}}\frac{{x_{14}^{{\Delta _t} - {\Delta _\phi }}x_{24}^{{\Delta _t} - {\Delta _\phi }}}}{{x_{13}^{{\Delta _t} - {\Delta _\phi }}x_{14}^{{\Delta _t} - {\Delta _\phi }}}}\left( {\begin{array}{*{20}{l}}
{\left( {\left( {{v_1} \cdot v} \right){{\left( {{v_2} \cdot v} \right)}^2},\left( {{v_3} \cdot v} \right){{\left( {{v_4} \cdot v} \right)}^2}} \right){z^{\frac{{{\Delta _\phi } + {\Delta _t}}}{2}}}{{\bar z}^{\frac{{{\Delta _\phi } + {\Delta _t}}}{2}}}}\\
{ + \left( {{v_1} \cdot {v_2}} \right)\left( {{v_3} \cdot {v_4}} \right)\left( {{v_2} \cdot {v_4}} \right){z^{\frac{{{\Delta _\phi } + {\Delta _t}}}{2}}}{{\bar z}^{\frac{{{\Delta _\phi } + {\Delta _t}}}{2}}}}
\end{array}} \right)~.
\end{align}
\item $\left\langle {t\phi s\phi } \right\rangle $: We have the crossing equation
\begin{align}
&{\mathbb{S}_{t\phi s\phi }}\sum\limits_{O = 1} {} {\hat{f}_{t\phi O}}{\hat{f}_{s\phi O}}g_{O}^{t\phi ,s\phi }(z,\bar z)\nonumber\\
&= {\mathbb{T}_{t\phi s\phi }}\sum\limits_{O' = 1} {} {\hat{f}_{s\phi O'}}{\hat{f}_{t\phi O'}}g_{O'}^{s\phi ,t\phi }(1 - z,1 - \bar z)\nonumber\\
&= {\mathbb{U}_{t\phi s\phi }}\sum\limits_{O' = 2} {} {\hat{f}_{\phi \phi O'}}{\hat{f}_{tsO'}}g_{O'}^{\phi \phi ,ts}(1/z,1/\bar z)~.
\end{align}
This correlator vanishes in MFT.
\end{itemize}

\subsection{Crossing equations involving $\chi$}
Here we summarize the crossing relations we will use involving $\chi$. 

\begin{itemize}
\item $\langle \phi \chi \phi \chi \rangle $: We have the crossing equation
\begin{align}
&{\mathbb{S}_{\phi \chi \phi \chi }}\sum\limits_{O = 4} {{}{\hat{f}_{\phi \chi O}}{\hat{f}_{\phi \chi O}}g_{O}^{\phi \chi ,\phi \chi }(z,\bar z)} \nonumber\\
&= {\mathbb{T}_{\phi \chi \phi \chi }}\sum\limits_{O' = 2} {{}{\hat{f}_{\phi \chi O'}}{\hat{f}_{\phi \chi O'}}g_{O'}^{\phi \chi ,\phi \chi }(1 - z,1 - \bar z)} \nonumber\\
&= {\mathbb{U}_{\phi \chi \phi \chi }}\sum\limits_{O' = {0^ + }} {{}{\hat{f}_{\chi \chi O'}}{\hat{f}_{\phi \phi O'}}g_{O'}^{\chi \chi ,\phi \phi }(1/z,1/\bar z)} \nonumber\\
&+ {\mathbb{U}_{\phi \chi \phi \chi }}\sum\limits_{O' = {0^ - }} {{}{\hat{f}_{\chi \chi O'}}{\hat{f}_{\phi \phi O'}}g_{O'}^{\chi \chi ,\phi \phi }(1/z,1/\bar z)} ~,
\end{align}
\begin{align}
&{\mathbb{S}_{\phi \chi \phi \chi }}\sum\limits_{O = 2} {{}{\hat{f}_{\phi \chi O}}{\hat{f}_{\phi \chi O}}g_{O}^{\phi \chi ,\phi \chi }(z,\bar z)}  \nonumber\\
&= {\mathbb{T}_{\phi \chi \phi \chi }}\sum\limits_{O' = 4} {{}{\hat{f}_{\phi \chi O'}}{\hat{f}_{\phi \chi O'}}g_{O'}^{\phi \chi ,\phi \chi }(1 - z,1 - \bar z)}  \nonumber\\
&= {\mathbb{U}_{\phi \chi \phi \chi }}\sum\limits_{O' = {0^ + }} {{}{\hat{f}_{\chi \chi O'}}{\hat{f}_{\phi \phi O'}}g_{O'}^{\chi \chi ,\phi \phi }(1/z,1/\bar z)}  \nonumber\\
&- {\mathbb{U}_{\phi \chi \phi \chi }}\sum\limits_{O' = {0^ - }} {{}{\hat{f}_{\chi \chi O'}}{\hat{f}_{\phi \phi O'}}g_{O'}^{\chi \chi ,\phi \phi }(1/z,1/\bar z)} ~,
\end{align}
and for MFT we have
\begin{align}
&\langle \phi \chi \phi \chi \rangle  = \frac{{\left( {{v_1} \cdot {v_3}} \right){{\left( {{v_2} \cdot {v_4}} \right)}^3}}}{{x_{12}^{{\Delta _\phi } + {\Delta _\chi }}x_{34}^{{\Delta _\phi } + {\Delta _\chi }}}}\frac{{x_{14}^{{\Delta _\chi } - {\Delta _\phi }}x_{24}^{{\Delta _\chi } - {\Delta _\phi }}}}{{x_{13}^{{\Delta _\chi } - {\Delta _\phi }}x_{14}^{{\Delta _\chi } - {\Delta _\phi }}}}{z^{\frac{{{\Delta _\phi } + {\Delta _\chi }}}{2}}}{{\bar z}^{\frac{{{\Delta _\phi } + {\Delta _\chi }}}{2}}}\nonumber\\
&= \frac{1}{{x_{12}^{{\Delta _\phi } + {\Delta _\chi }}x_{34}^{{\Delta _\phi } + {\Delta _\chi }}}}\frac{{x_{14}^{{\Delta _\chi } - {\Delta _\phi }}x_{24}^{{\Delta _\chi } - {\Delta _\phi }}}}{{x_{13}^{{\Delta _\chi } - {\Delta _\phi }}x_{14}^{{\Delta _\chi } - {\Delta _\phi }}}}\left( {\begin{array}{*{20}{c}}
{\left( {\left( {{v_1} \cdot v} \right){{\left( {{v_2} \cdot v} \right)}^3},\left( {{v_3} \cdot v} \right){{\left( {{v_4} \cdot v} \right)}^3}} \right){z^{\frac{{{\Delta _\phi } + {\Delta _\chi }}}{2}}}{{\bar z}^{\frac{{{\Delta _\phi } + {\Delta _\chi }}}{2}}}}\\
{ + \left( {{v_1} \cdot {v_2}} \right)\left( {{v_3} \cdot {v_4}} \right){{\left( {{v_2} \cdot {v_4}} \right)}^2}{z^{\frac{{{\Delta _\phi } + {\Delta _\chi }}}{2}}}{{\bar z}^{\frac{{{\Delta _\phi } + {\Delta _\chi }}}{2}}}}
\end{array}} \right)~.
\end{align}
\item $\left\langle  \phi \chi t t \right\rangle$: We have the crossing equation
\begin{align}
&{\mathbb{S}_{\phi \chi tt}}\sum\limits_{O = 4} {{}{\hat{f}_{\phi \chi O}}{\hat{f}_{ttO}}g_{O}^{\phi \chi ,tt}(z,\bar z)} \nonumber\\
&= {\mathbb{T}_{\phi \chi tt}}\sum\limits_{O' = 1} {{}{\hat{f}_{t\chi O'}}{\hat{f}_{\phi tO'}}g_{O'}^{t\chi ,\phi t}(1 - z,1 - \bar z)} {\text{ }}\nonumber\\
&= {\mathbb{U}_{\phi \chi tt}}\sum\limits_{O' = 1} {{}{\hat{f}_{t\chi O'}}{\hat{f}_{\phi tO'}}g_{O'}^{t\chi ,\phi t}(1/z,1/\bar z)} ~.
\end{align}
\item $\left\langle {\chi s\phi t} \right\rangle$: We have the crossing equation
\begin{align}
&{\mathbb{S}_{\chi s\phi t}}\sum\limits_{O = 3} {{}{\hat{f}_{\chi sO}}{\hat{f}_{\phi tO}}g_{O}^{\chi s,\phi t}(z,\bar z)} \nonumber\\
&= {\mathbb{T}_{\chi s\phi t}}\sum\limits_{O' = 1} {{}{\hat{f}_{\phi sO'}}{\hat{f}_{\chi tO'}}g_{O'}^{\phi s,\chi t}(1 - z,1 - \bar z)} {\text{ }}\nonumber\\
&= {\mathbb{U}_{\chi s\phi t}}\sum\limits_{O' = 2} {{}{\hat{f}_{tsO'}}{\hat{f}_{\chi \phi O'}}g_{O'}^{ts,\chi \phi }(1/z,1/\bar z)} ~.
\end{align}
\end{itemize}

\subsection{Crossing matrices}
\label{app:Crossing_Matrices}
With the above results, taking into account sign factors from $f_{ijk}=(-1)^{\ell_k}f_{jik}$, we can write down the crossing matrices used in the O(2) model inversion formulas such as Eq.~(\ref{eq:Cffff}):
\begin{align}
&\mathcal{M}^{\f\f\f\f}=\left(
\begin{array}{ccc}
 \frac{1}{2} & -\frac{1}{2} & \frac{1}{2} \\
 -\frac{1}{2} & \frac{1}{2} & \frac{1}{2} \\
 1 & 1 & 0 \\
\end{array}
\right)~, \quad
&\mathcal{M}^{\f\f tt}=\left(
\begin{array}{cc}
 \frac{1}{2} & \frac{1}{2} \\
 -\frac{1}{2} & \frac{1}{2} \\
\end{array}
\right)~,
\quad  
& \mathcal{M}^{\phi t \phi t}=\left(
\begin{array}{cc}
 0 & 1 \\
 1 & 0 \\
\end{array}
\right)~, \nonumber
\\
&\mathcal{M}^{\phi\chi\phi\chi}=\left(
\begin{array}{cc}
 0 & 1 \\
 1 & 0 \\
\end{array}
\right)~,
\quad
&\mathcal{M}^{\phi\chi\chi\phi}=\left(
\begin{array}{cc}
 1 & -1 \\
 1 & 1 \\
\end{array}
\right)~,
\quad
&\mathcal{M}^{tttt}=\left(
\begin{array}{ccc}
 \frac{1}{2} & -\frac{1}{2} & \frac{1}{2} \\
 -\frac{1}{2} & \frac{1}{2} & \frac{1}{2} \\
 1 & 1 & 0 \\
\end{array}
\right)~.
\end{align}
To explain the notation we have:
\begin{eqnarray*}
&\mathcal{M}^{\f\f\f\f}_{r_s,r_t} \quad &\text{with} \quad  r_s,r_t=\{0^+,0^-,2\}~,
\\
&\mathcal{M}^{\f\f tt}_{r_s,r_t} \quad &\text{with} \quad  r_s=\{0^+,0^-,2\}, \ r_t=\{1,3\}~,
\\
&\mathcal{M}^{\f t\f t}_{r_s,r_t} \quad &\text{with} \quad  r_s,r_t=\{1,3\}~,
\\
&\mathcal{M}^{\f \chi\f \chi}_{r_s,r_t} \quad &\text{with} \quad  r_s,r_t=\{2,4\}~,
\\
&\mathcal{M}^{\f \chi\chi\f }_{r_s,r_t} \quad &\text{with} \quad  r_s=\{2,4\}, r_t=\{0^+,0^-\}~,
\\
&\mathcal{M}^{tttt}_{r_s,r_t}  \quad &\text{with} \quad  r_s,r_t=\{0^+,0^-,4\}~.
\end{eqnarray*} 
All the other crossing matrices we need are either related by symmetry or trivial:
\begin{align}
\cM^{\f\f st}=\cM^{\phi s\f t}=\cM^{\f\f\f\chi}=\cM^{st\f\chi}=\cM^{st\chi\f}=\cM^{tt\chi\phi}=\cM^{t sst}=\cM^{tsts}=1~.
\end{align}
One can notice that the crossing matrices are exactly the same for $\left\langle { \phi  \phi \phi \phi} \right\rangle$ and $\left\langle {tttt} \right\rangle$. In fact, in the O(2) model for scalar operator $O_q$ with charge $q$ ($q\in \mathbb{Z}_{\ge 1}$), the corresponding matrices are always the same for four-point functions of identical operators, in the basis of charge $\{0^+, 0^-, 2q\}$ representations and the index-free notation defined above.

\section{Conformal block expansions in three dimensions}\label{expansion}
Three-dimensional conformal blocks are relatively inconvenient to perform inversion directly on. Thus we consider decomposing three-dimensional conformal blocks down to two-dimensional blocks. In this paper, we will mainly use the following two methods, which we call the $\text{SL}_2$ expansion and dimensional reduction. 
\subsection{$\text{SL}_2$ expansion}
The idea of the $\text{SL}_2$ expansion is basically the following (see for instance \cite{Simmons-Duffin:2016wlq}). One can expand three-dimensional conformal blocks in the small $z$ limit,
\begin{align}
g_{h,\bar{h}}^{r,s}(z,\bar z) = \sum\limits_{n = 0}^\infty  {\sum\limits_{j =  - n}^n {A_{n,j}^{r,s}} } (h,\bar h){z^{h + n}}k_{\bar h + j}^{r,s}(\bar z)~.
\end{align}
Here, we use $r=\Delta_{12}/2$, $s=\Delta_{34}/2$, $h=(\Delta-\ell)/2$ and $\bar{h}=(\Delta+\ell)/2$ to parametrize the conformal block $g$. Finally $k$ is the $\text{SL}_2$ block
\begin{align}
k^{r,s}_{h}(z)=z^h{}_2F_1(h-r,h+s,2h,z)
\end{align}

The coefficient $A$ can be derived from the quadratic Casimir equation
\begin{align}
{{\cal D}_2}g_{\Delta,\ell}^{r,s}(z,\bar z) = {C_{h,\bar h}}g_{\Delta,\ell}^{r,s}(z,\bar z)~,
\end{align}
where
\begin{align}
&{C_{h,\bar h}} = h(h -2) + \bar h(\bar h - 1)\;,\nonumber\\
&{D_2} = {D_z} + {D_{\bar z}} + \frac{{z\bar z}}{{z - \bar z}}\left( {(1 - z){\partial _z} - (1 - \bar z){\partial _{\bar z}}} \right)\;,\nonumber\\
&{D_z} = {z^2}(1 - z)\partial _z^2 - ( - r + s + 1){z^2}{\partial _z} + rsz\;.
\end{align}
One can determine the solution iteratively starting from $A_{0,0}=1$ and performing the expansion around $0<z \ll \bar{z}\ll 1$. For generic external scalar operators, the formulas for $A$ don't have known closed forms. Some examples are:
\begin{align}
&A_{0,0}^{r,s}(h,\bar h) = 1~,\nonumber\\
&A_{1, - 1}^{r,s}(h,\bar h) = \frac{{(\bar h - h)}}{{2\bar h - 2h - 1}}~,\nonumber\\
&A_{1,0}^{r,s}(h,\bar h) = \frac{{s + r - h}}{2} - \frac{{rs\left( {2{{\bar h}^2} - 2\bar h - h + 1} \right)}}{{2(\bar h - 1)\bar h(2h - 1)}} - r + h~,\nonumber\\
&A_{1,1}^{r,s}(h,\bar h) = \frac{{(h + \bar h - 1)(\bar h - r)(\bar h + r)(\bar h - s)(\bar h + s)}}{{4{{\bar h}^2}(2\bar h - 1)(2\bar h + 1)(2h + 2\bar h - 1)}}~.
\end{align}
The above coefficients are mostly applied in the context of the correlation function in the $t$-channel. For expansions in the Weyl-reflected block, we will instead use
\begin{align}
\left( {\frac{{\bar z - z}}{{z\bar z}}} \right)\frac{1}{z}g_{2-h,\bar{h}}^{{h_{12}},{h_{34}}}(z,\bar z) = \sum\limits_{m = 0}^\infty  {{z^{ - h + m}}} \sum\limits_{j =  - m}^m {\mathcal{C}_{m,j}^{{h_{12}},{h_{34}}}} (h,\bar h)k_{\bar h + j}^{{h_{12}},{h_{34}}}(\bar z)~.
\end{align}
The coefficients $\mathcal{C}$ can be derived completely in a similar way. We again list a few examples:
\begin{align}
&\mathcal{C}_{0,0}^{r,s}(h,\bar h) = 1~,\nonumber\\
&\mathcal{C}_{1, - 1}^{r,s}(h,\bar h) =  - \frac{{h + \bar h - 3}}{{2h + 2\bar h - 5}}~,\nonumber\\
&\mathcal{C}_{1,0}^{r,s}(h,\bar h) = \nonumber\\
&\frac{{ - r\left( {s\left( {h - 2{{\bar h}^2} + 2\bar h - 2} \right) + (2h - 3)(\bar h - 1)\bar h} \right) + (2h - 3)(\bar h - 1)\bar h(s - h + 1)}}{{2(2h - 3)(\bar h - 1)\bar h}}~,\nonumber\\
&\mathcal{C}_{1,1}^{r,s}(h,\bar h) =  - \frac{{\left( {{{\bar h}^2} - {r^2}} \right)\left( {{{\bar h}^2} - {s^2}} \right)(h - \bar h - 2)}}{{4{{\bar h}^2}\left( {4{{\bar h}^2} - 1} \right)(2h - 2\bar h - 3)}}~.
\end{align}

\subsection{Dimensional reduction}
Here we describe an alternative expansion that has been used in \cite{Albayrak:2019gnz,Hogervorst:2016hal}. We expand the three-dimensional conformal block as
\begin{align}
g_{h,\bar{h}}^{r,s}(z,\bar z) = \sum\limits_{p = 0}^\infty  {\sum\limits_{q =  - p}^p {{\cal A}_{p,q}^{r,s}} } (h,\bar h)k_{h + p}^{r,s}(z)k_{\bar h + q}^{r,s}(\bar z)~.
\end{align}
This expansion amounts to reducing a three-dimensional block to a sum of chiral two-dimensional blocks. One can derive the coefficients $\mathcal{A}$ in a very similar way. To simplify the recursion relations we make use of the inner product formula of the $\text{SL}_2$ blocks
\begin{align}
\left( {k_{h + p}^{r,s},k_{h + q}^{r,s}} \right) = \oint_{{C_0}} {\frac{{dz}}{{2\pi i}}} \frac{1}{{{z^2}}}k_{1 - (h + p))}^{ - r, - s}(z)k_{h + q}^{r,s}(z) = {\delta _{p,q}}~,
\end{align}
where $C_0$ is a contour encircling $z=0$. Since we now have an orthogonality relation for the $\SL _2$ blocks, it is natural to express all the objects produced in the Casimir equation in terms of them. Specifically, we will use the following expansions:
\begin{align}
&{z^q}k_h^{r,s}(z) = \sum\limits_{n = 0}^\infty  {W_{q,n}^{r,s}} (h)k_{h + n + q}^{r,s}(z)~,\nonumber\\
&{z^{q + 1}}{\partial _z}k_h^{r,s}(z) = \sum\limits_{n = 0}^\infty  {Y_{q,n}^{r,s}} (h)k_{h +n + q}^{r,s}(z)~,
\end{align}
where
\begin{align}
&W_{q,n}^{r,s}(h) =\nonumber\\
& \frac{{{{( - r + h)}_n}{{(s + h)}_n}}}{{\Gamma (n + 1){{(2h)}_n}}}{ \times _4}{F_3}\left( {\begin{array}{*{20}{c}}
{ - 2h - n + 1, - n,r - h - n - q + 1, - s - h - n - q + 1}\\
{r - h - n + 1, - s - h - n + 1, - 2h - 2q + 2}
\end{array};1} \right)~,\nonumber\\
&Y_{q,n}^{r,s}(h) = hW_{q,n}^{r,s} + \frac{{( - r + h)(s + h)}}{{2h}}Z_{q,n}^{r,s}(h)~,\nonumber\\
&Z_{q,n}^{r,s} =
\sum\limits_{m = 0}^{n - 1} {\frac{{{{( - r + h + 1)}_{ - m + n - 1}}{{(s + h + 1)}_{ - m + n - 1}}{{(r - h - n - q + 1)}_m}{{( - s - h - n - q + 1)}_m}}}{{m!\Gamma (n - m){{(2h + 1)}_{ - m + n - 1}}{{( - 2(h + n + q - 1))}_m}}}} ~.
\end{align}
Based on these formulas, one can expand all terms in the quadratic Casimir equation in terms of the $\SL _2$ hypergeometrics. Then one can derive a recursion relation by using the orthogonality property of these functions given above.

We can perform this calculation easily by computer algebra. The first few leading terms of $\mathcal{A}$ are:
\begin{align}
&{\cal A}_{0,0}^{r,s}(h,\bar h) = 1~,\nonumber\\
&{\cal A}_{1, - 1}^{r,s}(h,\bar h) = \frac{{h - \bar h}}{{ - 2\bar h + 2h + 1}}~,\nonumber\\
&{\cal A}_{1,0}^{r,s}(h,\bar h) =  - \frac{{rs\left( {\left( {\bar h - 1} \right)\bar h - {h^2} + h} \right)}}{{2h(2h - 1)\left( {\bar h - 1} \right)\bar h}}~,\nonumber\\
&{\cal A}_{1,1}^{r,s}(h,\bar h) = \frac{{\left( {\bar h + h - 1} \right)\left( {{{\bar h}^2} - {r^2}} \right)\left( {{{\bar h}^2} - {s^2}} \right)}}{{4{{\bar h}^2}\left( {2\bar h - 1} \right)\left( {2\bar h + 1} \right)\left( {2\bar h + 2h - 1} \right)}}~.
\end{align}

In \cite{Hogervorst:2016hal}, a closed-form formula for $\mathcal{A}$ was discovered for equal external operators $r=s=0$. While we have not found a closed form expression for $\mathcal{A}$ for general $r$ and $s$, we are able to use the recursion relation to go to very high orders in this expansion.
\subsection{Connections}
Those two expansions are connected to each other by the following. Using the defining sum for the hypergeometric function we have
\begin{align}
k_{h}^{r,s}(z) = {z^h}\sum\limits_{\beta  = 0}^\infty  {\frac{{{{(h - r)}_\beta }{{(h + s)}_\beta }}}{{{{(2h)}_\beta }}}} \frac{{{z^\beta }}}{{\beta !}}~,
\end{align}
we can easily derive a relation between $A$ and $\mathcal{A}$, 
\begin{align}
&A_{n,q}^{r,s}(h,\bar h) = \sum\limits_{p = |q|}^n {V_{n,p}^{r,s}} (h){\cal A}_{p,q}^{r,s}(h,\bar h)~,\nonumber\\
&{\cal A}_{p,q}^{r,s}(h,\bar h) = \sum\limits_{n = |q|}^p {Q_{p,n}^{r,s}} (h)A_{n,q}^{r,s}(h,\bar h)~.
\end{align}
The coefficient $V$ is directly related to the Taylor expansion of the dimensional reduction formula in $\bar{z}$,
\begin{align}
V_{n, p}^{r, s}(h)=\frac{(h+p-r)_{n-p}(h+p+s)_{n-p}}{(2 h+2 p)_{n-p}(n-p) !}~.
\end{align}
We can also obtain $Q$ by inverting $V$. We list the first few leading coefficients:
\begin{align}
&V_{0,0}^{r,s} = V_{1,1}^{r,s} = V_{2,2}^{r,s} = 1~,~~~~~Q_{0,0}^{r,s}{\text{ }} = Q_{1,1}^{r,s} = Q_{2,2}^{r,s} = 1~,\nonumber\\
&V_{1,0}^{r,s} = \frac{{(h - r)(h + s)}}{{2h}}~,~~~~~Q_{1,0}^{r,s}{\text{ }} =  - \frac{{(h - r)(h + s)}}{{2h}}~.
\end{align}

\section{Some comparisons of computational performance}
In this part, we will include comparisons between different methods. This will include a discussion about different expansions, a discussion about choosing different values of $z_0$, the effect of mixing in the twist Hamiltonian, and a discussion about double-twist sums.

\subsection{Comparison using different expansions}\label{performance}
\begin{figure}[H]
  \centering
  \includegraphics[width=0.8\textwidth]{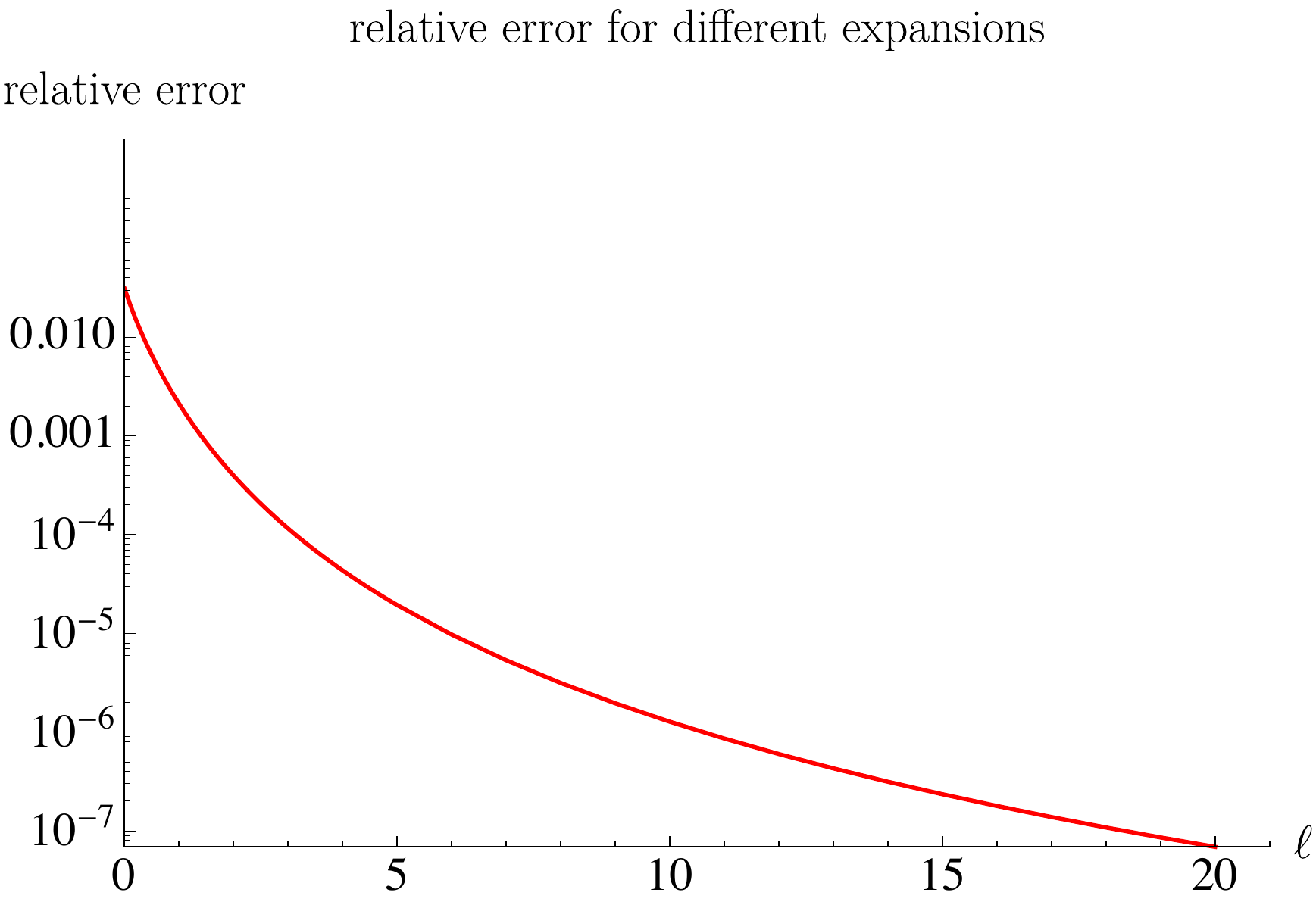}
  \caption{\label{expancompare} A comparison between two different expansions in charge 3 even-spin sector. We evaluate the relative error by $\left| {\frac{{({\text{dimensional reduction results for }}{\tau _{[\phi t]_0^3}}) - ({\text{S}}{{\text{L}}_2}{\text{ results for }}{\tau _{[\phi t]_0^3}})}}{{({\text{dimensional reduction results for }}{\tau _{[\phi t]_0^3}})}}} \right|$.}
\end{figure}

Here, we show a typical example of the comparison between two different expansions: the $\text{SL}_2$ expansion and dimensional reduction. In figure \ref{expancompare}, we compare the predictions for the leading twist charge 3 operators of even spin using the $\text{SL}_2$ expansion and dimensional reduction for fixed $z_0=0.1$ and inverting isolated operators only. We include both zeroth and first order and we plot the relative error between the expansions. This analysis shows that even at very low orders, generically, we expect a very small difference between two expansions at a small $z_0$. For higher spins, the error is relatively smaller. Computationally, the $\text{SL}_2$ expansion is cheaper when performing the inversion formula, since generically evaluating $_3 F_2$ is easier than evaluating $_4 F_3$. On the other hand, dimensional reduction captures more information about descendants, so it contains a higher amount of non-perturbative information and has higher complexity. In most parts of this paper, we will use the dimensional reduction method. 

\subsection{Comparison using different values of $z_0$}\label{z0}
\begin{figure}
  \centering
  \includegraphics[width=0.8\textwidth]{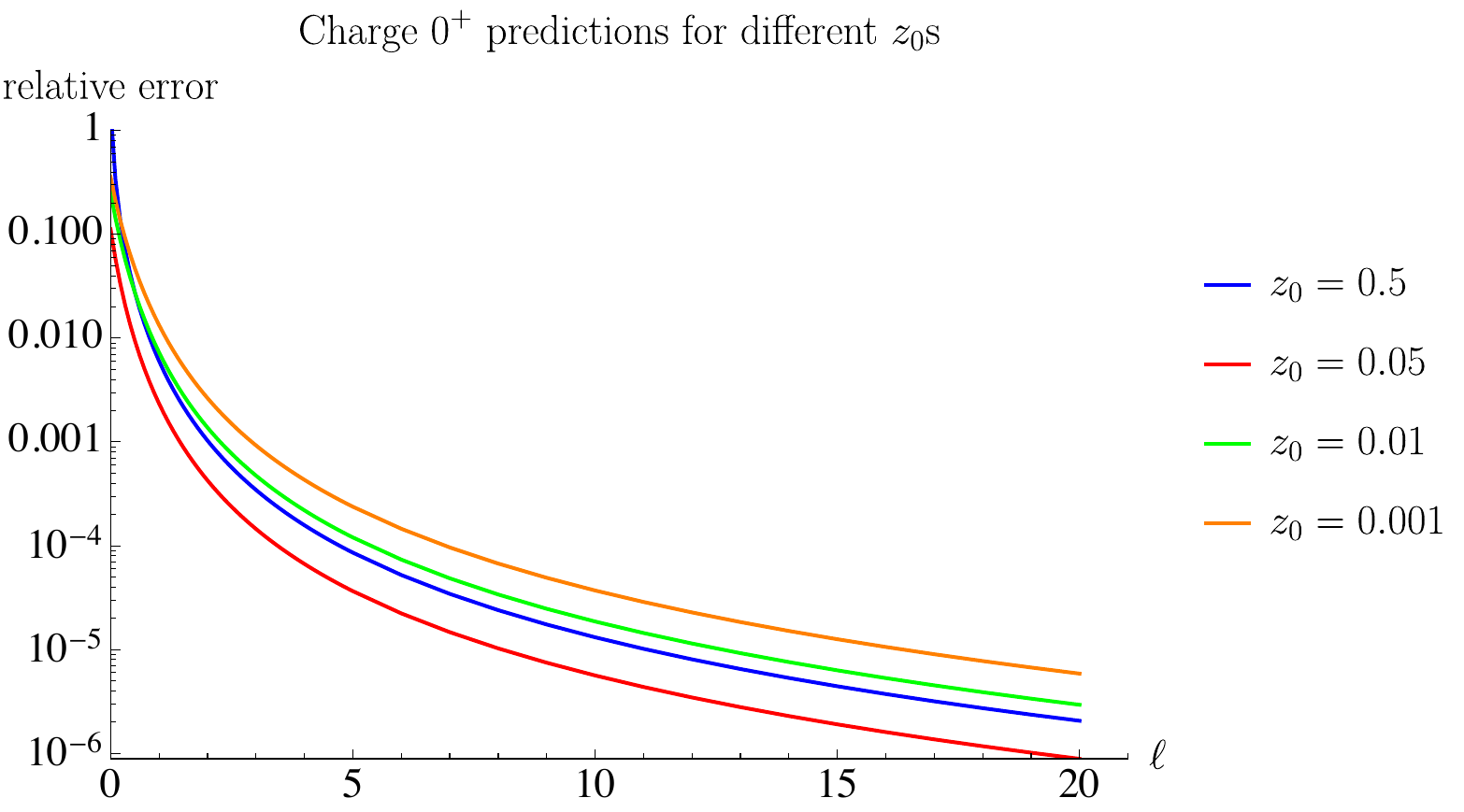}
  \caption{\label{z0pic} A comparison between using different values of $z_0$ in the charge $0^+$ sector. We evaluate the relative error by $\left| {\frac{{({z_0} = \# {\text{ results for }}{\tau _{[\phi \phi ]_0^{{0^ + }}}}) - ({z_0} = 0.1{\text{ results for }}{\tau _{[\phi \phi ]_0^{{0^ + }}}})}}{{({z_0} = 0.1{\text{ results for }}{\tau _{[\phi \phi ]_0^{{0^ + }}}})}}} \right|$.}
\end{figure}

\begin{figure}
  \centering
  \includegraphics[width=0.8\textwidth]{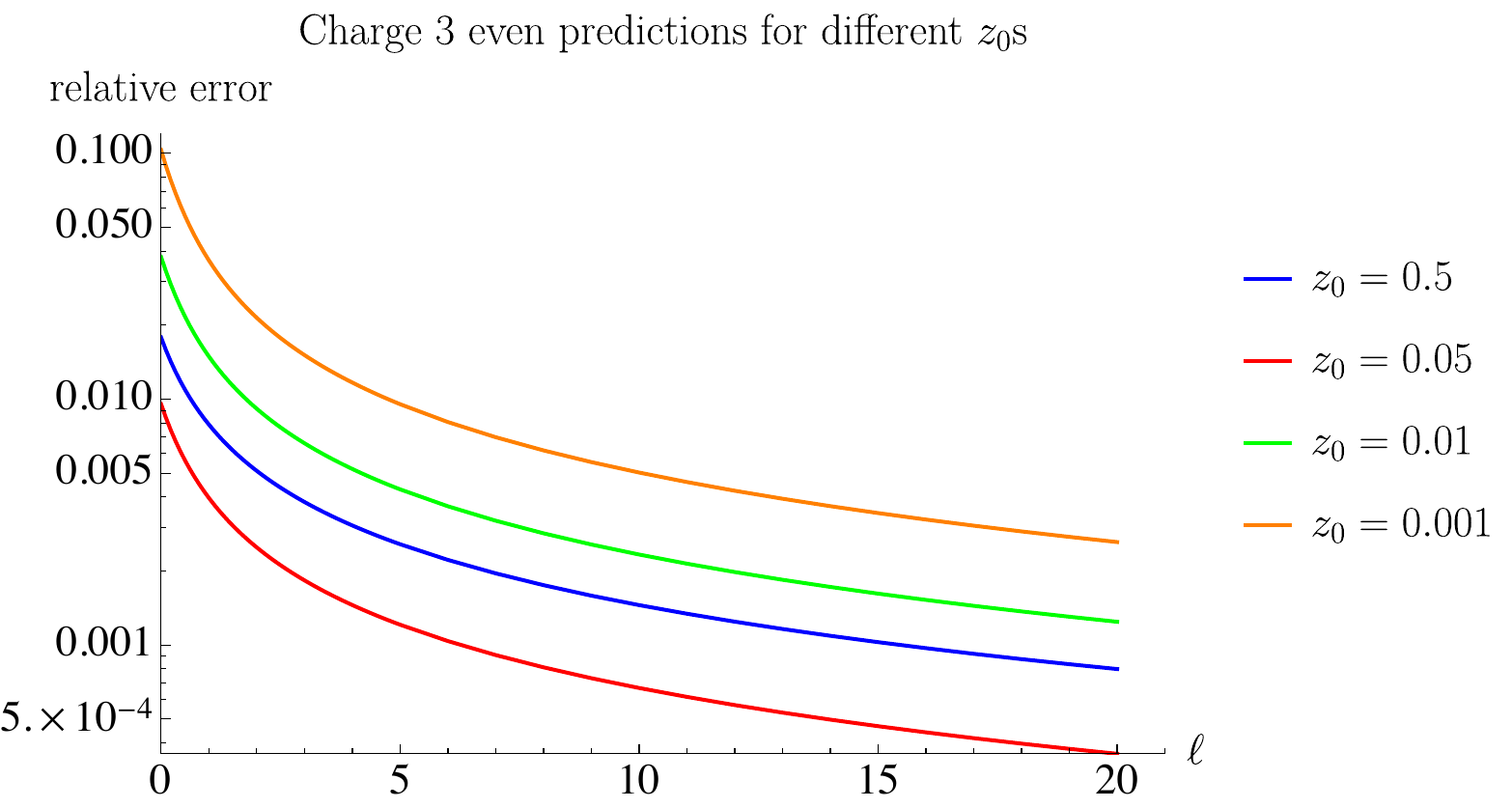}
  \caption{\label{z0pic2} A comparison between using different values of $z_0$ in the charge 3 even-spin sector. We evaluate the relative error by $\left| {\frac{{({z_0} = \# {\text{ results for }}{\tau _{[\phi t]_0^3}}) - ({z_0} = 0.1{\text{ results for }}{\tau _{[\phi t]_0^3}})}}{{({z_0} = 0.1{\text{ results for }}{\tau _{[\phi t]_0^3}})}}} \right|$. }
\end{figure}

Here we show some examples of how the analytic predictions for anomalous dimensions depend on the choice of $z_0$. In figures \ref{z0pic} and \ref{z0pic2}, we computed the relative error between different values of $z_0$ in the charge $0^+$ and charge 3 even sectors. We expect that $z_0=0.1$ is a reasonably good value for all predictions used in this paper, but the question remains if there is a universal principle to determine the optimal value of $z_0$. We know that there are large logs when we take $z_0\to 0$ while for large $z_0$ we have to consider heavier operators in the $s$-channel. In general, the optimal choice of $z_0$ might be different for different CFTs, different sectors, and even different spins. In our model, figures \ref{z0pic} and \ref{z0pic2} show that the result is relatively stable against some ranges of $z_0$s, but we think it will be important to find a general algorithm to determine $z_0$.

\newpage

\subsection{Effects of the twist Hamiltonian}\label{twistcompare}
\begin{figure}[H]
  \centering
  \includegraphics[width=0.8\textwidth]{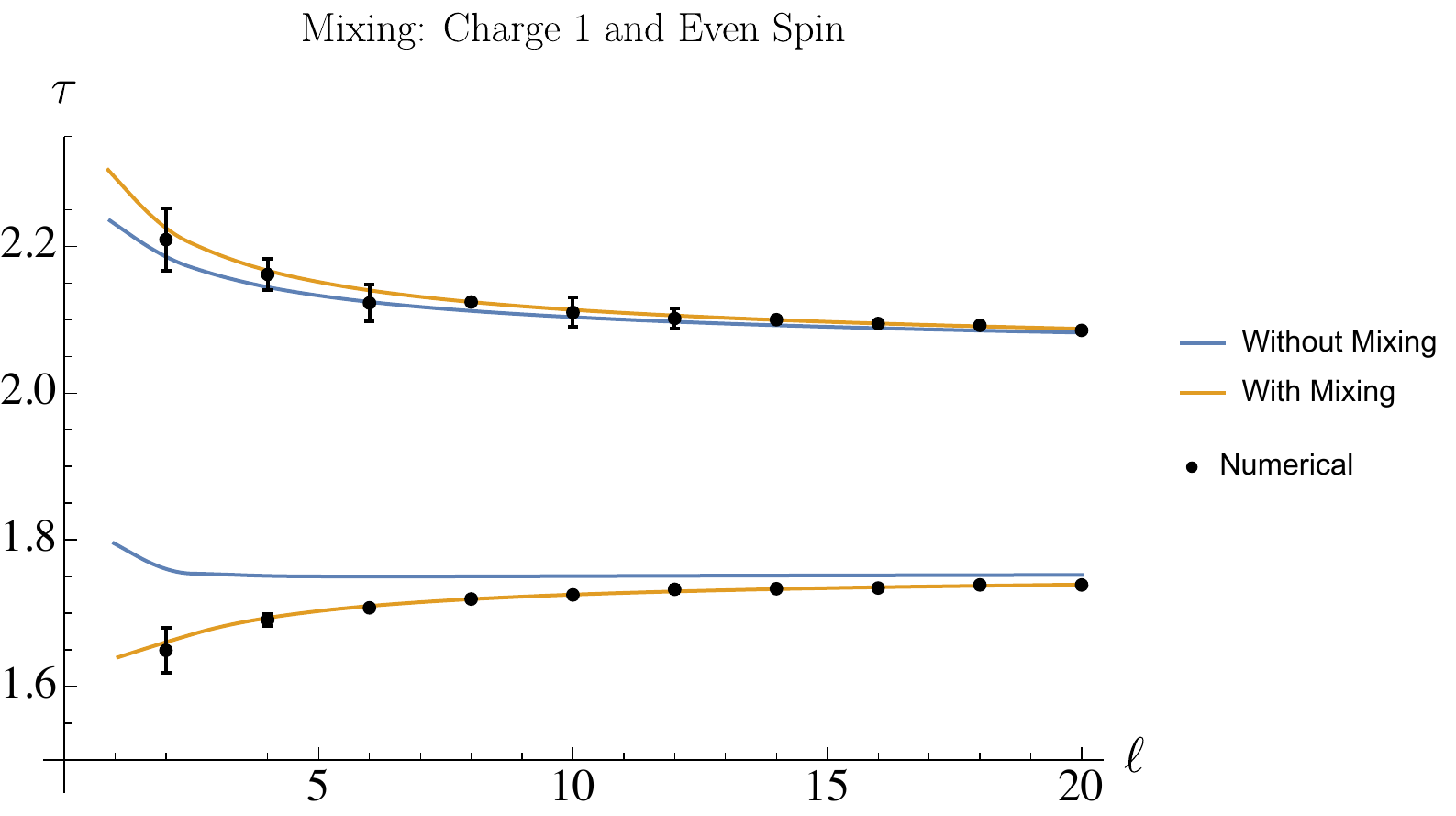}
  \caption{\label{twistcomparefig} A comparison between using or not using the twist Hamiltonian for the leading trajectories in the charge 1 even-spin sector.}
\end{figure}

\begin{figure}[H]
  \centering
  \includegraphics[width=0.8\textwidth]{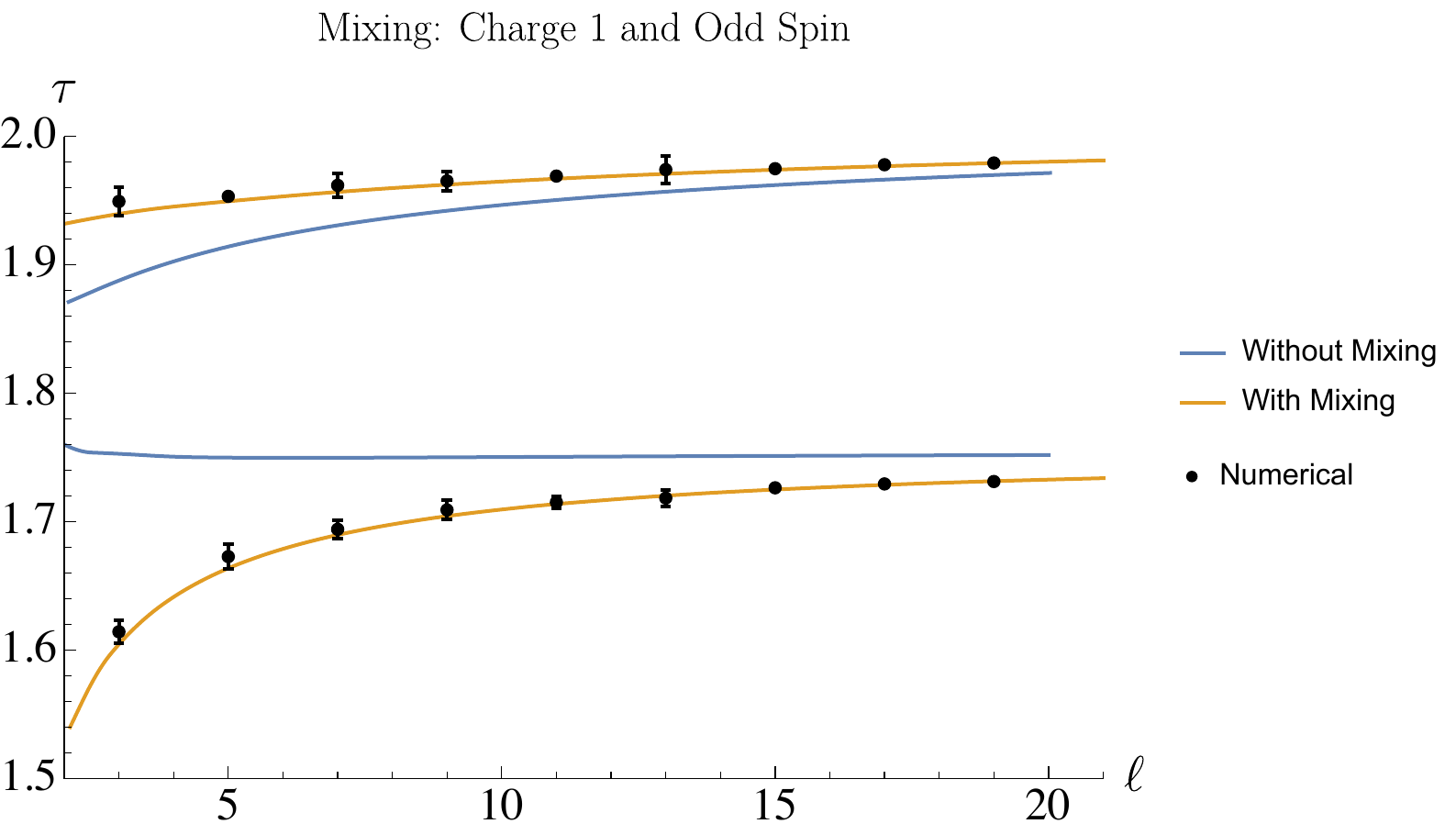}
  \caption{\label{twistcomparefig2} A comparison between using or not using the twist Hamiltonian for the leading trajectories in the charge 1 odd-spin sector.}
\end{figure}

In this section, we will show that using the twist Hamiltonian is important for obtaining accurate results in certain sectors. In figures \ref{twistcomparefig} and \ref{twistcomparefig2}, we show the effects of including or not including mixing in the charge 1 sector. Specifically, we compare the results when we diagonalize the twist Hamiltonian with the results when we ignore mixing by just studying the diagonal elements of $M_1(z,\bar{h})$. This results show clearly that it is important to resolve the mixing effects, particularly at low spins, and therefore one should use the twist Hamiltonian approach.

\subsection{DTI versus non-DTI}\label{DTIcompare}
\begin{figure}[H]
  \centering
  \includegraphics[width=0.8\textwidth]{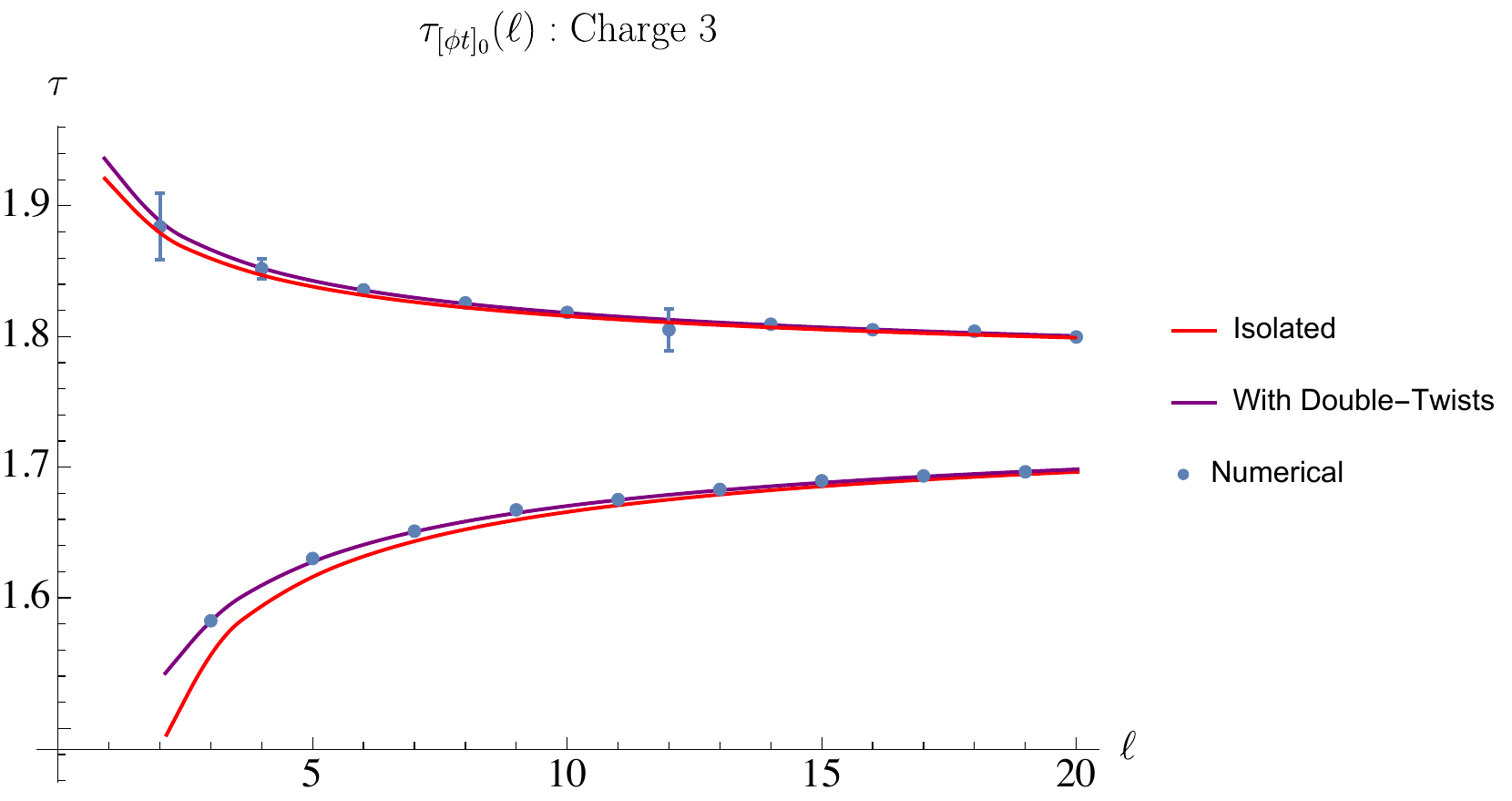}
  \caption{\label{dticompare1} The effects of DTIs for leading double twists in the charge $3$ $[\phi t]_0^3$ sector. The top curve corresponds to even spin and the lower curve corresponds to odd spin.}
\end{figure}

\begin{figure}[H]
  \centering
  \includegraphics[width=0.8\textwidth]{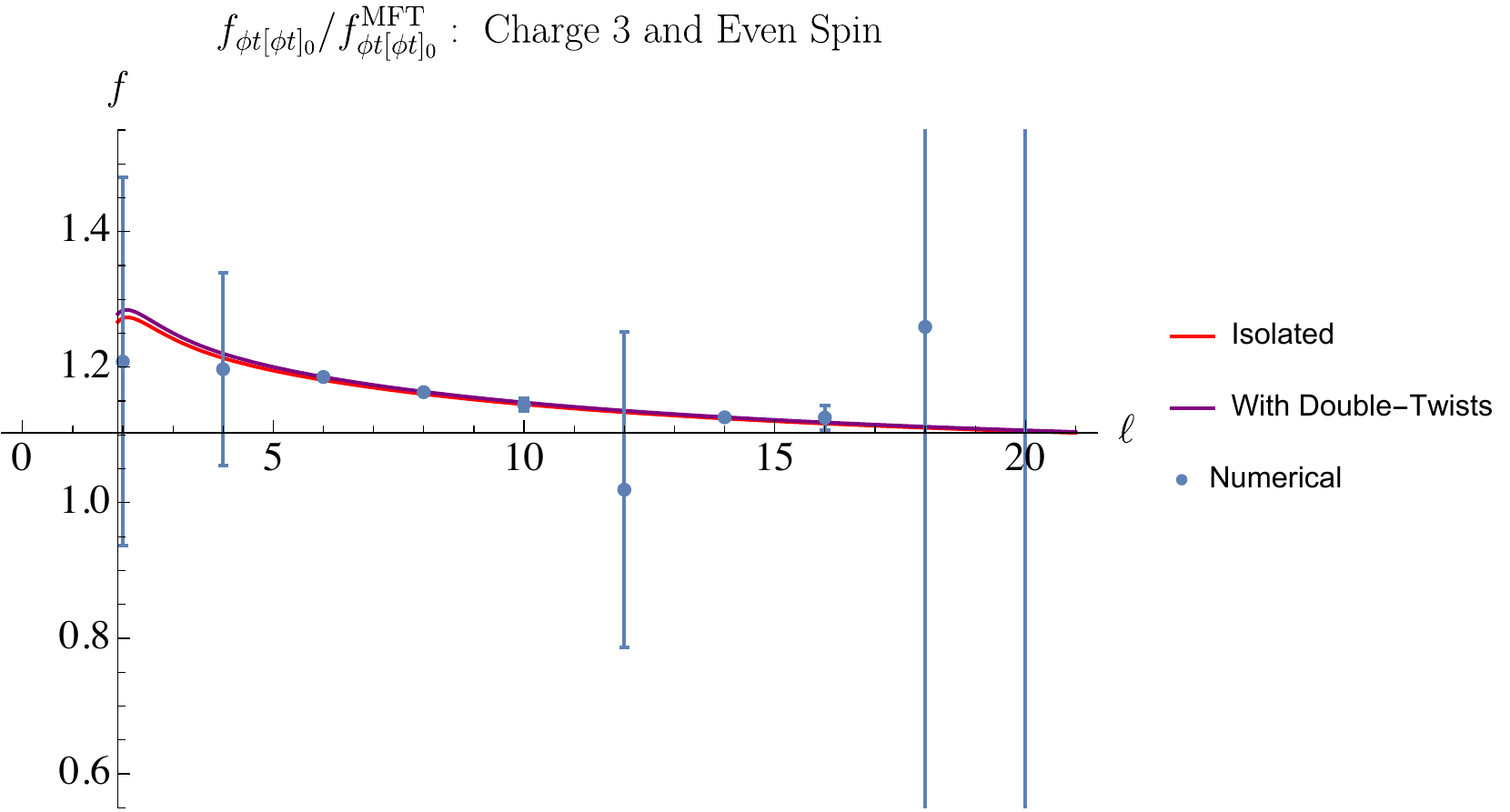}
  \caption{\label{dticompare21} The effects of DTIs for OPEs of leading double twists in the charge $3$ $[\phi t]_0^3$ even-spin sector.}
\end{figure}

\begin{figure}[H]
  \centering
  \includegraphics[width=0.8\textwidth]{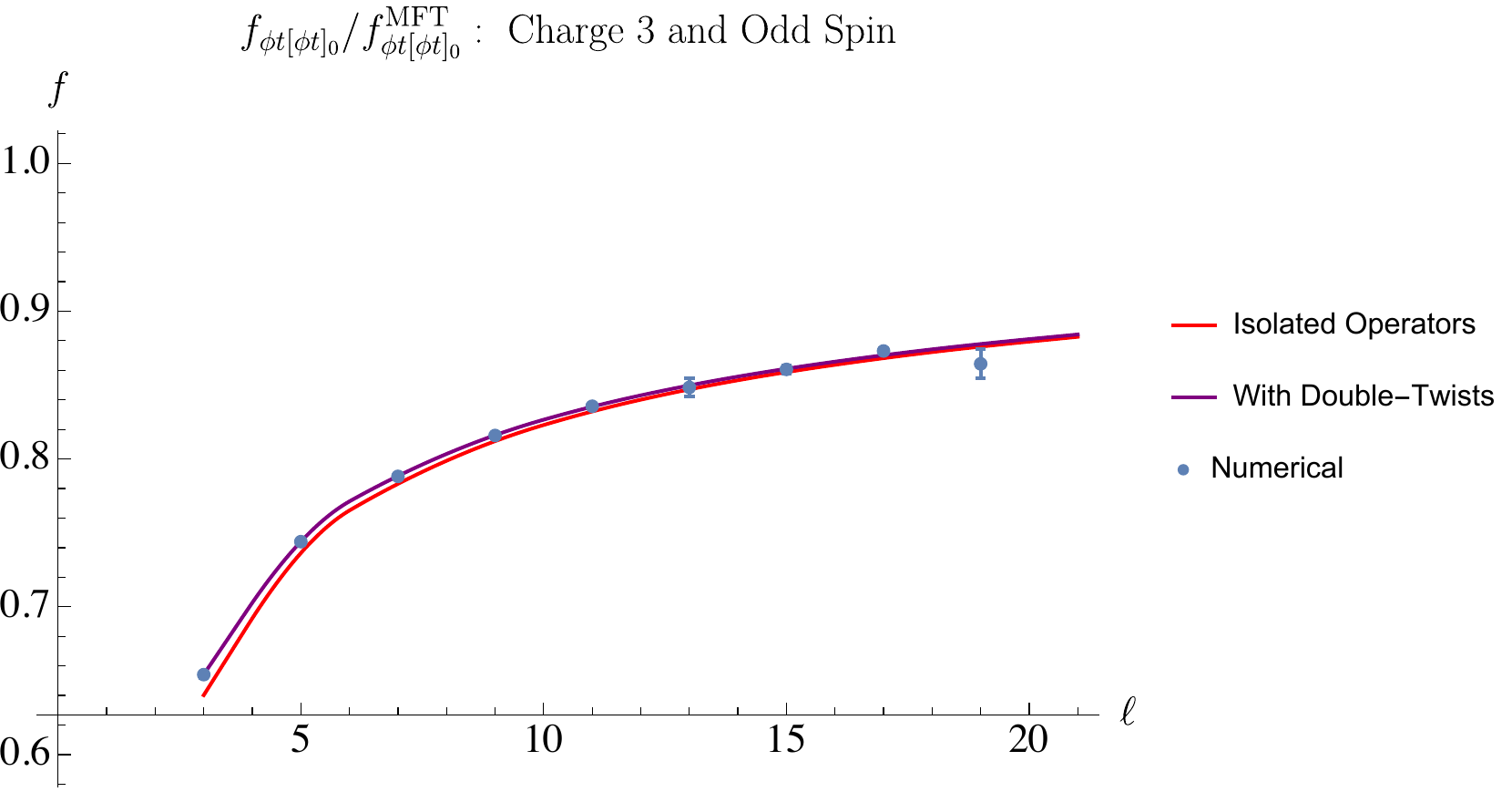}
  \caption{\label{dticompare22} The effects of DTIs for OPEs of leading double twists in the charge $3$ $[\phi t]_0^3$ odd-spin sector.}
\end{figure}

\begin{figure}[H]
  \centering
  \includegraphics[width=0.8\textwidth]{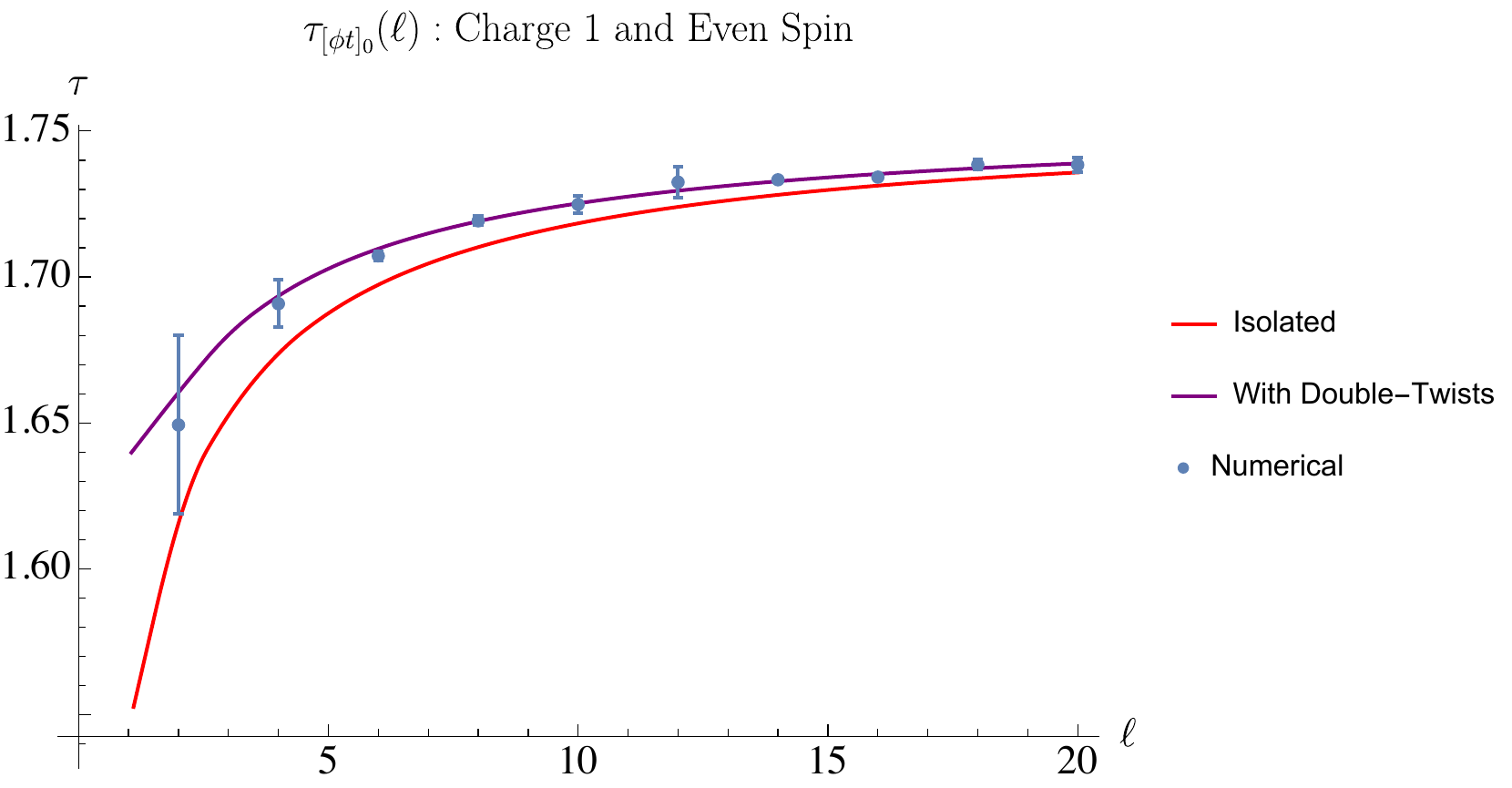}
  \caption{\label{dticompare31} The effects of DTIs for low-lying double twists in the charge $1$ $[\phi t]_0^1$ even-spin sector.}
\end{figure}

\begin{figure}[H]
  \centering
  \includegraphics[width=0.8\textwidth]{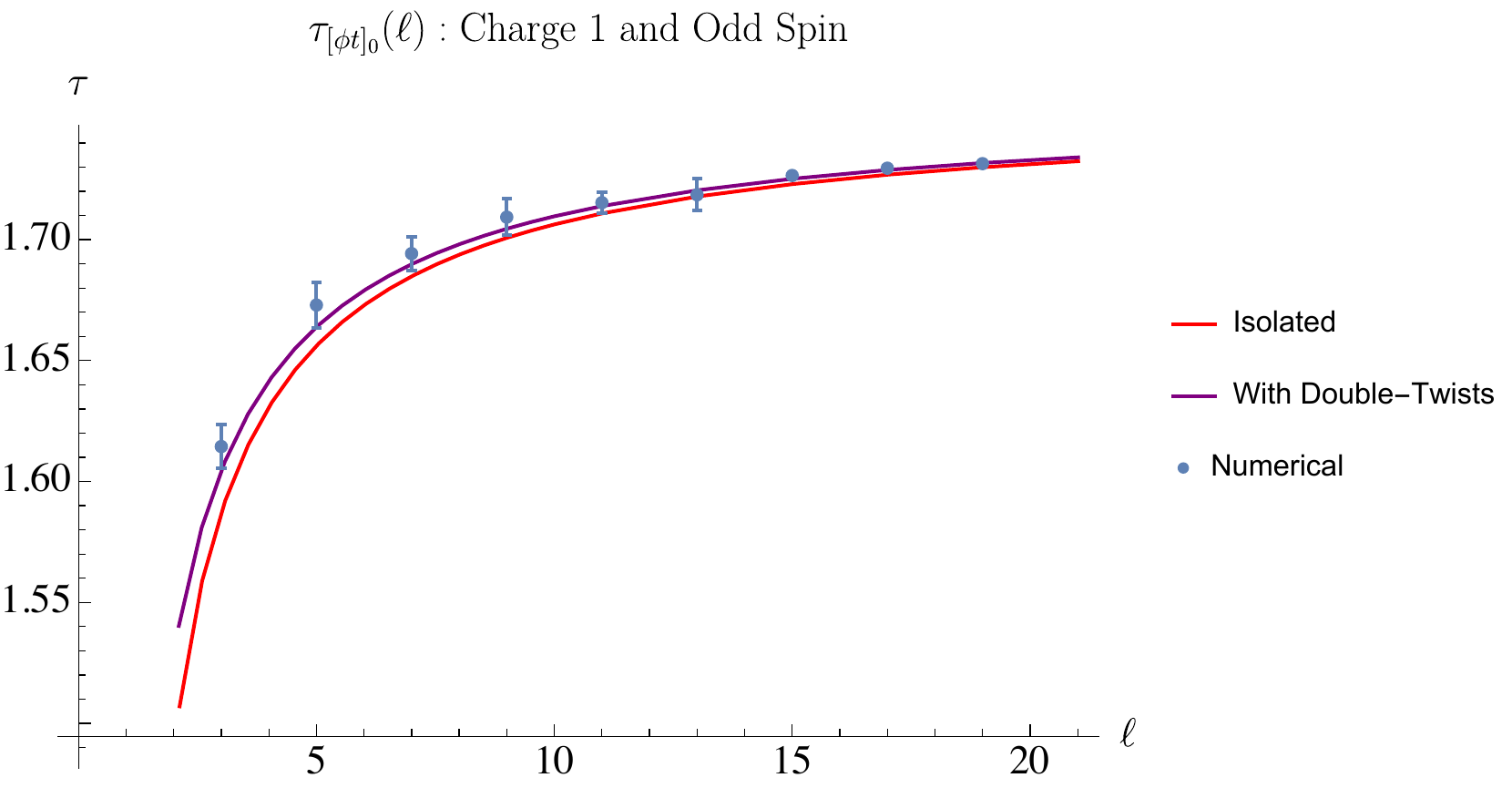}
  \caption{\label{dticompare32} The effects of DTIs for low-lying double twists in the charge $1$ odd-spin sector.}
\end{figure}

\begin{figure}[H]
  \centering
  \includegraphics[width=0.8\textwidth]{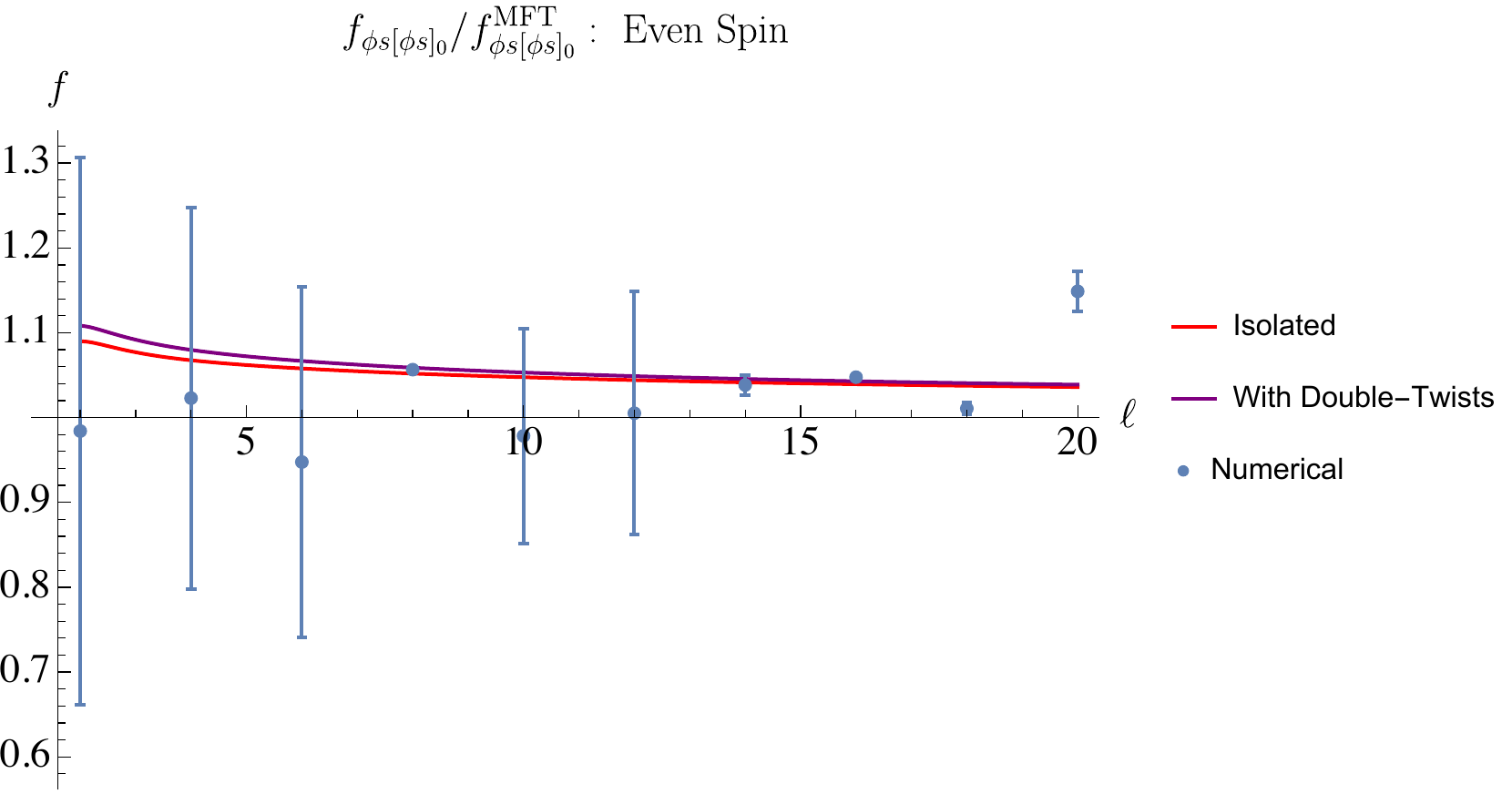}
  \caption{\label{dticompare41} The effects of DTIs for OPEs of different low-lying double twists in the charge $1$ $[\phi s]_0^1$ even-spin sector.}
\end{figure}

\begin{figure}[H]
  \centering
  \includegraphics[width=0.8\textwidth]{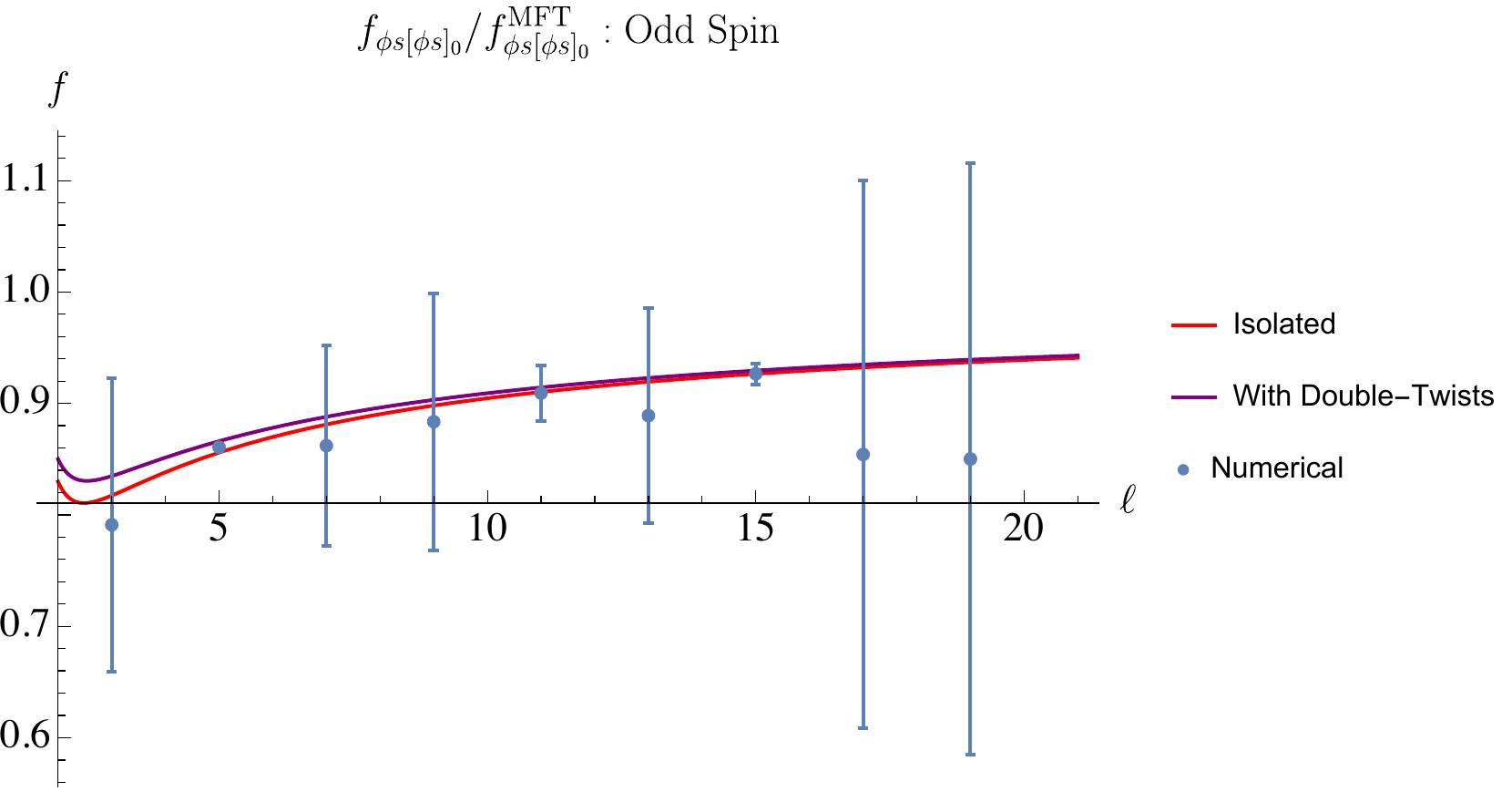}
  \caption{\label{dticompare42} The effects of DTIs for OPEs of different low-lying double twists in the charge $1$ $[\phi s]_0^1$ odd-spin sector.}
\end{figure}

\begin{figure}[H]
  \centering
  \includegraphics[width=0.8\textwidth]{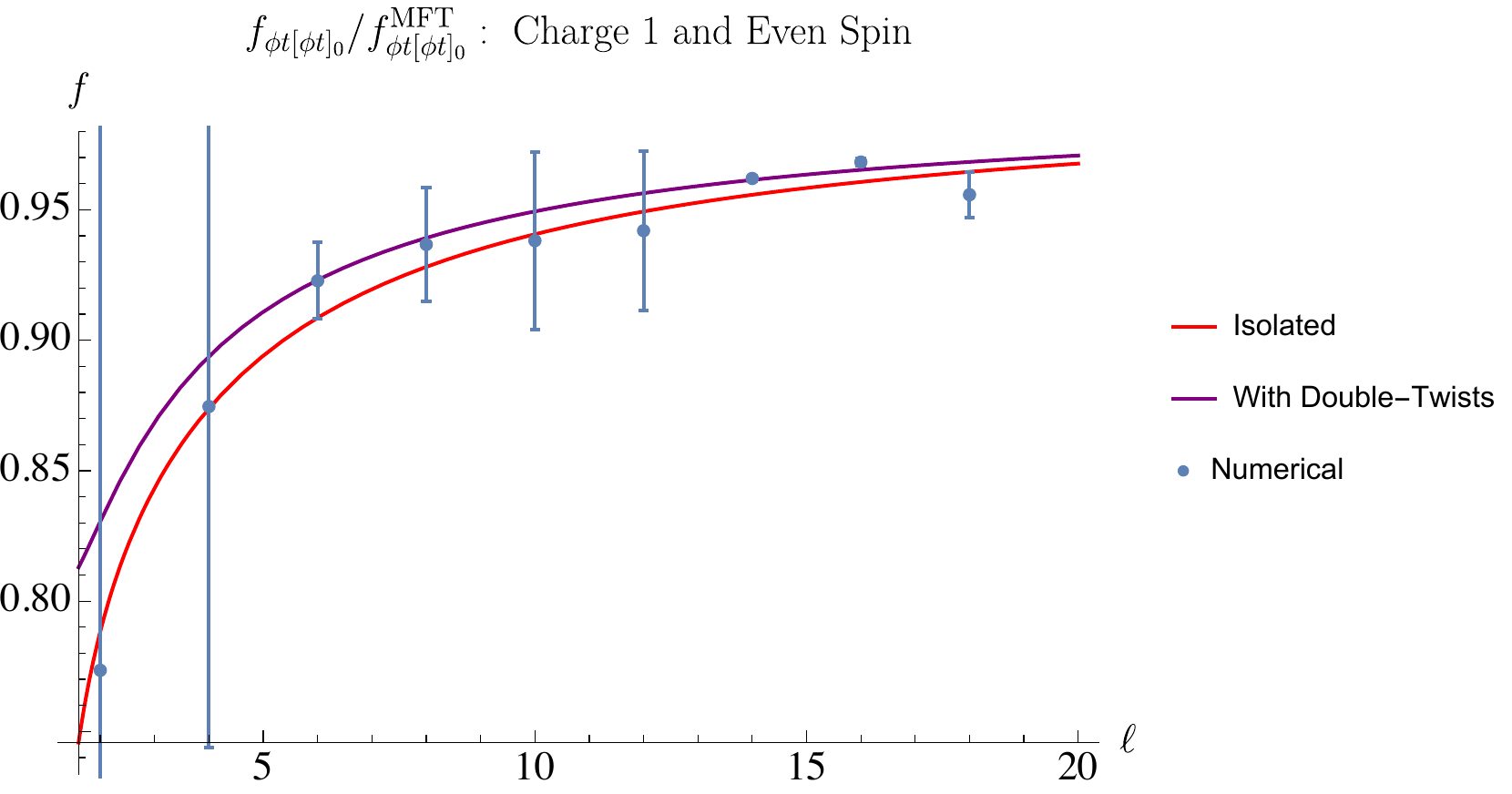}
  \caption{\label{dticompare43} The effects of DTIs for OPEs of different low-lying double twists in the charge $1$ $[\phi t]_0^1$ even-spin sector.}
\end{figure}

\begin{figure}[H]
  \centering
  \includegraphics[width=0.8\textwidth]{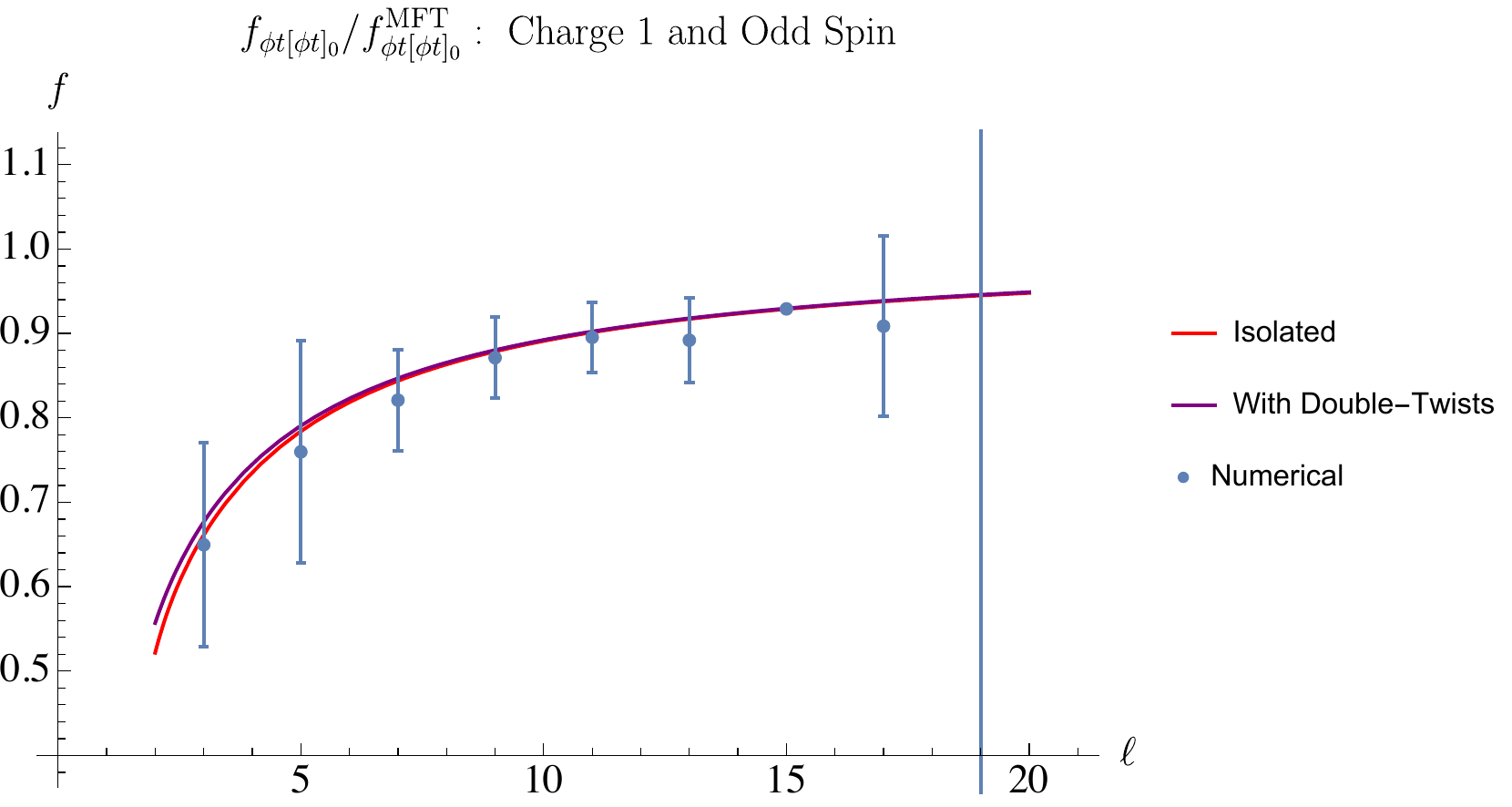}
  \caption{\label{dticompare44} The effects of DTIs for OPEs of different low-lying double twists in the charge $1$ $[\phi t]_0^1$ odd-spin sector.}
\end{figure}

In this section we compare the results that include or exclude the infinite sum of double-twist operators, see figures \ref{dticompare1}, \ref{dticompare21}, \ref{dticompare22}, \ref{dticompare31}, \ref{dticompare32}, \ref{dticompare41}, \ref{dticompare42}, \ref{dticompare43} and \ref{dticompare44}. We see that the inclusion of double-twist operators typically yields very small effects for the OPE coefficients, but in some cases can improve the accuracy for the spectrum. This can be seen most clearly for the charge 3 odd-spin operators, see the lower curve of figure \ref{dticompare1} and the charge 1 even-spin operators, see figures \ref{dticompare31} and \ref{dticompare32}.

\section{Conformal block conventions}\label{block}
Here we give a summary of the conventions related to conformal block expansions that we will use in this paper. 

One can consider the three-dimensional conformal block in the following limit,
\begin{align}
g_{\Delta ,\ell }^{{\Delta _{12}},{\Delta _{34}}}(z,\bar z)\mathop \sim \limits_{z,\bar z \to 0} {\mathcal{N}_{d = 3,\ell }}{(z\bar z)^{\frac{\Delta }{2}}}{\text{Ge}}{{\text{g}}_\ell }\left( {\frac{{z + \bar z}}{{2\sqrt {z\bar z} }}} \right)~,
\end{align}
where we take the dimension $d=3$. Here $\operatorname{Geg}_{\ell}(x)$ is the Gegenbauer polynomial. Now in $d=3$ we have
\begin{align}
{{\mathop{\text{Geg}}\nolimits} _\ell}(x)  = {P_\ell}(x)~,
\end{align}
where $P$ is the Legendre polynomial. The definition of the coefficient $\mathcal{N}$ will fix the normalization of the conformal blocks. Different convention choices are summarized in table 1 of \cite{Poland:2018epd}. 

In the analytic computations of this paper, we will use the convention in \cite{Simmons-Duffin:2016wlq}. The definition of conformal blocks in \cite{Simmons-Duffin:2016wlq} demands the behavior $ z^h\bar{z}^{\bar{h}} $ in the limit $0<z \ll \bar{z} \ll1$, which fixes the coefficient $\mathcal{N}$ to be
\begin{align}
\mathcal{N}_{d = 3,\ell }^{{\text{lightcone}}} = \frac{{\ell !}}{{{{(1/2)}_\ell }}} = \frac{{\ell !\Gamma (1/2)}}{{\Gamma (1/2 + \ell )}}~.
\end{align}

In the numerical computations we have performed, we used a different convention. The convention of our code follows the convention of the \texttt{Mathematica} code published in \cite{blocks}, which uses the following convention for $\mathcal{N}$. Let us consider the coordinates for the radial expansion
\begin{align}
\rho=\frac{z}{(1-\sqrt{1-z})^{2}}~~,~~~~\quad z=\frac{4 \rho}{(1+\rho)^{2}}~,~~~~~\rho=r e^{i\theta}~,~~~~~\cos\theta=\eta~,
\end{align}
for $z$ and similarly for $\bar{z}$. One can then expand
\begin{align}
g_{\Delta, \ell}^{\Delta_{12}, \Delta_{34}}(u, v)=r^{\Delta} \sum_{m=0}^{\infty} r^{m} \sum_{j} w(m, j) \operatorname{Geg}_{j}(\eta)~.
\end{align}
In the small $z$ limit, we take $\rho=\frac{1}{4}z$, so we get
\begin{align}
g_{\Delta, \ell}^{\Delta_{12}, \Delta_{34}}(r, \eta) \underset{r \rightarrow 0}{\sim} \mathcal{N}_{d, \ell}(4 r)^{\Delta} \operatorname{Geg}_{\ell}(\eta)~,
\end{align}
which matches the behavior defined by $\mathcal{N}$ as 
\begin{align}
w(0, \ell)=\mathcal{N}_{d=3, \ell} 4^{\Delta}~.
\end{align}
The \texttt{Mathematica} code uses the convention
\begin{align}
w(0,\ell ) = 1
\end{align}
and hence
\begin{align}
{\cal N}_{d = 3,\ell }^{{\text{code}}} = \frac{{\ell !}}{{{4^\Delta }{{(1)}_\ell }}} = {4^{ - \Delta }} ~.
\end{align}
It is then different from the lightcone convention \cite{Simmons-Duffin:2016wlq} by the factor
\begin{align}
\frac{{\mathcal{N}_{d = 3,\ell }^{{\text{lightcone}}}}}{{\mathcal{N}_{d = 3,\ell }^{{\text{code}}}}} = \frac{{{{( - 1)}^\ell }\ell !\Gamma (1/2){4^\Delta }}}{{\Gamma (1/2 + \ell )}}~.
\end{align}

\section{Technical details about $\text{SL}_2$ sums}\label{sum}
Here we briefly summarize some $\text{SL}_2$ sums which are used in the inversion formula. These are a generalization of the sum given in (\ref{eq:sum2dBlocks}) and were derived in \cite{Simmons-Duffin:2016wlq}.
\subsection{Asymptotic form for $\Omega$}
From (\ref{eq:inversionOmega}) we see all the non-trivial $\bar{h}$ dependence comes from the function $\Omega$. To compare with previous results in the lightcone bootstrap, we then need the large $\bar{h}$ expansion of this function. We find that the relation is:
\begin{align}
\kappa _{2{h_5}}\Omega _{{h_5},{h_6},{h_2} + {h_3}}^{1234}\sim \sum\limits_{m = 0}^\infty  {\eta _{56}^{1234}} (m)S_{{h_6} - {h_1} - {h_3} + m}^{{h_{12}},{h_{34}}}\left( {{h_5}} \right)~,
\end{align}
where
\begin{align}
&\eta _{56}^{1234}(m) = \frac{{{{( - 1)}^m}\Gamma \left( {2{h_6}} \right)}}{{2m!\Gamma \left( {2{h_6} + m} \right)\Gamma \left( { - {h_1} + {h_4} + {h_6}} \right)\Gamma \left( {{h_2} - {h_3} + {h_6}} \right)}}\nonumber\\
&\times \frac{{\Gamma \left( { - {h_1} + {h_4} + {h_6} + m} \right)\Gamma \left( {{h_2} - {h_3} + {h_6} + m} \right)}}{{\sin \left( {\pi \left( {{h_1} + {h_4} - {h_6}} \right)} \right)\sin \left( {\pi \left( {{h_2} + {h_3} - {h_6}} \right)} \right)}}~,
\\
&S^{r,s}_{a}(h)\equiv \frac{\Gamma(h-r)\Gamma(h-s)\Gamma(h-a-1)}{\Gamma(-a-r)\Gamma(-a-s)\Gamma(2h-1)\Gamma(h+a+1)}~.
\end{align}
In all cases, we will only need the $m=0$ term.

\subsection{General $\text{SL}_2$ sums}
\label{app:SL2Sums}
Here we will summarize some of the $\text{SL}_2$ sums used to calculate the contributions of double-twist operators. For more details on their derivation see \cite{Simmons-Duffin:2016wlq}.

For general correlation functions of external scalars, $\<\f_1\f_2\f_3\f_4\>$ we have:
\begin{align}
&\sum\limits_{\substack{h = {h_0} + \ell \\ \ell  = 0,1, \ldots} } {S_a^{r,s}(h){z^{ - r}}k_h^{ r,s}(1 - z)}  = {y^a} + \sum\limits_{k = 0}^\infty  {\left( {C_{a,k}^{r,s}\mathcal{B}_{a,r-k-1}(h_0) {y^{k - r}}+ (s \leftrightarrow r)} \right)}~, \label{eq:SL2SumGen}\nonumber\\
&C_{a,k}^{r,s} =\frac{\pi}{\sin(\pi(s-r))}\frac{\Gamma(-a)^{2}}{\Gamma(-a-r)\Gamma(-a-s)}\frac{\Gamma(k+1-r)^{2}}{\Gamma(k+1+s-r)k!}~,
\end{align}
where $y=\frac{z}{1-z}$. By taking the limit $s\rightarrow r$ and $r\rightarrow 0$ we recover the sum given in (\ref{eq:sum2dBlocks}). To perform sums involving $\lim\limits_{\epsilon\rightarrow 0}\Gamma(-\epsilon)^2S_{\epsilon}(h)$ when $r=s=r$ or $\lim\limits_{\epsilon\rightarrow 0}\Gamma(-\epsilon)S_{\epsilon}(h)$ for non-zero $r$ and $s$ we first perform the sum and then take $\epsilon\rightarrow 0$. This will produce $\log(z)$ terms which correspond to expanding $z^{h+\delta h}$ in $\delta h$.

We also need to consider sums where the signs are alternating:
\begin{align}
&\sum\limits_{\substack{h = {h_0} + \ell \\ \ell  = 0,1, \ldots} } {(-1)^\ell S_a^{r,s}(h){z^{ - r}}k_h^{ r,s}(1 - z)} =  \sum\limits_{k = 0}^\infty  {\left( {J_{a,k}^{r,s}\mathcal{B}^-_{a,r-k-1}(h_0) {y^{k - r}}+ (s \leftrightarrow r)} \right)} ~,\label{eq:SL2Sumalt}
\\
&J_{a,k}^{r,s} =\frac{\pi  \Gamma (-a)^2 \Gamma (k-r+1)^2 }{\sin (\pi  (s-r))\Gamma (k+1) \Gamma (-a-r) \Gamma (-a-s) \Gamma (k-r+s+1)}~, 
\\
&\mathcal{B}^{-}_{a,b}(h_0)=\sum\limits_{k=0}^{K}t_{a,b}(k)\mathcal{B}^{-}_{a+b+k+1,-1}(h_0)
\nonumber \\
&\hspace{.62in}\sum\limits_{\substack{h=h_0+\ell \\ \ell=0,1,...}}(-1)^{\ell}(1-2h)\left(T_{a}(h)T_{b}(h)-\sum\limits_{k=0}^{K}t_{a,b}(k)T_{a+b+k+1}(h)\right)~,\label{eq:sumSLaltB}
\\
&\mathcal{B}^{-}_{a,-1}(h_0)=-(h_0+a)T_{a}(h_0)~.
\end{align}
To compute $\mathcal{B}^{-}_{a,b}(h_0)$ for general $a$ and $b$ we take $K>-a-b-5/2$ so the sum over $\ell$ in (\ref{eq:sumSLaltB}) converges and can be computed numerically. To perform sums that involve operators of only even or odd spin we add or subtract the sums (\ref{eq:SL2SumGen}) and (\ref{eq:SL2Sumalt}).

\bibliographystyle{utphys}
\bibliography{Biblio}
\end{document}